# Machine Learning and Data Analytics for Design and Manufacturing of High-Entropy Materials Exhibiting Mechanical or Fatigue Properties of Interest[1]

| | |
|---|---|
| Baldur Steingrimsson | Portland State University and Imagars LLC |
| Xuesong Fan | University of Tennessee, Knoxville |
| Anand Kulkarni | Siemens Corporation, Corporate Technology |
| Michael C. Gao | National Energy Transportation Laboratory |
| Peter K. Liaw | University of Tennessee, Knoxville |

This chapter presents an innovative framework for the application of machine learning and data analytics for the identification of alloys or composites exhibiting certain desired properties of interest. The main focus is on alloys and composites with large composition spaces for structural materials. Such alloys or composites are referred to as high-entropy materials (HEMs) and are here presented primarily in context of structural applications. For each output property of interest, the corresponding driving (input) factors are identified. These input factors may include the material composition, heat treatment, manufacturing process, microstructure, temperature, strain rate, environment, or testing mode. The framework assumes the selection of an optimization technique suitable for the application at hand and the data available. Physics-based models are presented, such as for predicting the ultimate tensile strength (UTS) or fatigue resistance. We devise models capable of accounting for physics-based dependencies. We factor such dependencies into the models as *a priori* information. In case that an artificial neural network (ANN) is deemed suitable for the applications at hand, it is suggested to employ custom kernel functions consistent with the underlying physics, for the purpose of attaining tighter coupling, better prediction, and for extracting the most out of the - usually limited - input data available.

---

[1] The chapter is dedicated to the loving memory of Dr. Steingrimur Baldursson, Professor Emeritus at the University of Iceland (1930 – 2020). In 1958, prior to returning back to Iceland to take on a position within the Faculty of Sciences at the University of Iceland, Dr. Baldursson completed a doctorate from the University of Chicago, under guidance of Prof. Joseph E. Mayer and Prof. Maria G. Mayer, a 1963 Nobel Laureate in Physics.

## Table of Contents

## 1. Introduction

Artificial intelligence (AI) is defined here as the use of computers to mimic the cognitive functions of humans. When machines carry out tasks based on algorithms in an "intelligent" manner, it is referred to as AI. Machine learning (ML) is a subset of AI that focuses on the ability of machines to receive a set of data and learn for themselves, and change algorithms as they learn more about the information that they are processing.

This section introduces areas of material design and manufacturing, which can benefit from AI or ML, and offers relevant perspectives. While the primary emphasis is on the application of ML and data analytics to expedite the identification of high-entropy alloys (HEAs) exhibiting mechanical or fatigue properties of interest, an effort is also placed on investigating and understanding structure-property-performance relationships for polymeric matrix composites (PMCs) and ceramic matrix composites (CMCs), to accelerate materials design for next-generation multi-functional composites and polymers, such as for high-temperature applications. The section concludes with an exposition of the benefits, provided by AI or ML in this context, and of selected background literature.

### 1.1. High-Entropy Materials: Materials with Large Composition Space

The history of multi-component and high-entropy crystalline alloys only extends over 16 years, since the first publications on the concept in 2004 [1, 2]. Despite this relatively short history, the field has inspired new ideas and has expanded the vast composition space offered by the new concept of multi-principal element alloys (MPEAs). This new design approach has opened up a huge unexplored realm of alloy compositions. It seems to have the potential to influence the solid-solution phase stability by adjusting the configurational entropy [3]. The discovery of said concepts has catalyzed significant excitement in the materials community, due to the attractive properties of high-entropy materials (HEMs). Through efforts over the course of the past decade, HEMs have been established as potentially suitable applications for structural and functional materials [3-7]. Extensive research has been carried out on numerous HEMs, and many attractive properties have been achieved, such as high hardness and strength or good ductility [8-23], high fatigue resistance and fracture toughness [24-27], high-temperature oxidation resistance [28-32], good corrosion resistance [33-40], and unique electrical and magnetic properties [41-54]. Another intriguing characteristic of HEMs involves the exceptional fracture toughness at cryogenic temperatures, making them a candidate for low-temperature structural materials [55].

Design strategies for conventional materials usually assume one principal element as the base and then add up to a dozen of other elements with relatively small amounts for relatively minor modifications of given properties of interest. With such an approach, the characteristic properties of the base element can usually be retained, due to its majority in atomic fraction. The additional elements are introduced to optimize or enhance the material properties for certain purpose. HEMs, however, usually have more than four base elements with equal or near-equal atomic ratio. Each combination of the elements usually results in a unique material system, with the material properties tending to differ significantly. This strategy for designing HEMs drastically increases the combination of materials systems possible, material systems that have not been studied yet [56]. For example, there are more than 37 elements that have been used in MPEAs so far [3]. With only one dominant element, the 37 elements yield 37 different conventional material systems to be explored. But if 5 elements were selected from the 37 elements, there would be 435,897 possible combinations of HEM systems. The same 37 elements would also give rise to 66,045 possible 4-element combinations, 2,324,784 systems with 6-base elements, and a total of more than 3 billion combinations with 3 – 12 elements! Despite those material systems showing properties with less interest, there is still a very large number of combinations that are yet to be explored. As noted by Miracle & Senkov [3], a bold and expansive exploration of the vastness of the composition space of HEAs is essential. This is a compelling prospect in the field of complex, concentrated alloys. It is safe to say that the potential for discoveries associated with this concept has hardly been scratched. Roughly 400 MPEAs have been reported, and many of those are non-equimolar variations of the same elements, giving only 112 different element combinations considered, as of 2017 [3].

To accelerate the exploration of complex compositions and microstructures, Miracle & Senkov [3] encourage the complex, concentrated alloy community to develop and apply high-throughput computation and experiments, and to include data on complex, concentrated materials from other fields. For this purpose, the application of ML may accelerate the discovery of new materials by leveraging as input existing data on potential alloy compositions and processing. Figure 1 presents a high-level summary of the primary input factors that impact the mechanical properties of alloys, such as high-entropy alloys. The mechanical properties considered include the hardness, yield strength, ultimate strength, ductility/plasticity, fatigue resistance, fracture toughness, and creep resistance.

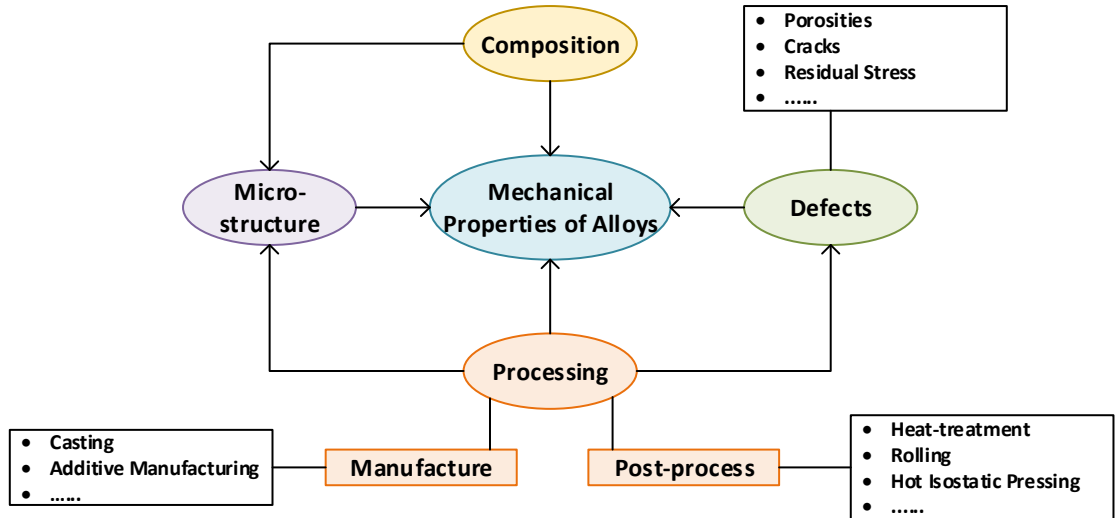

**Figure 1**: The primary factors impacting the mechanical properties of alloys.

## 1.2. Additive Manufacturing of HEA Metallic Components

1. Introduction

Additive manufacturing (AM) is considered as a revolutionary technology to fabricate lightweight, flight- and marine-critical metallic components. AM has the ability to produce near-net-shape components (components with high buy-to-fly ratio), which can prove important, when using expensive alloys, such as refractory HEAs (RHEAs). Here, you may be able to develop a final part using grinding, as opposed to machining. Further, AM overcomes the geometric limits of components produced with traditional subtractive methods, and so provides the opportunity to manufacture highly complex shapes. The ability to produce complex and tailored structure designs opens the door for improved efficiency in existing products and can function as a key enabler to new uses like hypersonic applications. Many merits, such as high efficiency, flexibility, and cost saving, give AM the potential to become a widely utilized fabrication process for industrial applications.

2. Primary Categories of Metal AM Technologies

Figure 2 presents a generic overview over metal AM technologies [57]. Metal AM processes can be grouped into three categories: powder-bed fusion (PBF), direct energy deposition (DED) and sheet lamination (SL) [58]. Each process has its own advantages and disadvantages. The suitable manufacturing technologies usually depend on the design that the designer is trying to make. They, for example, tend to depend on the size of the components desired, on the fineness of details that the design is expected to contain, on the brittleness or ductility of the alloy that the designer is trying to make, on the machinability of the alloy, and on the cost of the alloy. Due to relative popularity, the primary focus here is on PBF and DED.

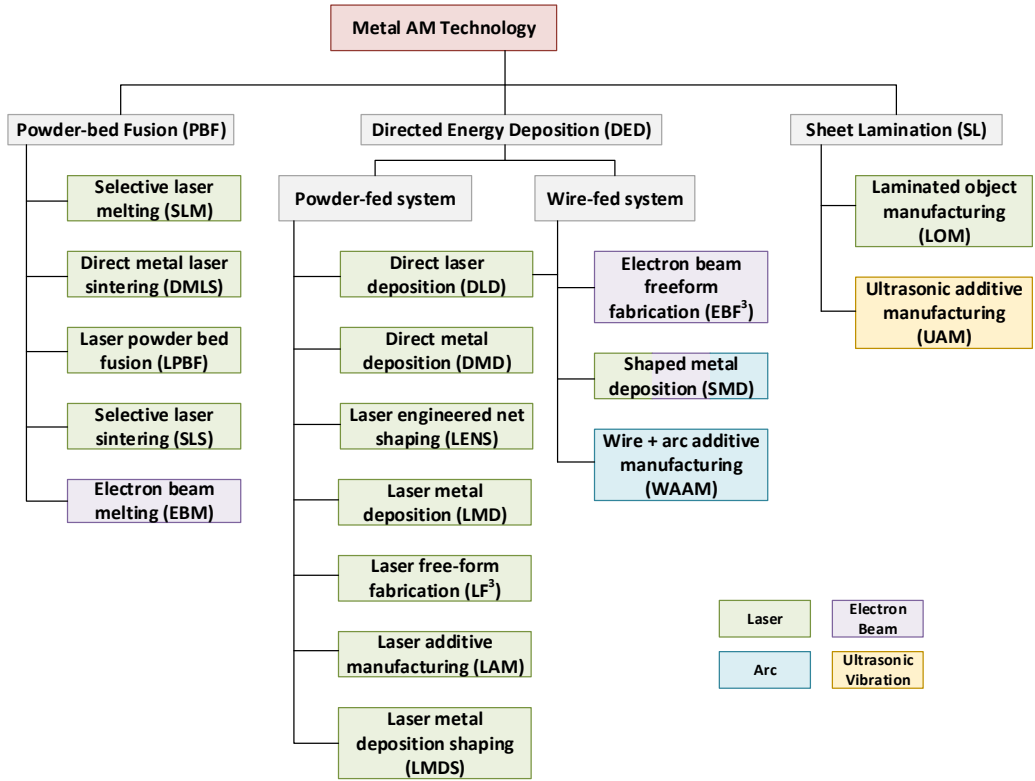

**Figure 2**: The Figure presents an overview over metal AM technologies.

*2.1 Powder-Bed Fusion*

The PBF process includes the selective laser melting (SLM), direct metal laser sintering (DMLS), selective laser sintering (SLS), and electron beam melting (EBM).

| No. | Parameter | Description | Configuration |
|---|---|---|---|
| **Laser and scanning parameters** | | | |
| 1 | Average power | Measure of total energy output of a laser | Controlled |
| 2 | Mode | Continuous wave or pulsed | Predefined |
| 3 | Peak power | Maximum powder in a laser purse | Predefined |
| 4 | Pulse width | Length of a laser pulse when operating in pulsed mode | Predefined |
| 5 | Wavelength | Distance between crests in laser electromagnetic waves | Predefined |
| 6 | Frequency | Pulses per unit time | Predefined |
| 7 | Polarization | Orientation of electromagnetic waves in laser beam | Predefined |
| 8 | Beam quality | Related to intensity profile and used to predict how well beam can be focused and determine min. theoretical spot size | Predefined |
| 9 | Intensity profile | Determination of how much energy added at a specific location | Predefined |
| 10 | Spot size | Length and width of elliptical spot | Controlled |
| 11 | Scan velocity | Velocity at which laser moves across build surface | Controlled |
| 12 | Scan spacing | Distance between neighboring laser passes | Controlled |
| 13 | Scan strategy | Pattern in which the laser is scanned across the build surface (hatches, zig-zags, spirals, etc.) and associated parameters | Controlled |
| **Powder material properties** | | | |
| 14 | Bulk density | Material density, limits maximum density of final component | Predefined |
| 15 | Thermal conductivity | Measure of material's ability to conduct heat | Predefined |
| 16 | Heat capacity | Measure of energy required to raise temp. of material | Predefined |
| 17 | Latent heat fusion | Energy required for solid-liquid and liquid-solid phase change | Predefined |
| 18 | Melt pool viscosity | Measure of resistance of melt to flow | Predefined |
| 19 | Melting temperature | Temperature at which material melts | Predefined |
| 20 | Boiling temperature | Temperature at which material vaporizes | Predefined |
| 21 | Coefficient of thermal | Measure of volume change of material on heating or cooling | Predefined |
| | expansion | | |
| 22 | Surface free energy | Free energy required to form new unit area on heating or cooling | Predefined |
| 23 | Vapor pressure | Measure of the tendency of material to vaporize | Predefined |
| 24 | Heat (enthalpy) of reaction | Energy associated with a chemical reaction of material | Predefined |
| 25 | Material absorptivity | Measure of laser energy absorbed by the material | Predefined |
| 26 | Diffusivity | Important for solid state sintering, not critical for melting | Predefined |
| 27 | Solubility | Solubility of solid material in liquid melt | Predefined |
| 28 | Particle morphology | Measures of shape of individual particles and their distribution | Predefined |
| 29 | Surface roughness | Arithmetic mean of the surface profile | Predefined |
| 30 | Particle size distribution | Distribution of particle sizes, usually diameter, in a powder sample | Predefined |
| 31 | Pollution | Ill-defined factor describing change in properties of powder due to reuse | Predefined |
| **Powder bed and recoat parameters** | | | |
| 32 | Density | Measure of packing density of powder particles, influence heat balance | Predefined |
| 33 | Thermal conductivity | Measure of powder bed's ability to conduct heat | Predefined |
| 34 | Heat capacity | Measure of energy required to raise temp. of powder bed | Predefined |
| 35 | Emissivity | Ratio of energy radiated to that of black body | Predefined |
| 36 | Absorpivity | Measure of laser energy absorbed | Predefined |
| 37 | Deposition system parameters | Recoater velocity, pressure, recoater type, dosing | Controlled |
| 38 | Layer thickness | Height of a single powder layer | Controlled |
| 39 | Powder bed temp. | Bulk temperature of the powder bed | Controlled |
| **Build environment parameters** | | | |
| 40 | Shield gas | Usually Ar or $N_2$, but may also be He or something else | Predefined |
| 41 | Oxygen level | Probably most important environment parameter; oxygen can lead to oxide formation in metal, change wettability, energy required for welding, etc. | Controlled |
| 42 | Shield gas molecular weight | Influences heat balance, diffusivity into and out of part | Predefined |
| 43 | Shield gas viscosity | May influence free surface activity of melt pool, convective heat balance | Predefined |
| 44 | Heat capacity of gas | Term in heat balance | Predefined |
| 45 | Thermal conductivity | Term in heat balance | Predefined |
| 46 | Pressure | Influences vaporization of metal as well as oxygen content | Controlled |
| 47 | Gas flow velocity | Influences convective cooling, removal of condensate | Controlled |
| 48 | Convective heat transfer coef. | Convective cooling of just melted part by gas flowing over the surface | Predefined |
| 49 | Ambient temperature | Appears in heat balance, may impact powder preheat and residual stress | Controlled |
| 50 | Surface free energy | Between liquid and surround gas influence melt pool shape | Predefined |

**Table 1**: Summary of key process parameters of SLM and SLS [59].

There typically are *over 50 process parameters* that impact the ultimate quality of a finished part from a PBF process. This places significant importance on understanding the underlying process physics. Broadly speaking, these parameters can be classified into four categories: laser and scanning parameters, powder material properties, powder bed properties and recoat parameters, and build environment parameters. The key properties (parameters) of SLM and SLS are summarized in Table 1 [59].

*2.2 Direct Energy Deposition*

The DED process can be further classified with respect to the use of powder or wire as feedstock. The powder-based DED process includes laser engineered net shaping (LENS), shown in Figure 3, direct metal deposition (DMD), laser metal deposition (LMD), and laser free-form fabrication (LF3). Meanwhile, the wire-based DED process includes electron beam free-form fabrication (EBF[3]) and shaped metal deposition (SMD). SMD includes wire & arc AM (WAAM). The WAAM process can be further categorized into three types, one based on metal inert gas (MIG), another based on tungsten inert gas (TIG), and a third plasma-based. Table 2 presents the comparison between processing parameters of LENS and laser PBF. Table 3 presents further comparison between the SLM, EBM and WAAM processes for metal AM.

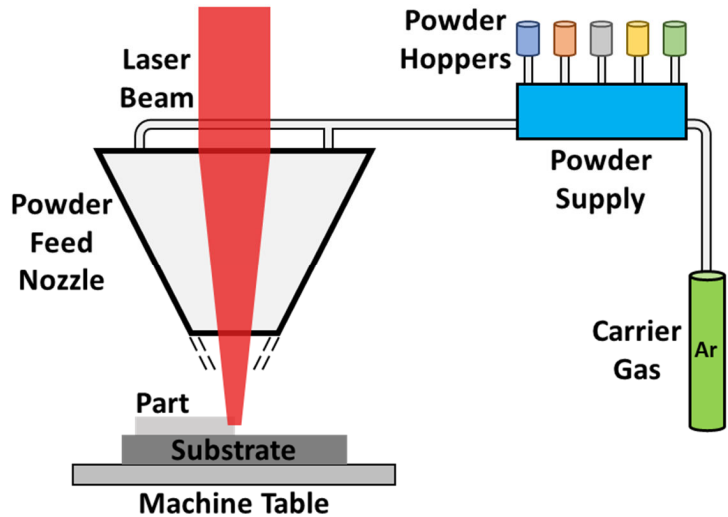

**Figure 3**: Schematic illustration of the operation of a combinatorial, 5-hopper LENS system. [60]

| | LENS (e.g., OPTOMEC 450) | Laser PBF (e.g., EOS M290) |
|---|---|---|
| **Maximum laser power** | 400W | 400W (actual use, not exceed 370W) |
| **Maximum scan speed** | 1016 mm/min | 7000 mm/s |
| **Focus diameter** | 400 μm | 100 μm |
| **Layer thickness** | 200-300 μm | 20-50 μm |
| **Hatch spacing** | 200-400 μm | 50-150 μm |
| **Powder feed rate** | 5-15 g/min | N / A |
| **Oxygen concentration** | Less than 10 ppm | Less than 0.1% (1000 ppm) |
| **Particle size** | 36-150 microns | 15-53 microns |
| **Pre-heating** | Can be heated to 600℃ or cooled to 20℃ by circulating coolant | Can be preheated to 200℃ |
| **Scan strategy** | Continuous, island, etc. | Continuous, island, etc. |

**Table 2**: Comparison of processing parameters for LENS and laser PBF [61].

| Category | Sub-Category | Selective Laser Melting | Electron Beam Melting | Wire + Arc Additive Manufacturing |
|---|---|---|---|---|
| **System** | Energy source | - Laser | - Electron beam | - TIG-based<br>- MIG-based<br>- Plasma-based |
| | Max. Fabricated component size | 400×400×400 mm³ | 200×200×180 mm³ | 610×610×5,182 mm³ |
| | Material deposition rate | 5-35 cm³/hr. | 80 cm³/hr. | Up to 10 kg/hr. |
| | Functionally-graded material | Difficult | Difficult | Available |
| **Material** | Type | Powder | Powder | Wire |
| | Available materials | - Titanium alloy<br>- Carbon steels<br>- Stainless steels<br>- Inconel alloys<br>- …… | - Titanium alloy<br>- Carbon steels<br>- Stainless steels<br>- Inconel alloys<br>- …… | - Stainless steel<br>- Carbon steels<br>- Cobalt alloys<br>- Nickel alloys<br>- Aluminum alloys<br>- Titanium alloy<br>- …… |
| **Overall** | Advantage | - Can support very fine details<br>- Availability of data for analysis | - Stress free part when done | - Suitable for making large components (fast) |
| | Disadvantage | - Limitations on max. component size<br>- Relatively slow | - Powder kept hot through entire process | - Hard to make components with very fine details |

**Table 3**: Comparison between SLM, EBM, and WAAM metallic AM processes [61].

### 3. Key Differentiating Characteristics between Metal AM and Conventional Manufacturing

AM differs considerably from traditional manufacturing of alloys. The main difference relates to the speed of manufacturing. AM can be quite fast. The speed can impact aspects, such as the microstructure and kinetics, to name a few, and can contribute to defects.

### 4. Main Limitations of Metal AM Processes

Despite the high potential of additive manufacturing, it has been found that the fatigue life of as-deposited AM components is often low, compared to wrought components produced by conventional manufacturing technologies. Further, the repeatability of the metal AM manufacturing processes has not been reliable enough to warrant utilization in mass production. The main undesirable features of metal AM processes often involve non-equilibrium thermal cycles. The metal AM processes are usually comprised of numerous cycles of material addition and rapid heating and cooling (rapid melting and solidification). These steps can lead to solid-melting crystallization and solid-remelting recrystallization under fast heating and cooling conditions, which can result in anisotropic microstructures and defects [57]. The microstructures and defects can significantly affect the performance of AM metallic parts, such as the fatigue life. The fatigue performance in AM parts has been attributed to a complex combination of material and process-induced imperfections. For critical components, like those in airframe applications, developing a better understanding of fatigue performance may be essential for further adoption of the technology.

The artifacts impacting the properties and performance of AM metallic components can be classified into macro, micro and nano-scale effects, as shown in Table 4.

| Scale | Artifact |
|---|---|
| Nano | Twin boundary, grain boundary |
| Micro | Micro-voids, micro-cracks, micro-structure size, micro-structure orientation, segregation, intermetallic |
| Macro | Residual stress, surface roughness / waviness, voids, crack, geometric shape |

**Table 4**: Identification of the scale at which different artifacts apply [61].

### 5. Review of Sources Impacting Properties and Performance of AM Metallic Components

Since our focus is placed on achieving good mechanical and fatigue properties, Figure 4 lists some of the primary sources contributing to the mechanical properties of additively manufactured metallic components. Other parameters, which can also impact the mechanical properties, include the powder characteristics (say, the powder size and distribution), weld pool, and cooling speed [62, 63].

With numerous, *sometimes over 100, sources affecting the properties of additively manufactured components* [63], the application of parametric models becomes infeasible. Traditional, parametric models can account for key sources, but not for all the 100+ sources. ML can help in terms of accounting for *all* the sources that impact the properties of AM components.

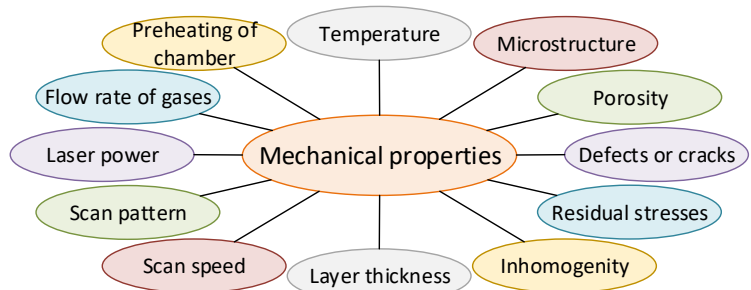

**Figure 4**: High-level overview of primary sources contributing to the mechanical properties of additively manufactured metallic components [61].

6. <u>Interrelationships between Sources Impacting Properties of AM Components and the Underlying Physics</u>

FIG. 6 and FIG. 7 of [61] provide insight into the relationship between the input parameters and the underlying physics [64]. The sources listed there, such as residual stress, may exhibit dependence on other parameters, such as on the layer thickness or temperature. Understanding the root causes of defects and inhomogeneity (properly accounting for the underlying sources of variation) is one of the keys to successfully predicting the properties and performance of additively manufactured components.

7. <u>More on Relationship of AM Processes with Material and Geometry Properties and Component Level Properties</u>

To gain understanding into the root causes of defects and inhomogeneity, one needs to account for the interrelationship between AM process parameters, material & geometry properties, and properties at the component level. In the case of AM, the material properties are largely determined by the solidification microstructure, which is essentially determined by the alloy composition, the thermal gradient (G), and the solidification front velocity (R). For a given alloy, G and R determine its microstructure and mechanical properties. Generally, G and R during PBF are significantly higher than that during LENS, as shown in Figure 5, giving rise to a much higher cooling rate (GxR), and hence a more refined microstructure in PBF-processed materials, which often exhibit better mechanical properties.

Together with the microstructure, processing defects during additive manufacturing mainly consist of lack-of-fusion and keyholes and residual stresses. These processing defects can significantly impact the mechanical properties of the fabricated components.

Lack of fusion and key holing defects have been studied by a physical model of normalized volumetric energy density, as mentioned in [65]. These defects occur not due to one single processing parameter, but a combination of improper processing variables. Reference [66] presents related work that focuses on modeling of three key defects resulting from an AM process: key holing, lack-of-fusion and balling.

Residual stresses arise from the rapid melting and cooling/solidification during layer-by-layer builds. Residual stresses can be classified into three types (Types I, II, and III), depending on the length scale at which they occur. Type I residual stresses are component-level stresses, Types II and III residual stresses are inter-granular and intra-granular level microscale stresses. Type I stresses can cause part distortion, cracking, or delamination. Type-II and III stresses can impact yield strength and work hardening behavior.

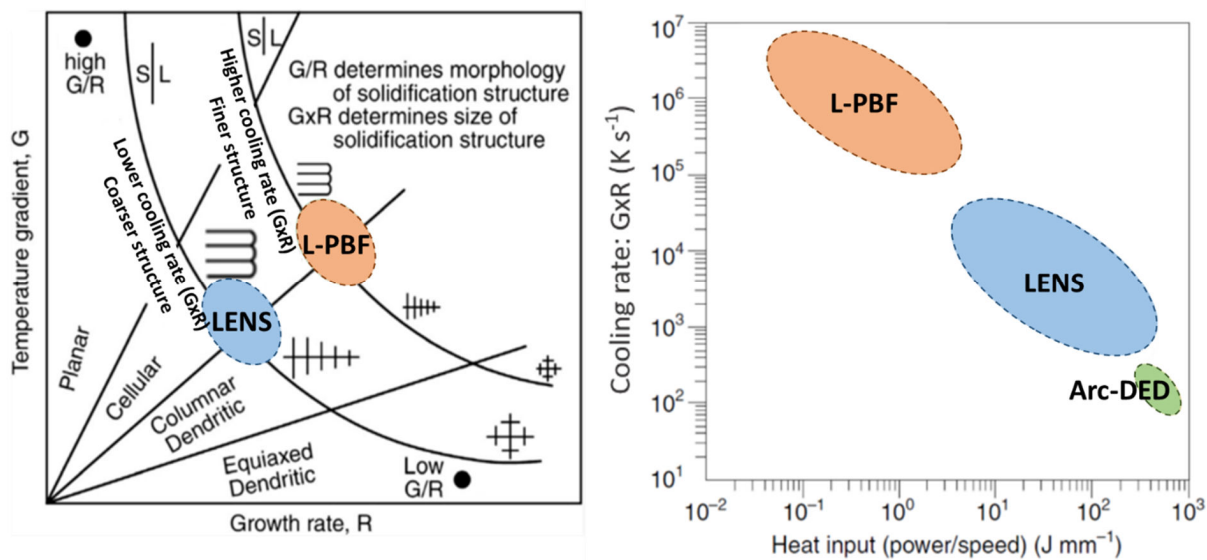

**Figure 5**: Schematic diagrams of solidification microstructure morphology and microstructure feature size of AM metals. Left: solid/liquid (S|L) interfacial morphology during solidification is affected by the thermal gradient, G, and the solidification front growth rate, R (closely related to laser scan speed). Right: G and R in L-PBF processes are substantially higher than those in LENS, giving rise to much higher cooling rates of L-PBF than LENS and arc-based directed energy deposition (Arc-DED).

## 1.3. Phase Diagrams and Integrated Computational Materials Engineering

As noted by Miracle & Senkov [3], phase diagrams serve as roadmaps for materials design. They provide essential information, for a given alloy composition and temperature, on the phases present, their compositions, volume fractions, and transformation temperatures. Most binary and some ternary phase diagrams have been measured experimentally, but multi-component systems remain mostly unexplored. Experimental definition of multi-component phase diagrams is impractical, due to the tremendous amount of work involved. In recent years, the integration of the Computer Coupling of Phase Diagrams and Thermochemistry (CALPHAD) approach with key experiments has been demonstrated as an effective approach to determine complicated multi-component phase diagrams [3, 39, 67, 68].

Integrated Computational Materials Engineering (ICME) tool sets, based on the CALPHAD methodology, include Thermo-Calc [69], Pandat [70], and FactSage [71] software. Figure 6 provides an overview of CALPHAD. Existing toolsets for ICME cannot predict how AM affects material properties of additively manufactured parts. While ML or data analytics are yet to be incorporated into CALPHAD, to our knowledge, Reference [72] lists patents addressing alloys developed by QuesTek Innovations LLC using ICME tools.

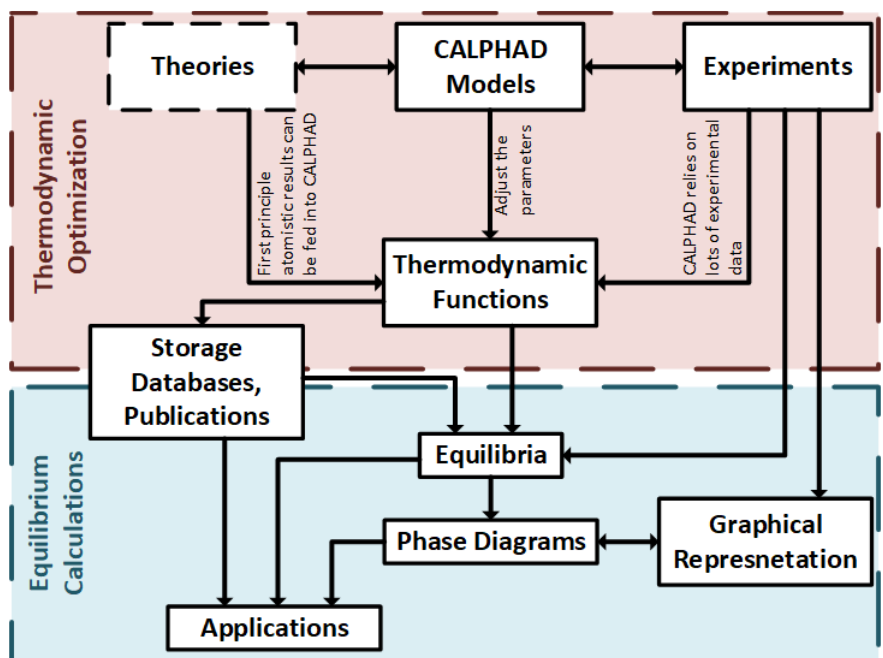

**Figure 6**: Specific information on the internal structure of CALPHAD [61, 73].

## 1.4. High-Temperature Applications

1. <u>Motivation</u>

Electrical power generation is becoming increasingly reliant on gas turbines with multiple fuel capability, with research advances keep focusing on increased efficiency/power output and reduced emissions. Increasing gas turbine efficiency primarily requires higher operating temperatures and reduced coolant flow in the turbine flow path, which makes it challenging to increase component performance as these will encounter high stresses and large temperature gradients. Current Nickel base alloys operate at elevated temperatures/pressures (600 - 1,100ºC / 1 to 16 bar) for prolonged periods of time (around 100,000 hours) leading to significant evolution of microstructure and mechanical properties. To achieve higher efficiency, an increase in the combustor firing temperature of up to 250 - 300 °C is required. In materials terms, this requirement translates to a 200 °C increase in materials limits. As a result, a disruptive and transformative materials solution is needed for improved prime reliance of turbine components. HEAs offer a higher materials temperature capability with enhanced creep, fatigue, and oxidation resistance.

Figure 7 provides glimpse into the "status quo", i.e., the benefits that RHEAs presently offer over conventional alloys, for high temperature applications [74]. Per Figure 7, the refractory HEAs of Type 3 (HEA-3) exhibit $2x - 4x$ *higher yield strength* at high temperature ($T > 1,200$ K), compared to the conventional alloys. The higher yield strength of the RHEAs may allow turbine vendors to *decrease the weight* of the turbo blades, and hence *reduce the cost*.

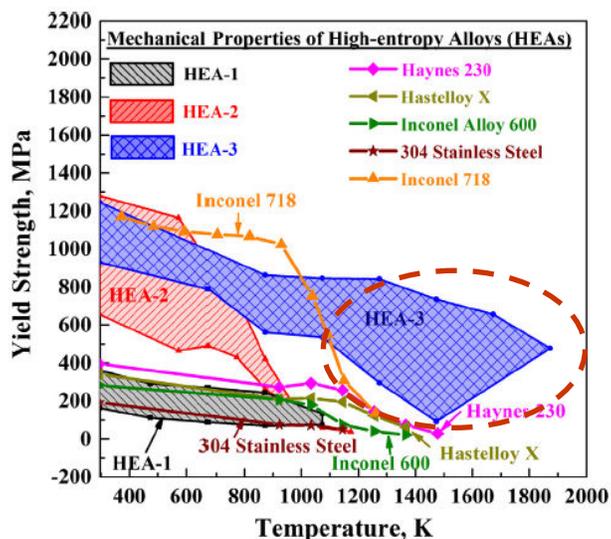

**Figure 7:** The figure illustrates that refractory metals in category HEA-3, yield much superior yield strength at high temperature, compared to alloys conventionally used in turbine blades [74]. Reproduced with permission.

### 2. Role of AI and ML for Development of Alloys Suitable for High-Temperature Applications

Overall, there is need for refractory structural alloys yielding superior strength at higher temperature, together with favorable oxidation properties, while still exhibiting reasonable ductility at room temperature. To that end, AI or ML can help with rapid screening of candidate compositions.

### 3. Role of AM for High-Temperature Applications

Increases beyond 62% for the gas turbine engine combined cycle efficiency (CCE) are hindered by material-property limitations of the current cast Ni-based superalloy stage-one blades. In order to achieve CCE in excess of 65%, and >300 °C increase in the combustor-firing temperature will likely be required. AM technologies have an immense potential to allow for gas turbines of higher efficiency (up to 7% - 9% increase in efficiency) and more cost-effective generator and steam turbine components (up to 15 - 20% cost reduction), by taking advantage of AM innovative component geometries. FIG. 9 of [61] shows that AM can result in lower manufacturing cost, compared to conventional manufacturing, in case of designs consisting of a few number of parts or of very complex components. Hence, AM can be one of the solutions to fabricate the RHEAs for energy conversion applications.

### 4. Environmental (Corrosion) Resistance

AI and ML can be utilized to detect data patterns and characteristic trends, learn from accumulated data, and evolve distinguishing characteristics between different types of hot corrosion attacks, such as calcium-magnesium-alumino-silicate (CMAS) and calcium sulfate attacks, with or without the influence of sea salt, in order to develop suitable coatings resistant to hot corrosion.

## 1.5. Applications of AI or ML to Material Design or Manufacturing

### 1. For Rapid Screening of Material Data Sets – For Accelerated Identification of HEMs with Desired Properties

It can be quite time-consuming and cost-intensive to fabricate alloy structures, esp. HEA structures via additive manufacturing, using a trial & error approach. Data analysis methods using AI and ML have been applied to rapidly screen material property data sets (databases) for desirable material functions, as motivated by Figure 8. By necessitating fewer experiments, AI and ML techniques offer the promise of reducing development times for new energy material manufacturing from discovery to marketable product to about 4 - 5 years, down from the current 15 - 20 years, with a corresponding reduction in development costs. To this effect, General Electric has been able to cut their jet engine alloy development cycle from 15 years to 9 years, by using computational approaches [75]. General Electric hopes to cut the time by half again using improved models and data [75].

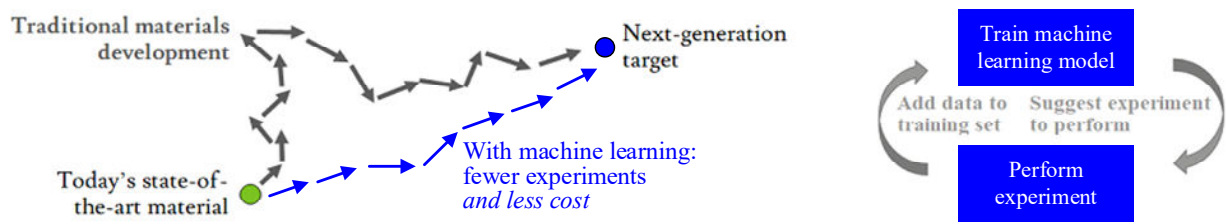

**Figure 8**: Sequential learning with ML can result in fewer experiments and *less cost* [76].

Several material property data sets are currently available to the public. Citrine [77] is aggregating data from multiple databases and making them available to integrated search and data mining. Open Databases Integration for Materials Design (OPTiMaDe) [78] is making a single portal to many databases of computed properties. For other pertinent material databases, refer to NIST CALPHAD Data Informatics databases [79], the Center for Hierarchical Materials Design (CHIMAD) Polymer Property Predictor Database [80], or the databases associated with the Materials Genome Initiative (MGI) [75].

The accuracy of prediction, resulting from the aforementioned screening of material property data sets, tends to be highly dependent upon quality and completeness of the input data. Oftentimes, the input data tends to be less complete than desired. In this case, it is of paramount importance to make the most of the limited data available. However, similar to expert knowledge integration, physics-based models tend to help here. As opposed to "blindly" extrapolating into sparsely populated, or even unpopulated, sections of the parameter space, it helps to account for the underlying physics along the way. Thus, utilization of physics-based models may prove significant for improving the prediction accuracy.

2. <u>To Account for All Sources of Variations in AM Processes – or for Real-Time *In-Situ* AM Quality Control</u>

AI and ML approaches leading to the next generation of commercially viable energy materials manufacturing will require new models of materials and processes, based on experimental and computational data, and the use of models to predict material compositions of optimal material functionality correlated to manufacturing process conditions. With numerous, *sometimes over 100, sources affecting the properties of additively manufactured components* [63], such as the fatigue life, the application of parametric models becomes infeasible. AI or ML can help in terms of accounting for *all* the sources, as noted above.

Further, despite continued progress in AM technologies, AM parts still require several trial and error runs with post-processing treatments and machining to optimize builds, reduce defects and residual stresses, and meet tolerances. AM still lacks a stable process that can produce consistent, defect-free parts on a first time basis, due to inability to reliably predict the optimal trajectory in the multidimensional process parameter space, which can be attributed to the inherent spatiotemporal variability in process parameters and chaotic nature of the AM process. Per Figure 8, Application of AI or ML through sequential learning may necessitate fewer experiments and result in less cost.

AI and ML can also be used to better distribute, monitor and control the processing energy in a laser metal powder bed fusion AM systems, for the purpose of real-time process monitoring and control towards producing high-quality, defect-free AM parts with build periods comparable or shorter than present ones.

3. <u>Practical Approach to Material Design</u>

The subsequent sections contain pedagogical presentation of machine learning in context with material design and manufacturing. In practice, the alloy designer may first identify compositions of interest, on basis of their phase properties, and then design the alloys around the predicted compositions to optimize their properties. For reference to research work on application of ML to optimization of phase properties of HEAs, refer to Section 6.4.

4. <u>Practical Approach to Additive Manufacturing</u>

As opposed to a brute-force optimization of all the parameters impacting the quality of AM parts, the optimization may in practice be guided by expert knowledge. The parameters to be optimized can be prioritized, based on a priori physics-based insights.

## 1.6. Review of Selected Background Work

Liu et al. [81] reported that ML could yield time reduction of 72% - 88%, over exhaustive search (eSearch) or smarter-than-random guided search (gSearch), when applied to optimization of five design problems, involving a recently discovered alloy, Galfenol. Here both the eSearch and gSearch were applied as a part of a ML-based structure-property optimization route. Liu et al. addressed how one could identify the complete space (or as much of it as possible) of microstructures that were theoretically predicted to yield the desired combination of properties demanded by a selected application. The authors presented a problem involving the design of the magnetoelastic Fe-Ga alloy microstructure for the enhanced elastic, plastic, and magnetostrictive properties. While theoretical models for computing properties given the microstructure were known for this alloy, the inversion of these relationships, to obtain microstructures that lead to desired properties, was challenging, primarily due to the high dimensionality of the microstructure space, due to multi-objective design requirements, and non-uniqueness of solutions. These challenges rendered traditional search-based optimization methods incompetent in terms of both searching for efficiency and result optimality. In this chapter, ML is presented as a route to address these challenges. A systematic framework, consisting of random data generation, feature selection and classification algorithms, is presented. Experiments with five design problems, that involve the identification of microstructures, that satisfy both linear and nonlinear property constraints, show that the proposed framework outperforms traditional optimization methods with the average running time reduced by as much as 80% and with the optimality that would not be achieved otherwise.

Agrawal et al. [82] used ML methods to predict the fatigue strength of steels, based on their composition and processing. Oliynyk et al. [83] showed that a high-throughput machine learning approach could be used to screen for potential Heusler alloys. Xue et al. [84] used machine learning to predict the transformation temperature of shape memory alloys using computational results. Conduit et al. [85] built a neural network to design new nickel super-alloys and showed that if given composition and processing steps, they could build a model for several mechanical properties. DeCost et al. [86] used a convolutional neural network to classify the microstructures of steel processed with different heat treatments. These and other studies demonstrate the applicability of machine learning algorithms to predict quantities of interest for both structural and functional alloys.

The platform of [87] uses ensemble ML algorithms, which means that instead of building a single model to predict a given quantity of interest, many models (often hundreds) are built on different random subsets of the training data. Each model makes a prediction for any new test point, and the final ensemble prediction is given by the average value of all the individual model predictions. Such ensemble models have been shown to give high performance and to be robust to noise in the training data [88]. These ensemble models are implemented in such a way as to be capable of operating on sparse data sets (for which data on all the inputs and all the properties of interest are not available for every test point).

For further exposition of the background literature, refer to Section 6.

## 2. Standard Machine Learning in Context with a Generic System Model

### 2.1. Key Take-Aways

1. Machine learning can be thought of as a non-linear correlation (association) technique.
2. As opposed to applying machine learning, narrowly defined in terms of neural networks (single-layer or multi-layer), Bayesian graphical models, support vector machines or decision trees, to the identification of alloys or composites of interest, we reformulate the task of deriving the system model in the broader context of engineering optimization.
3. For determining the association between the source factors and observed properties, *we recommend picking an optimization technique suitable for the application at hand and the data available*.

### 2.2. Fundamental Principles

1. <u>Data Filtering</u> (Curation)

*1.1. Key Principles*

1. The data filtering, also referred to as data curation, is independent of the inference (machine learning) algorithm employed. These are unrelated.
2. The purpose of the data filtering (curation) is to ensure the input data to the ML inference is of highest quality possible.
3. How one ought to filter the data depends on the application. In general, one is looking for relevance. Generally, it is recommended to filter out data that has no relevance with the application domain and task at hand. The intent is to look for outliers, suspected cases of discrepancy or incorrect data (data that one may not fully trust).
4. There is no universal definition for how to fully assess the quality of the data filtered. In addition to the general rules, there is also a level of subjectivity (human-based interpretation) involved.
5. Due to application specific rules and subjective interpretation, it is hard-to-impossible to automate the data filtering. If one could fully specify the filtering mechanism, without subjectivity, then a computer could handle the job.
6. Proper data filtering reduces the inference problem at hand to a manageable complexity (to something one can solve with a computer), because one is left with data that has some degree of relevance with the application domain and task at hand.
7. Data filtering (curation) is separate from data mining, where one is looking for patterns (correlations) in the data.

*1.2 General Guidelines for Devising the Data Filtering*

When determining what type of data filtering to employ, it is of importance to consider the properties listed below [89]:

1. Anomaly detection (outlier/change/deviation detection): This refers to the identification of unusual data records, that might be interesting or data errors that require further investigation.
2. Redundancy: For some cases of material science, repeated logs of fully identical measurements may not convey new (or relevant) information. There may be other cases where the frequency information is of value.
3. Association rule learning (dependency modeling): Association rule learning searches for relationships between variables.

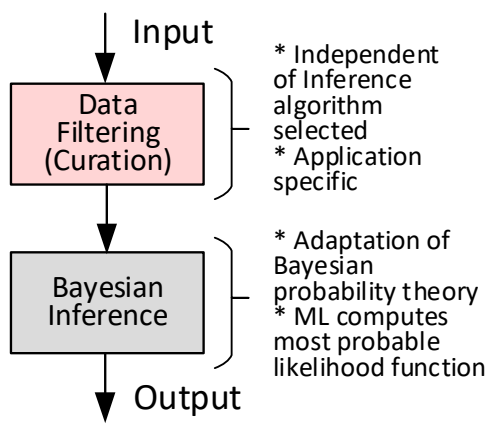

**Figure 9**: Abstracted structure of a machine learning algorithm, in context with fundamental theory.

4. Clustering: Clustering is the task of discovering groups and structures in the data that are in some way or another "similar", without using known structures in the data.
5. Classification: Classification is the task of generalizing known structure to apply to new data. For example, an e-mail program might attempt to classify an e-mail as "legitimate" or as "spam".
6. Regression: Regression attempts to find a function that models the data with the least error, i.e., for estimating the relationships among data or datasets.
7. Summarization: Summarization involves providing a more compact representation of the data set; it can include visualization and report generation.

*1.3 Simplistic Illustration of Application Specific Nature of Data Filtering*

To develop intuition for the level of subjectivity involved in data filtering, one can consider for illustration a simplified example involving image recognition of cats and dogs. Each image is expected to contain only one object, either a cat or a dog. Assuming the image classification algorithm is expected to return a single decision per image, then an image with a cat and a dog is considered unacceptable. But if the algorithm operates on localized subsections of the input images, then an image with a cat and a dog may be considered acceptable. Cases of ingrained subjectivity may involve the following:

1. Is an image with an occluded cat or a dog considered a good image for classification?
2. What if only a small part of a tail of a cat is occluded? Is that considered a good data point or not?
3. How does one quantify how much occlusion for an image to be considered good for classification?
4. What if there is one cat in an image, but there is also a mirror with reflections, so the cat looks like two cats? Is such an image good for classification or not?
5. What if there is a water pool in the image and there is a reflection from a dog or a cat from the water pool?
6. What if there is a shadow covering a part of a dog or a cat?
7. What if there is a shadow covering a part of a dog or a cat, and the shadow is quite strong?
8. What if there is a shadow covering a part of a dog or a cat, but the shadow is quite weak? Should such an image be considered good for classification or not?

2. <u>Main Categories</u>

Generally, ML can be divided into four main categories, namely, supervised learning, unsupervised learning, reinforcement learning and intrinsic motivational learning, as shown in Table 5 [90]:

1. Supervised learning is a data mining task that involves inference of a function from labeled training data.
2. Unsupervised learning is a type of machine learning algorithm used to draw inferences from datasets consisting of input data without labeled responses.
3. Reinforcement learning (RL) is an area of machine learning concerned with how software agents ought to take actions in an environment so as to maximize some notion of cumulative reward.
4. Intrinsic motivational learning exhibits considerable resemblance with reinforcement learning, except that it relies on a problem independent intrinsic reward, as opposed to a problem-specific extrinsic reward [91].

Reinforcement learning can, for example, be used effectively to teach computers how to play chess. It resembles strategic learning and can outperform a local strategy (tactical learning), say, based on exhaustive search 5 or 6 steps ahead. It focuses is on finding a balance between exploration (of uncharted territory) and exploitation (of current knowledge).

| | **With Teacher** | **Without Teacher** |
|---|---|---|
| **Passive** | Supervised Learning | Unsupervised Learning |
| **Active** | Reinforcement Learning / Active Learning | Intrinsic Motivational Learning / Exploration |

**Table 5**: Primary categories of machine learning [90].

3. <u>Brief Summary of Bayesian Inference</u>

Assuming the data is perfectly filtered, conceptually many ML algorithms can be considered an adaption of Bayesian probability theory (Bayesian inference). We are introducing Bayesian inference here, in order to facilitate theoretical analysis of ML models. The Bayesian theory, for example, can inform how much data is needed to achieve a given level of accuracy. For a more in-depth coverage of the fundamental concepts, we refer to [92].

*3.1. Bayes' Theorem*

In the context of ML (say, a neural network parametrized by $\mathbf{w}$ and the image classification problem above, i.e., binary hypothesis testing), Bayes' theorem seeks to evaluate the posterior probabilities, $P(\mathbf{w}|\{t\},\{\mathbf{x}\},\mathcal{H})$, as

$$P(\mathbf{w}|\{t\},\{\mathbf{x}\},\mathcal{H}) = \frac{P(\{t\}|\mathbf{w},\{\mathbf{x}\},\mathcal{H})}{P(\{t\}|\{\mathbf{x}\},\mathcal{H})} P(\mathbf{w}|\mathcal{H}). \tag{1}$$

By utilizing the definitions outlined below, Bayes' theorem can be written in a more accessible form as

$$\text{Posterior probability} \propto \text{Likelihood function} \times \text{Prior probability}. \tag{2}$$

This theorem, which is central to Bayesian Inference, can alternatively be written as

$$\text{Posterior} = \frac{\text{Likelihood}}{\text{Evidence}} \times \text{Prior}. \tag{3}$$

### 3.2. The Prior Probability Distribution

The prior probability $P(\mathbf{w}|\mathcal{H})$ captures the prior knowledge or assumptions about the weighting function, for example from previous batches of training, together with prior knowledge about the hypothesis $\mathcal{H}$ (here, about the presence of dogs or cats in the images observed).–Note the prior probability $P(\mathbf{w}|\mathcal{H})$ reflects your knowledge prior to receiving any data; the observations from the current batch of data, $\{t\}$ and $\{\mathbf{x}\}$, do not factor into the prior probability.

### 3.3. The Likelihood Function

The likelihood function, $P(\{t\}|\mathbf{w},\{\mathbf{x}\},\mathcal{H})$, captures the probability of the observations given the parametrized expression $\mathbf{w}$: This is the distribution of the observed data conditioned on parameters of the models. Note again that the observations factor into the likelihood function, but not the prior distribution. *ML algorithms usually compute the most probable likelihood function, for a given set of input data, $\{\mathbf{x}\}$, and parameters, $\mathbf{w}$.* We refer to the resulting parameter vector, $\mathbf{w}_{\text{MP}}$, as the (locally) most probable parameter vector (see Section 2.6).

### 3.4. The Posterior Probability Distribution

The posterior probability, $P(\mathbf{w}|\{t\},\{\mathbf{x}\},\mathcal{H})$, represents the probability of the parameter vector, $\mathbf{w}$, given the observations (the input images, $\{\mathbf{x}\}$, plus the observation of a dog, which may correspond to $\{t=0\}$, or of a cat, which may correspond to $\{t=1\}$). Since one does know the prior and likelihood functions, the ML algorithms can be viewed as providing the posterior distribution as output.

### 3.5. Salient Aspects of Bayesian Networks and Inference

Bayesian inference is a method of statistical inference in which Bayes' theorem is used to update the probability for a hypothesis as more evidence or information becomes available [93]. A Bayesian network consists of a directed graph with a random variable $X_i$ at each node, as shown in Figure 10 [94]. Assuming $\pi_i$ represent the parents of $X_i$, the Bayesian network defines a distribution of $\mathbf{X} = \{X_1, ..., X_n\}$ of the form

$$P(\mathbf{X}) = \prod_{i=1}^{n} P(X_i|\pi_i). \tag{4}$$

Inference in a Bayesian network corresponds to calculating the conditional probabilities, $P(Y|Z=z)$, where $Y$, $Z \subset \{X_1, \ldots, X_n\}$ represents sets of latent and observed variables, respectively [94]. The association of machine learning with Bayesian inference, outlined in Section 2.6, provides opportunity for extended theoretical analysis. Cooper showed that Bayesian inference is of exponential complexity in general (NP-hard) [95], meaning there is no polynomial time solution guaranteed to be available. Dagum and Luby showed that even approximate inference is also NP-hard [96].

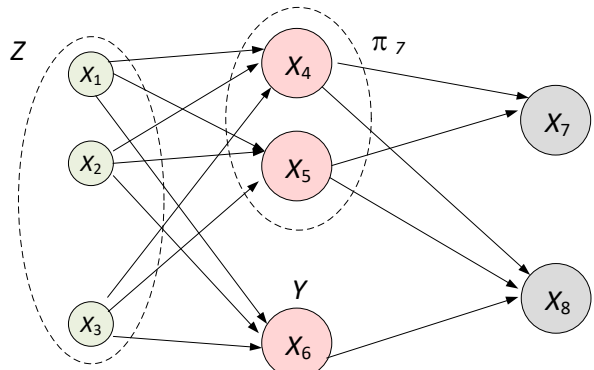

**Figure 10**: Simple example of inference in a Bayesian network.

### 4. Occam's Razor: Preference to Simple Models (Case of Overfitting)

In simple terms, Occam's razor presents a dilemma, where a very complex model yields a close-to-perfect fit to the observations available. However, is such a complex model really better than a simpler model, that describes the data very well, but with a little bit of error? In this case, how does one quantify overfitting? Note that, assuming the very complex model yielded perfect fit (no error), the error would still be zero, even after normalizing with the model size or the number of observations. This dilemma can, for example, be addressed by retaining 80% of the data available for a training set, but quantifying the accuracy of the model using the remaining 20% (i.e., unseen data). The Occam's razor explains why simple model is often preferred, on basis of superior ability to generalize, although a more complex model may yield greater accuracy to the training data. For a more detailed explanation, refer to [92].

5.  Estimation of How Much Input Data Is Needed for a Given ML Algorithm (Model) to Be Suitable

First, the complexity of a probabilistic model controls how much data is needed for the model to be suitable. The estimates for the data needed is usually presented in the form of a bound for *sample-complexity* [$m \leq O(\cdot)$]. The *sample-complexity* specifies the number of samples, that one needs to draw from a given probability density function (PDF), in order to construct an estimated PDF that approximates the true PDF within a given $\delta$ of accuracy [97]. This is quite different from the *algorithm complexity*, which specifies how the number of floating-point operations, that are needed to compute the output of an algorithm, scales with the size of the vector input provided to the algorithm. Second, it is important to understand what type of distribution the machine learning (Bayesian inference) algorithm is trying to approximate. Often times, it is a Gaussian conditional PDF, as we will see in Section 2.6. Since the Gaussian distribution is fully defined by a mean and a variance, one can employ well known results for the distribution and confidence interval for the mean and standard deviation of a Gaussian PDF, for the purpose of estimating the sample-complexity of Bayesian inferencing algorithms. For additional coverage (references) on sample-complexity of ML algorithms, refer to Section 6.1.

## 2.3.  The Generic System Model Assumed

We assume a generic system model of the form

$$\tilde{\mathbf{y}} = f(\tilde{\mathbf{x}}) \tag{5}$$

The input vector, $\tilde{\mathbf{x}}$, can be considered as the definition of a feature set comprising of parameters related to the compositions, post-processing or manufacturing, essentially all the source parameters impacting the output quantity (observed material properties) of interest. The transformation, $f(\cdot)$, can be a non-linear function of the input, $\tilde{\mathbf{x}}$. AI and supervised learning are presented as one of the alternatives for deriving the system model. Table 6 and Table 7 capture sample feature sets.

We present here a scalable solution for determining the system model, one that accounts for the application at hand and the input data available. In case of a small set of input data, regression analysis is presented as a suitable tool for deriving the system model. But for a large set of input data, say, hundreds, thousands, or millions of ($\tilde{\mathbf{x}}, \tilde{\mathbf{y}}$) duplets, we present neural networks as a suitable tool for determining the system model.

| Abbrev | Category | Details | Abbrev | Category | Details |
|---|---|---|---|---|---|
| Cr | | % Chromium | FCC | | Face centered cubic structure |
| Co | | % Cobalt | | | |
| Mn | | % Manganese | HCP | | Hexagonal closed packed structure |
| Mo | | % Molybdenum | | Microstructure | |
| Hf | | % Hafnium | BCC | | Body centered cubic structure |
| Fe | | % Iron | | | |
| Ni | | % Nickel | Precip | | Precipitate |
| Nb | Composition | % Niobium | Tension | | Tension |
| Re | | % Rhenium | Compres | Testing Mode | Compression |
| Ta | | % Tantalum | Bending | | Bending |
| Sn | | % Tin | Torsion | | Torsion |
| Ti | | % Titanium | PowdBT | | Powder bed temp |
| V | | % Vanadium | PowdFT | Temperature | Powder feeder temp |
| W | | % Tungsten | ElevatT | | Elevated temp |
| Zr | | % Zirconium | LaserPow | | Laser power |
| Al | | % Aluminum | SpotSize | | Spot size |
| HeatEnvir | | Heat treatment environment | ScanPattn | | Scan pattern |
| HeatTime | Heat Treatment | Treatment time | HatchDist | Process | Hatch distance |
| Temp | | Temperature | ScanSpd | | Scan speed |
| | | | PulseFreq | | Pulse frequency |
| LoadRate | Loading Rate | $10^{-5}$ sec$^{-1}$ – | PulseDur | | Pulse duration |

| | | $10^6$ sec$^{-1}$ | PartShape | | Particle shape |
|---|---|---|---|---|---|
| ColdRoll | Environment | Cold rolling | PartSize | | Particle size |
| HotRoll | | Hot rolling | PartDist | | Particle distribution |

**Table 6**: Sample feature set for the prediction of composition of RHEA and parameters of powder-bed AM [61].

| Abbrev. | Category | Details | Abbrev. | Category | Details |
|---|---|---|---|---|---|
| Ag | Composition | % Silver | W | Composition | % Tungsten |
| Al | | % Aluminum | Y | | % Yttrium |
| Au | | % Gold | Zn | | % Zinc |
| B | | % Boron | Zr | | % Zirconium |
| C | | % Carbon | Heat Envir | Heat Treatment | Heat treatment environment |
| Co | | % Cobolt | Heat Time | | Heat treatment time |
| Cr | | % Chromium | Temp | | Temperature |
| Cu | | % Copper | | | |
| Dy | | % Dysprosium | Load Rate | Loading Rate | $10^{-5}$ sec$^{-1}$ – $10^6$ sec$^{-1}$ |
| Fe | | % Iron | | | |
| Gd | | % Gadolinium | FCC | Micro-structure | Face centered cubic structure |
| Ge | | % Germanium | | | |
| Hf | | % Hafnium | HCP | | Hexagonal closed packed structure |
| Li | | % Lithium | | | |
| Lu | | % Lutetium | BCC | | Body centered cubic structure |
| Mn | | % Manganese | | | |
| Mo | | % Molybdenum | Precip | | Precipitate |
| Nb | | % Niobium | Tension | Testing Mode | Tension |
| Nd | | % Neodymium | Bending | | Bending |
| Ni | | % Nickel | Compres | | Compression |
| P | | % Phosphorus | Torsion | | Torsion |
| Pd | | % Palladium | ElevatT | Temperature | Elevated temp |
| Rh | | % Rhodium | RoomT | | Room temperature |
| Ru | | % Ruthenium | CryoT | | Cryogenic temp |
| S | | % Sulphur | AddMfg | Process | Additive manufacturing |
| Sc | | % Scandium | DropCast | | Drop casting |
| Si | | % Silicon | Powder | | Powder metallurgy |
| Sn | | % Tin | Sputter | | Sputtering |
| Ta | | % Tantalum | SizePrec | | Size of precipitate |
| Tb | | % Terbium | VolPreci | | Vol. % of precipitate |
| Ti | | % Titanium | ShapePre | | Shape of precipitate |
| Tm | | % Thulium | ColdRoll | Environment | Cold rolling |
| V | | % Vanadium | HotRoll | | Hot rolling |

**Table 7**: Sample feature set for prediction of fatigue endurance limit of HEAs [61].

## 2.4. Statistical Regression[2]

1. Model Structure

In case of relatively small-to-modest number, or even of moderately large number, of observations, we recommend applying a multi-variate linear-regression model: We model the relationship between predictors and observations, in this case, as

$$\tilde{y}_i = b_0 + \sum_{j=1}^{p} b_j \tilde{x}_{ij} + e_i, \tag{6}$$

for $i \in \{1,\dots,n\}$. Here,

- $\tilde{y}_i \in \mathbb{R}$ represents the $i$-th observation.
- $b_0 \in \mathbb{R}$ is the unknown regression coefficient for the intercept.
- $b_j \in \mathbb{R}$ is the unknown regression coefficient for the $j$-th predictor.
- $e_i \in \mathbb{R}$ represents error samples, which we assume independent and identically distributed (i.i.d.) with

$$E[e_i] = 0, \qquad \text{Var}(e_i) = \sigma^2, \qquad \text{Cov}(e_i, e_j) = 0 \ \forall \ i \neq j. \tag{7}$$

---

[2] This subsection is modeled in part after [98]. The analysis of the properties of estimators and residuals, confidence intervals and inferences is intended as a prelude to similar analysis for AI predictors presented in Section 2.6.

Note that, while the error samples may often be Gaussian distributed, we are not explicitly making that assumption. Given $n$ independent observations, we can organize the observations into a matrix form as follows:

$$\tilde{\mathbf{y}} = \tilde{\mathbf{X}}\,\mathbf{b} + \mathbf{e}. \tag{8}$$

Here, $\tilde{\mathbf{y}}$ represents a $\mathbb{R}^{nx1}$ vector, $\tilde{\mathbf{X}}$ a $\mathbb{R}^{nx(p+1)}$ matrix, $\mathbf{b}$ a $\mathbb{R}^{(p+1)x1}$ a vector, and $\mathbf{e}$ a $\mathbb{R}^{nx1}$ vector. The $n \times n$ covariance matrix for $\mathbf{e}$ is then given by $\mathrm{E}[\mathbf{e}\,\mathbf{e}^T] = \sigma^2\,\mathbf{I}$. Similarly,

$$E[\tilde{\mathbf{y}}] = \tilde{\mathbf{X}}\,\mathbf{b}, \qquad\qquad \mathrm{Cov}(\tilde{\mathbf{y}}) = \sigma^2\,\mathbf{I}. \tag{9}$$

## 2. Least-Squares Solution

The corresponding ordinary least-squares problem is

$$\min_{\mathbf{b}\in\mathbb{R}^{(p+1)x1}} \left\|\tilde{\mathbf{y}} - \tilde{\mathbf{X}}\mathbf{b}\right\|^2, \tag{10}$$

where $\|\cdot\|$ denotes the Frobenius norm [99].

For continuous inputs (predictors), the ordinary least-squares problem has the following, closed-form solution [99]:

$$\hat{\mathbf{b}} = \left(\tilde{\mathbf{X}}^T\tilde{\mathbf{X}}\right)^{-1}\tilde{\mathbf{X}}^T\,\tilde{\mathbf{y}} \tag{11}$$

A new set of predictors (feature set) will give rise to the input matrix, $\mathbf{X}$, from which the observation, $\hat{\mathbf{y}}$, is obtained as

$$\hat{\mathbf{y}} = \hat{\mathbf{b}}^T\,\mathbf{X}. \tag{12}$$

Note that the multi-variate regression offers a deterministic solution (no convergence problems).

## 3. Residuals

The residuals can be computed as [98]

$$\hat{\mathbf{e}} = \tilde{\mathbf{y}} - \hat{\mathbf{y}} = \tilde{\mathbf{y}} - \hat{\mathbf{b}}^T\,\mathbf{X} = (\mathbf{I} - \mathbf{H})\,\tilde{\mathbf{y}}, \tag{13}$$

where the matrix, $\mathbf{H} = \tilde{\mathbf{X}}(\tilde{\mathbf{X}}^T\tilde{\mathbf{X}})^{-1}\tilde{\mathbf{X}}^T$, is idempotent, meaning that $\mathbf{H}^T\mathbf{H} = \mathbf{H}\,\mathbf{H}^T = \mathbf{I}$. The residual sums of squares can now be obtained as

$$\hat{\mathbf{e}}^T\,\hat{\mathbf{e}} = \tilde{\mathbf{y}}^T\,(\mathbf{I} - \mathbf{H})\,\tilde{\mathbf{y}}. \tag{14}$$

## 4. Sum of Squares

One can partition variability in y into variability due to changes in predictors and variability due to random noise (effects other than the predictors). The sum of squares decomposition is obtained as [98]:

$$\underbrace{\sum_{i=1}^{n}(\tilde{y}_i - \bar{y})^2}_{\text{Total SS}} = \underbrace{\sum_{i=1}^{n}(\hat{y}_i - \bar{y})^2}_{\text{SS Reg.}} + \underbrace{\sum_{i=1}^{n}\hat{e}_i^2}_{\text{SS Error}} \tag{15}$$

## 5. Coefficient of Multiple Determination

The coefficient of multiple determination is then given by [98]

$$R^2 = \frac{\text{SS Reg.}}{\text{Total SS}} = 1 - \frac{\text{SS Error}}{\text{Total SS}} \tag{16}$$

The coefficient of multiple determination specifies the proportion of the variability in the observed responses that can be attributed to changes in the predictor variables.

## 6. Properties of Estimators and Residuals – Before Gaussian Assumption

For the regression model specified above and for $\hat{\mathbf{b}} = \left(\tilde{\mathbf{X}}^T\tilde{\mathbf{X}}\right)^{-1}\tilde{\mathbf{X}}^T\,\tilde{\mathbf{y}}$, we have [98]

$$E[\hat{\mathbf{b}}] = \mathbf{b}, \qquad\qquad \mathrm{Cov}(\hat{\mathbf{b}}) = \sigma^2\,(\mathbf{X}^T\,\mathbf{X})^{-1}. \tag{17}$$

For the residuals, we obtain [98]

$$E[\hat{\mathbf{e}}] = 0, \qquad \mathrm{Cov}(\hat{\mathbf{e}}) = \sigma^2\,(\mathbf{I} - \mathbf{H}), \qquad \mathrm{E}[\hat{\mathbf{e}}^T\hat{\mathbf{e}}] = (n - p - 1)\sigma^2. \tag{18}$$

An unbiased estimate of $\sigma^2$ is given by [98]

$$s^2 = \frac{\hat{\mathbf{e}}^T\hat{\mathbf{e}}}{n-(p+1)} = \frac{\hat{\mathbf{y}}^T\,(\mathbf{I}-\mathbf{H})\,\tilde{\mathbf{y}}}{n-(p+1)} = \frac{\text{SS Error}}{n-(p+1)}\,. \tag{19}$$

## 7. Properties of Estimators and Residuals – After Gaussian Assumption

If now assume that the $n \times 1$ vector, $\mathbf{e} \sim N_n(0, \sigma^2\,\mathbf{I})$, then it follows that [98]

$$\hat{\mathbf{y}} \sim N_n(\mathbf{b}^T\,\mathbf{X}, \sigma^2\,\mathbf{I}), \tag{20}$$

$$\hat{\mathbf{b}} \sim N_{p+1}(\mathbf{b}, \sigma^2\,(\mathbf{X}^T\mathbf{X})^{-1}). \tag{21}$$

The distribution of $\hat{\mathbf{e}}$ is independent of $\hat{\mathbf{b}}$ and is given by [98]

$$\tilde{\mathbf{e}} = (\mathbf{I} - \mathbf{H})\,\tilde{\mathbf{y}} \sim N(\,0, \sigma^2\,(\mathbf{I} - \mathbf{H})), \tag{22}$$

$$(n - p - 1)\,s^2 = \hat{\mathbf{e}}^T\,\hat{\mathbf{e}} \sim \sigma^2\,\chi^2_{n-p-1}. \tag{23}$$

## 8. Confidence Intervals

A simple way of estimating the confidence in the quantity predicted involves computing the variance of the training data, and using it as a measure of confidence for the quantity predicted. But a more sophisticated approach involves estimating the $100\,(1 - \alpha)\,\%$ confidence region for $\mathbf{b}$ as the set of values that satisfy [98]

$$\frac{1}{s^2}(\mathbf{b} - \hat{\mathbf{b}})^T X^T \mathbf{X}(\mathbf{b} - \hat{\mathbf{b}}) \le (r+1)F_{p+1,n-p-1}(\alpha), \tag{24}$$

where $p + 1$ is the rank of $\mathbf{X}$. Reference [98] also lists simultaneous confidence intervals for any number of linear combinations of the regression coefficients.

## 9. Predictors Limited to a Continuous Range

When one or more of the inputs is limited to a single, continuous range,

$$x_j \in [x_{j,min}, x_{j,max}], \tag{25}$$

the multi-variate linear regression problem can be formulated as

$$\min_{\substack{\mathbf{B} \in \mathbb{R}^{(p+1)xm} \\ \widetilde{\mathbf{X}} \in \text{Conv}(P) \subseteq \mathbb{R}^{(p+1)}}} \left\| \widetilde{\mathbf{Y}} - \widetilde{\mathbf{X}}\mathbf{B} \right\|^2. \tag{26}$$

Here, $\text{Conv}(P) \subseteq \mathbb{R}^{(p+1)}$ is the convex hull well defined by the $(p+1)$ elements comprising the vector, $\widetilde{\mathbf{X}}$.

It is important to recognize that when any given element of the input vector, $\widetilde{\mathbf{X}}$, is confined to a single, continuous range, the resulting vector subset forms a *convex polytope* in $\mathbb{R}^{(p+1)}$.

Figure 11 shows, for a simple case, that one can travel between any given points in the polytope, and yet stay within the set. While the optimization problem in Equation (10) may not have a closed-form solution, the fact that it is *convex* means that one can generate a solution using efficient interior-point solvers, with the polynomial worst-case algorithm complexity of [100]:

$$O[(p+1)^{3.5}]. \tag{27}$$

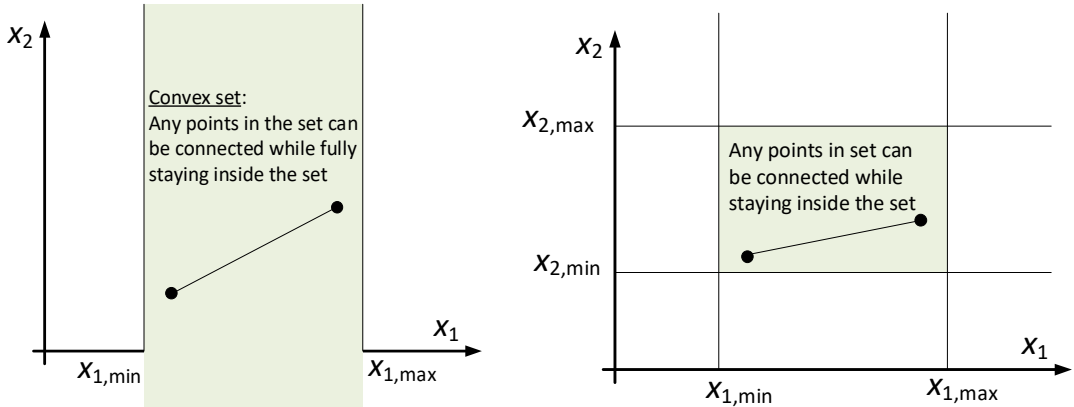

**Figure 11**: Visualization of the convex nature of the multivariate linear regression problem, for deriving the system model, when one or more input feature is limited to a single continuous range (a range limited by a minimum and a maximum values) [101].

## 10. Categorical Predictors

Once an input requirement (predictor) consists of categories, the optimization problem automatically becomes non-deterministic polynomial-time hard (NP-hard), meaning that the run time of a solver is no longer guaranteed to be deterministically polynomial in the size of the problem. For the categorical (discrete) input, there may be special cases that can be solved quickly. But in general the problem may be subjected to exponential worst-case complexity [94, 96]. Note that the predictors (features) in Table 6 and Table 7 may in general be categorical and/or limited to a range.

## 11. Priorities Assigned to Predictors (Features) in the Feature Set

The optimization problem in Equation (26) can be extended not only to support ranges, but also priorities assigned to specific features from the feature set:

$$\min_{\substack{\mathbf{b} \in \mathbb{R}^{(p+1)x1} \\ \widetilde{\mathbf{X}} \in \text{Conv}(P) \subseteq \mathbb{R}^{(p+1)}}} \left\| \widetilde{\mathbf{y}} - \widetilde{\mathbf{X}}\widetilde{\mathbf{W}}\mathbf{b} \right\|^2. \tag{28}$$

Here $\widetilde{\mathbf{W}}$ is a $p$ x $p$ diagonal matrix with an element, $w_{jj}$, specifying the priority associated with the input feature $j$, $x_j$. The features with a higher priority receive a greater weight. Even with the addition of $\widetilde{\mathbf{W}}$, the optimization problem is still linear and convex.

## 12. Variance Analysis

The variance in the system output, $\Delta \widetilde{\mathbf{y}}$, is in part determined by that in the system input, $\Delta \widetilde{\mathbf{x}}$, and in part by the model. In case of independent inputs, the variance in the system output can be modeled as

$$\Delta y_i = \sum_{j=1}^{p} \frac{\partial f}{\partial x_j} \Delta x_j. \tag{29}$$

In order to improve the prediction accuracy (decrease $\Delta y_i$), it may make sense to look at the input sources yielding the largest contributions to $\Delta y_i$, either through the large $\Delta x_j$ and/or large $\partial f / \partial x_j$.

### 13. Sensitivity Analysis

When it is of interest, the interior-point methods, which can be used to solve the optimization problems in Equations (10) and (26), can also inform the user of the relative contributions of given input features to the objective function. The sensitivity to a given input feature is defined as the derivative of the objective function with respect to that feature, $\partial f / \partial x_j$, per Equation (29). The sensitivities come about as Lagrange multipliers that are produced as a by-product from the interior-point solvers. While one already may be hitting a boundary, the derivatives may inform the user about variables that still can change within the feasible set.

### 14. Application of Regression Analysis for Optimization of a Feature Set (Input Predictors)

In order to maximize a given output, $\tilde{y}_{ik} \in \mathbb{R}$ , for the purpose of identifying a favorable composition or manufacturing process, it makes sense to increase the values of the input, $\tilde{x}_{ij} \in \mathbb{R}$ , corresponding to the largest (i.e., the most positive) values of the system model, $b_{jk} \in \mathbb{R}$. These values of the input contribute the most to the output, according to the system model. Correspondingly, in order to minimize a given output, $\tilde{y}_{ik}$, it makes sense to increase the values of the input, $\tilde{x}_{ij}$, corresponding to the smallest (i.e., the smallest positive or largest negative) values of the system model, $b_{jk}$. For an illustration, refer to Figure 29 - Figure 31.

### 15. Inferences Utilizing the Regression Model Derived

When $\mathbf{x} = \mathbf{x}_0$, the response has a conditional mean of $E[\mathbf{y}_0 | \mathbf{x}_0] = \mathbf{x}_0^T \mathbf{b}$ . An unbiased estimate is $\hat{\mathbf{y}}_0 = \mathbf{x}_0^T \hat{\mathbf{b}}$ which has variance of $\mathbf{x}_0^T (\mathbf{X}^T \mathbf{X})^{-1} \mathbf{x}_0 \, \sigma^2$. We might be interested in a confidence interval for the mean response at $\mathbf{x} = \mathbf{x}_0$ or in a prediction interval for a new observation $\mathbf{y}_0$ at $\mathbf{x} = \mathbf{x}_0$. A 100 (1 - $\alpha$)% confidence interval for $E[\mathbf{y}_0 | \mathbf{x}_0] = \mathbf{x}_0^T \mathbf{b}$, which is the expected response at $\mathbf{x} = \mathbf{x}_0$, is given by [98]

$$\mathbf{x}_0^T \hat{\mathbf{b}} \;\pm\; t_{n-p-1}\left(\frac{\alpha}{2}\right)\sqrt{\mathbf{x}_0^T (\mathbf{X}^T \mathbf{X})^{-1} \mathbf{x}_0 \, s^2}. \tag{30}$$

A 100 (1 - $\alpha$)% confidence interval for $E[\mathbf{y}_0 | \mathbf{x}_0] = \mathbf{x}_0^T \mathbf{b}$, for all $\mathbf{x}_0$ in some region is obtained from Scheffe method as [98]

$$\mathbf{x}_0^T \hat{\mathbf{b}} \;\pm\; \sqrt{(p+1)F_{p+1,n-p-1}(\alpha)}\sqrt{\mathbf{x}_0^T (\mathbf{X}^T \mathbf{X})^{-1} \mathbf{x}_0 \, s^2}. \tag{31}$$

If we wish to predict the value of a *future observation* $\mathbf{y}_0$, we need the variance of the prediction error $\mathbf{y}_0 - \mathbf{x}_0^T \hat{\mathbf{b}}$ [98]:

$$\text{Var}\left(\mathbf{y}_0 - \mathbf{x}_0^T \hat{\mathbf{b}}\right) = \sigma^2(1 + \mathbf{x}_0^T (\mathbf{X}^T \mathbf{X})^{-1} \mathbf{x}_0). \tag{32}$$

Note that the uncertainty is higher when predicting a future observation than when predicting the mean response at $\mathbf{x} = \mathbf{x}_0$. Then, a 100 (1 - $\alpha$)% prediction interval for a future observation at $\mathbf{x} = \mathbf{x}_0$ is given by [98]

$$\mathbf{x}_0^T \hat{\mathbf{b}} \pm t_{n-p-1}(\alpha/2)\sqrt{1 + \mathbf{x}_0^T (\mathbf{X}^T \mathbf{X})^{-1} \mathbf{x}_0 \, s^2}. \tag{33}$$

### 16. Alternative Regression Models

While multivariate linear regression may be a natural choice, in case that regression analysis is preferred for determining the system model, in particular for continuous and real-valued inputs, there are other regression options available, such as the quadratic regression, polynomial regression, logistic regression, multinomial logistic regression or ordinal regression.

### 17. Quadratic Regression

When the data justifies the use of a model more sophisticated than linear regression, one can, for example, employ an AI predictor (see the next section). Or one can consider adding a quadratic term to the linear-regression model. The quadratic term is the 2nd term from the Taylor expansion of the system model from Equation (5). Unless there is a major, a discrete jump in the data, i.e., as long as the underlying physics does not produce a discrete jump in the data (such as caused by the on-set of super-conductivity), one can expect the system model to lend itself to the Taylor expansion. A least-squared solution to the quadratic model

$$y = a\,x^2 + bx + c, \tag{34}$$

where $a \neq 0$, exists in a closed form, and is given by

$$\begin{bmatrix} \sum x_i^4 & \sum x_i^3 & \sum x_i^2 \\ \sum x_i^3 & \sum x_i^2 & \sum x_i \\ \sum x_i^2 & \sum x_i & n \end{bmatrix} \begin{bmatrix} a \\ b \\ c \end{bmatrix} = \begin{bmatrix} \sum x_i^2 y_i \\ \sum x_i y_i \\ \sum y_i \end{bmatrix}. \tag{35}$$

## 2.5. AI Predictor (Neural Network)

We define AI as the use of computers to mimic the cognitive functions of humans. We look at ML as a subset of AI, one that focuses on the ability of machines to receive a set of data, and learn for themselves (change algorithms as the machines learn more about the information that they are processing).

Neural networks lend themselves well to problems, such as the image recognition (say, shape recognition or face recognition), where the size of the input data set is very large, and where multi-variate linear regression is infeasible. Neural networks may similarly be suitable for the identification of events in the sampled audio, where the input may occupy a gigantic space of time and frequency. Certain data sets lend themselves well to an AI model. Neural networks are capable of learning complex hidden models. They can *learn a model*, i.e., estimate parameters, *without the knowledge of the underlying model structure* (with no a prior information and no functional description). There are also certain tasks, such as the fast and accurate counting of a large number of objects, which AI – at the present stage – can handle better than humans.

Even though there are certain tasks that AI can handle with more accuracy and faster than humans, AI in general – at the present stage – does not match human-based intelligence. The human brain consists of $10^{12} – 10^{13}$ neurons, which is a few-to-several orders of magnitude larger than what most neural networks possess today. Up to this point, AI can – with certain simplifications - be thought of as a tool capable of completing large-scale, generic functional regressions (correlations). Neural networks can compute correlations, at a large scale. But correlation is not the same as causation.

Neural networks currently used in ML include Feed-Forward Neural Networks, Radial Basis Function Neural Networks, Kohonen Self-Organizing Neural Networks, Recurrent Neural Networks, Convolutional Neural Networks, and Modular Neural Networks. For a reference on additional neural network types, we recommend Reference [102]. The main emphasis here is on Single and Two-Layer Feed-Forward Neural Networks. We show that Two-Layer Feed-Forward Neural Networks can approximate *any* system model. The Single-Layer Feed-Forward Neural Networks are preferred, on basis of simplicity, but the Two-Layer Networks, on basis of the approximation capability.

1. Single-Layer Feed-Forward Neural Network

In a single-layer feed-forward neural network, shown in Figure 12, the connections between the nodes do not form a cycle. The simplest kind of a neural network is a single-layer perceptron network, which consists of a single layer of output nodes, and where the inputs are fed directly to the outputs via a series of weights. The sum of the products of the weights and the inputs is calculated in each node. If the value is above a given threshold (typically 0), then the neuron fires and takes on the activated value (typically 1). Otherwise, the neuron takes on the deactivated value (typically -1). The perceptrons in the network can be trained by a simple learning algorithm, which is usually referred to as the Delta Rule [102]:

$$\Delta w_{ijx} = -\varepsilon \frac{\delta E}{\delta w_{ij}} = \varepsilon \; \delta a_{ix} \tag{36}$$

Here $\Delta w_{ijx}$ represents the update applied to the weight at the node (perceptron) between links, *i* and *j,* in a neural network. *E* denotes an error function over an entire set of training patterns (i.e., over one iteration, or epoch). $\varepsilon$ is a learning rate applied to this gradient descent learning. $a_{ix}$ denotes the actual activation for the node, *x,* in the output layer, *i*. The Delta Rule calculates the errors between the calculated output and sample output data, and uses this feature to create an adjustment to the weights, thus implementing a form of gradient descent.

A common choice for the activation function is the sigmoid (logistic) function:

$$f(z) = \frac{1}{1 + e^{-z}}. \tag{37}$$

With this choice, the single-layer network presented in Figure 12 is identical to logistic regression.

A single-layer neural network is only capable of learning linearly separable patterns. Yet, it has guaranteed convergence (equivalent to regression).

It may be noted that a straightforward application of a single-layer neural network model cannot accommodate all categories of features from a generic input feature set. To handle binary features (simple presence or absence), Exclusive OR (XOR)-like conditions, or features taking on assignments from pre-defined categories, a two-layer neural network may be necessary [102].

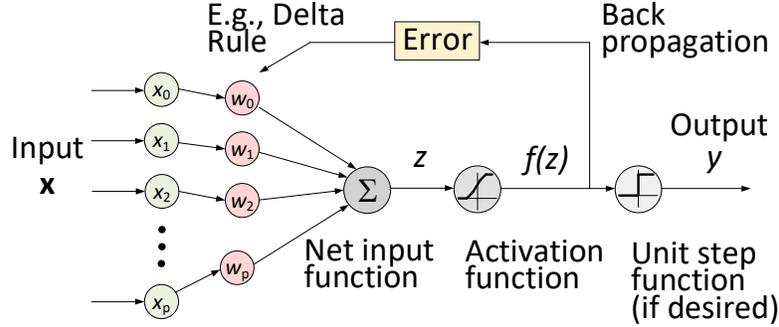

**Figure 12**: Schematics of a logistic regression classifier in the form of a single-layer feed-forward neural network [101].

## 2. Two-Layer Feed-Forward Neural Network

A two-layer feed-forward neural network, presented in Figure 13, consists of two layers of computational units, interconnected in a feed-forward fashion. Each neuron in a given layer has directed connections to the neurons of the subsequent layer. In many applications, the units of these networks apply the sigmoid function [Eq. (37)] as the activation function [102]. Similar to the single-layer case, an activation is defined as

Hidden layer:  $a_j^{(2)} = f^{(2)}\left(\sum_p w_{jp} x_p + b_j\right)$, where $j = 0, 1, 2, \ldots, m;$ (38)

Output layer:  $y_l = a_l^{(3)} = f^{(3)}\left(\sum_q w_{lq} x_q + b_q\right)$, where $l = 0, 1, 2, \ldots, k.$ (39)

Convergence of two-layer (and multi-layer) neural networks involves non-convex optimization, and hence is not guaranteed. The algorithm can converge to a local minimum, and become stuck there.

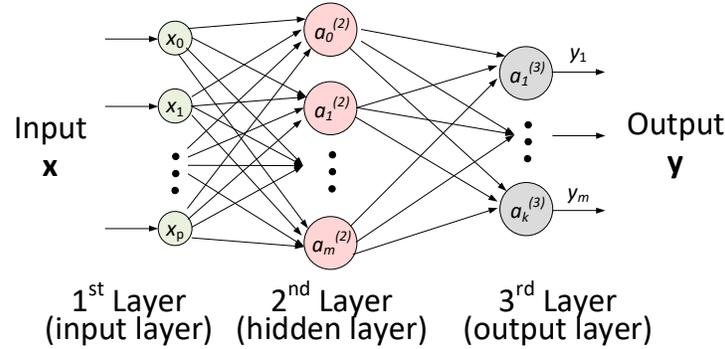

**Figure 13**: Schematics of a two-layer perceptron network [101].

## 3. Training of Two-Layer Feed-Forward Neural Network

A two-layer (or a multi-layer) neural network can be trained using a data set $D = \{\mathbf{x}^{(m)}, \mathbf{t}^{(m)}\}$ by adjusting $\mathbf{w}$ to minimize an error function, e.g., [92]

$$E_D(\mathbf{w}) = \frac{1}{2}\sum_m \sum_i (t_i^{(m)} - y_i(\mathbf{x}^{(m)}; \mathbf{w}))^2.$$ (40)

This objective function is a sum of terms, one for each input/target pair, $\{\mathbf{x}, \mathbf{t}\}$, measuring how close the output $\mathbf{y}(\mathbf{x}; \mathbf{w})$ is to the target $\mathbf{t}$ [92]. This minimization is often based on repeated evaluation of the gradient of $E_D$ using a backpropagation rule, similar to the one in Eq. (36). When so called regulation is employed, the objective function is modified to

$$M(\mathbf{w}) = \beta\, E_D(\mathbf{w}) + \alpha\, E_W(\mathbf{w}),$$ (41)

where, for example, $E_W(\mathbf{w}) = \frac{1}{2}\sum_i w_i^2$ [92].

## 4. Approximation Capability of Two-Layer Feed-Forward Neural Network

According to Kolmogorov, *any* continuous, real-valued function can be modeled in the form of a two-layer neural network [103]. More specifically, Kolmogorov showed in 1957 that any continuous real-valued function, $f$ $(x_1, x_2, \ldots, x_n)$, defined on $[0,1]$ $n$, with $n \geq 2$, can be represented as [103]

$$y = f(x_1, x_2, \cdots, x_n) = \sum_{j=1}^{2n+1} g_j\left(\sum_{i=1}^n \phi_{ij}(x_i)\right),$$ (42)

where the $g_j$'s are properly chosen functions with a single input variable, and the $\phi_{ij}$'s are continuously monotonically increasing functions independent of $f$.

Figure 14 offers a neural-network representation of the Kolmogorov's theorem [103]. Note that most physical functions are nice, continuous, and with derivatives. Hence, Equation (42) applies. One can expand most physical functions over basis functions (say, Gaussian, sin, or cos kernels), estimate the possibly large number of parameters in the decomposition, and account for the sources of error. The neurons in the neural networks model how the input and output react to specific basis functions. The coefficients learned by the neural networks are the coefficients in the expansion.

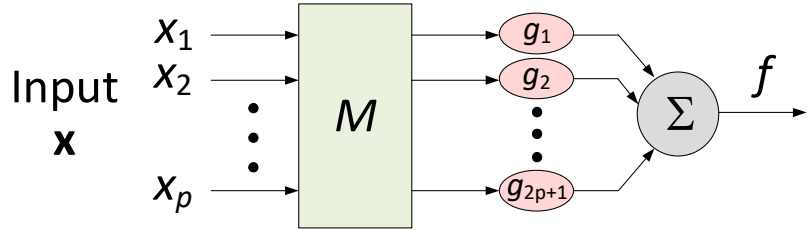

**Figure 14**: Neural network representation of Kolmogorov's theorem [101].

## 2.6. ML in Context with Bayesian Inference Revisited

1. <u>Neural Network Learning (Training) in Context with Bayesian Inference</u>

The neural network learning (training) process can be given the probabilistic interpretation outlined below. The error function in Eq. (41) can be interpreted as minus the log likelihood for a noise model [92]:

$$P(D|\mathbf{w}, \beta, \mathcal{H}) = \frac{1}{Z_D(\beta)} \exp(-\beta E_D), \tag{43}$$

where $Z_D(\beta)$ represents a normalization factor. Thus, the use of the sum-squared error $E_D(\mathbf{w})$ in Eq. (40) corresponds to an assumption of Gaussian noise on the target variables, and the parameter $\beta$ defines a noise level $\sigma_v^2 = 1/\beta$. Similarly, the regularizer in Eq. (41) can be interpreted in terms of a log prior probability distribution over the parameters [92]:

$$P(\mathbf{w}|\alpha, \mathcal{H}) = \frac{1}{Z_W(\alpha)} \exp(-\alpha E_W) \tag{44}$$

If $E_W$ is quadratic as defined above, then the corresponding prior distribution is a Gaussian with the variance, $\sigma_W^2 = 1/\alpha$. The probabilistic $\mathcal{H}$ specifies the functional form $A$ of the network, the likelihood distribution [Eq. (43)] and the prior distribution (Eq. (44)). The objective function $M(\mathbf{w})$ then corresponds to the *inference* of the parameters $\mathbf{w}$, given the data [92]:

$$P(\mathbf{w}|D, \alpha, \beta, \mathcal{H}) = \frac{P(D|\mathbf{w}, \beta, \mathcal{H}) \ P(\mathbf{w}|\alpha, \mathcal{H})}{P(D|\alpha, \beta, \mathcal{H})} = \frac{1}{Z_M} \exp(-M(\mathbf{w})). \tag{45}$$

The $\mathbf{w}$ found by (locally) minimizing $M(\mathbf{w})$ is then interpreted as the (locally) most probable parameter vector, $\mathbf{w}_{MP}$ [92].

2. <u>Approach to Estimating the Number of Samples Needed to Enable ML Models to Approximate Underlying Distributions with Given Accuracy</u>

Broadly speaking, this is a matter of first identifying the type of probability distributions the machine learning algorithms have been designed to model. Typically, these are Gaussian conditional PDFs, which are fully defined by a mean and a variance. To estimate such distributions, one needs to draw large enough number of samples estimate the mean and variance within a given degree of accuracy. Eq. (24) specifies a 100 (1 - $\alpha$)% confidence interval for the mean of a Gaussian distribution, given $n$ observations drawn, $p$ model parameters and an unknown standard deviation. Obviously, as larger number of samples are drawn from the distribution, the bound becomes tighter.

For in-depth analysis of sample-complexity of convolutional neural networks (CNNs) and recurrent neural networks (RNNs), refer to [104]. CNNs and RNNs tend to require fewer training samples to accurately estimate their parameters, purportedly due to a more compact parametric representation compared to their Fully-Connected Neural Network (FNN) counterparts. In [104], the authors initiate the study of rigorously characterizing the sample-complexity of estimating CNNs and RNNs. The authors show that the sample-complexity to learn CNNs and RNNs scales linearly with their intrinsic dimension, and that the sample-complexity is much smaller than for their FNN counterparts. For both CNNs and RNNs, the authors also present lower bounds showing that their sample complexities are tight up to logarithmic factors.

## 3. "Inverse" Design Representations Accomplished through "Forward" and "Backward" Prediction

### 3.1. Key Take-Aways

1. For accurate prediction, it is of primary importance to properly understand and account for, the sources contributing to variations in the quantity predicted.
2. We present a framework for "forward" predicting the properties observed.
3. "Backward" prediction (identification of candidate compositions) is accomplished through an "inverse" design framework, one that identifies the candidate compositions to test next, based on a set of property specifications and design goals.

### 3.2. "Inverse" Design Representations

In an "inverse" design representation, the product design, $\mathbf{x}_{design}$, is effectively expressed as an explicit function of the problem statement (the target outcome), $\mathbf{y}_{target}$, ideally with no iterations involved:

$$\mathbf{x}_{design} \approx f^{-1}(\mathbf{y}_{target}). \tag{46}$$

Here we assume the inverse problem is well-posed (that a unique inverse exists). This may require imposing requirements on the data sets, such as in terms of encoding or convexity. Through a sequential learning workflow, an "inverse" design framework can be used to identify candidate compositions. Data corresponding to these candidate compositions can then be used to retrain the model, presumably leading to iterative refinements and convergence, per Figure 8.

### 3.3. Overall Approach to "Forward" and "Backward" Prediction

Figure 15 summarizes the overall approach. We "forward" predict the observed properties, such as the properties listed in Figure 1 or Figure 16. Correspondingly, we "backward" predict the properties comprising the feature set, such as the sample feature sets in Table 6 and Table 7.

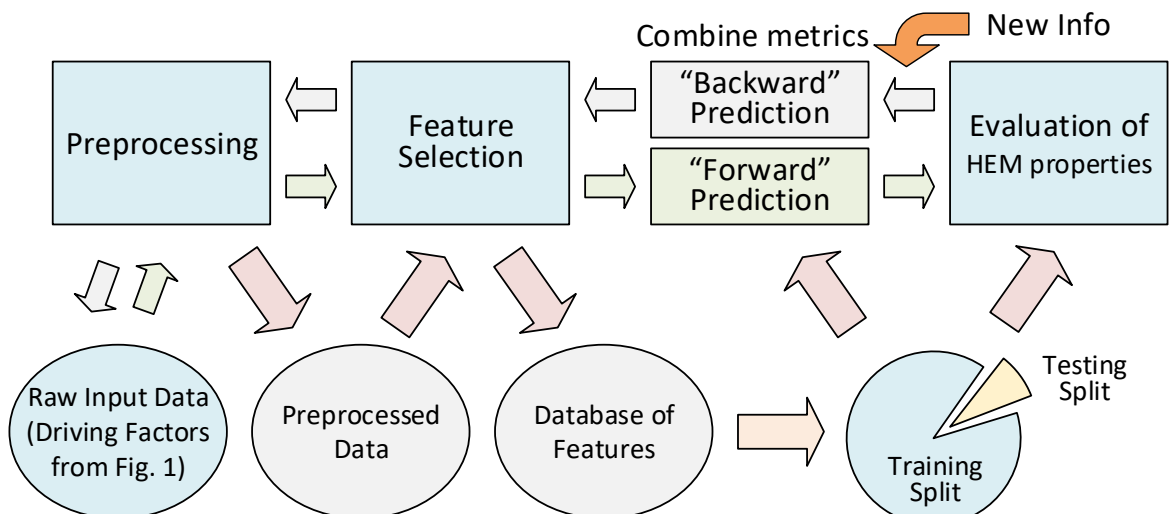

**Figure 15**: High-level approach to machine learning with knowledge discovery and data mining [61]. "New Info" refers to new application-specific information that is added in each iteration round, to prevent the optimization from fitting the model to the data available, and to help with convergence. "New Info" can, for example, consist of a new measurement added to the data set, per sequential learning, or of a theoretical computation of properties using first-principle, CALPHAD or another physics-based model.

### 3.4. Preprocessing

For fair comparison, we normalize the input data, as appropriate. For example, in case of the endurance limit, $S_e$, we can normalize with the ultimate tensile strength (UTS) [61]:

$$S_{e,norm} = \frac{S_e}{UTS} \tag{47}$$

### 3.5. Feature Selection

We derive the features selected from the driving factors listed in Figure 1. Table 6 presents a sample feature selection suitable for identification of RHEAs for high-temperature applications. Note the feature set also captures the AM processing parameters. Hence, the feature set correlates material functionality both with material compositions and manufacturing process conditions. Table 7 presents another representative sample of feature selection suitable for prediction of fatigue endurance limit of HEAs.

### 3.6. "Forward" Prediction

1. <u>Approach to Building a Generic System Model through "Forward" Prediction</u>

We offer a scalable solution for deriving the system model, one that accounts for the application at hand and the input data available. In the case of a small set of input data, we present regression as a suitable tool for deriving (constructing) the system model. But for a large set of input data, say, hundreds, thousands, or millions of $(\tilde{\mathbf{x}}, \tilde{\mathbf{y}})$ duplets, feed-forward neural networks represent a suitable tool for constructing the system model.

The recommended approach is founded in part on observations of Agrawal et al. [82]. Table 2 and Figure 5 of [82] illustrate that *there is at most a few percentage difference between the techniques applied to predict the fatigue strength of stainless steel*. Table 2 of [82] shows that both the simple linear regression and pace regression yield the coefficient of determination, $R^2$, of 0.963, while the artificial neural network, a traditional ML approach, results in $R^2$ of 0.972.

Through application of Occam's razor and sample-complexity, theoretical foundation can be provided to the recommended approach.

2. <u>Review of Modeling Techniques for "Forward" Prediction</u>

The modeling techniques for "forward" prediction considered include, but are not limited to, linear regression, pace regression, robust fit regression, multivariate polynomial regression (which can include quadratic regression), decision tables, support vector machines, artificial neural networks, reduced error pruning trees and M5 model trees [61, 101].

*2.1. Statistical Regression*

The statistical regression is covered in Section 2.4.

*2.2. k-Nearest Neighbor Averaging*

Figure 16 present a simple framework for predicting the observed properties for alloys or composites of interest. For each observed property of interest, we identify the corresponding driving factors. These input factors may include the material composition, heat treatment, process, microstructure, temperature, strain rate, environment, and testing mode. We then carry out the prediction through a customized averaging process in the space comprising the input parameters.

Figure 17 explains, at a high level, how we are going to populate the input parameter space and carry out the prediction. In case of a continuous-valued parameter space, one can apply $k$ nearest neighbor ($k$-NN) averaging for determining the parameters corresponding to point A in Figure 17, $\mathbf{p}_A$:

$$\mathbf{p}_A = \frac{1}{k}\sum_{i=1}^{k-NN} \mathbf{p}_i. \tag{48}$$

Here, $\mathbf{p}_i$ represents the parameter vector corresponding to the $i$-th nearest neighbor of point A in Figure 17 [61].

*2.3. Feed-Forward Neural Network*

The feed-forward neural networks are covered in Section 2.5.

*2.4 Other Modeling Techniques*

*1.* Multi-class ML or multi-class neural networks: More appropriate for classification problems.

*2.* Evolutionary methods: May exhibit problems with "curse of dimensionality" in high dimensional search spaces.

*3.* Combinatorial search methods: Do not provide insight desired into the physics and numerical aspects.

3. <u>Example: "Forward" Prediction of Fatigue Endurance Limit</u>

For a simple example, one can consider "forward" predicting the fatigue endurance limit:

$$\text{endurance limit} = f(\text{ UTS, process, defect(process), grain(process), microstructure (process)}, T. \ldots), \tag{49}$$

where

$$\text{UTS} = h(\text{composition, heat treatment process, defect level (process), grain size}, T). \tag{50}$$

Herein, the composition is specified in terms of the % strength of constituent elements, and UTS represents the ultimate tensile strength. In general, we may be looking at a complex, combinatorial optimization problem [61].

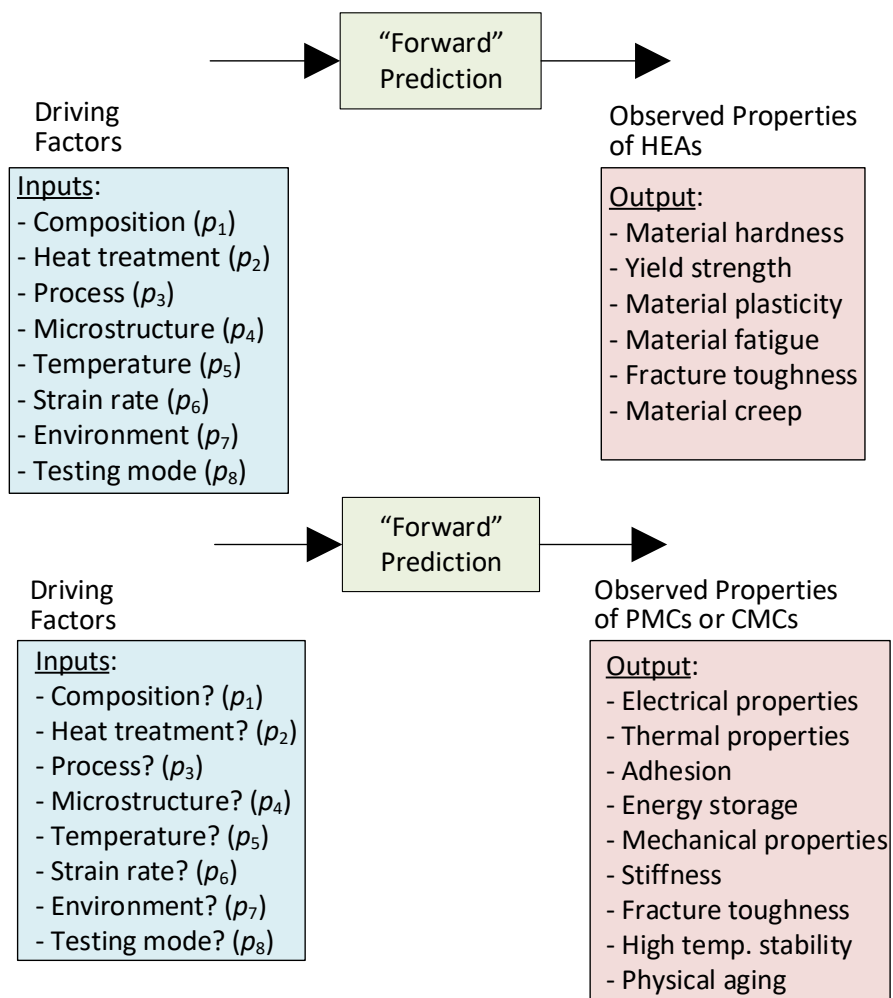

**Figure 16**: The Figure outlines the high-level approach to "forward" prediction of material properties of alloys, such as HEAs. The Figure also illustrates how prediction of material properties of alloys can be extended to matrix composites [61].

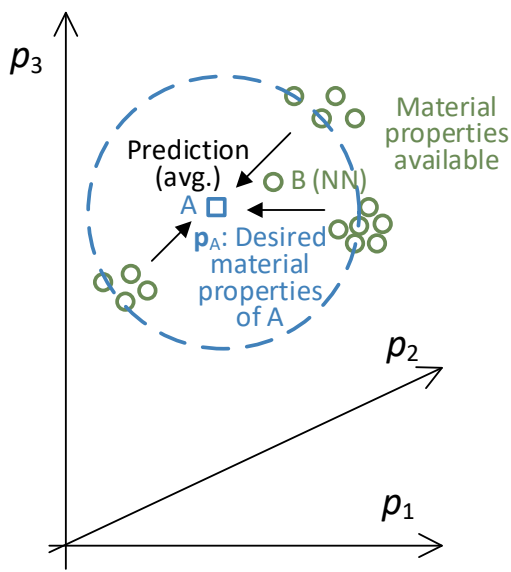

**Figure 17**: The Figure presents the essentials of averaging in the space comprising the input parameters [61].

### 3.7. "Backward" Prediction

1. <u>General Approach to Inferring the Feature Set (e.g., Composition) through "Backward" Prediction</u>

"Backward prediction" is accomplished through an inverse design framework, one that suggests the next input feature (e.g., candidate composition) to consider, based on the property specifications and design goals provided. Through a sequential learning workflow, the inverse design tool is used to suggest input test candidates, and the data from those input test candidates are then used to retrain the model, presumably leading to iterative refinements and convergence. For efficiency, we allow for polynomial fit, during the iterative refinements, when the underlying physics are not expected to result in major nonlinearity [61].

<u>Specific Approaches for "Backward" Prediction</u>

For the "backward" prediction, one can consider a few approaches.

*2.1 Starting Point: Microstructure of the Nearest Neighbor*

The simplest approach consists of identifying the neighbor, B, to the desired HEA, A, as shown in Figure 17, and simply using the microstructure of B as a starting point [61].

*2.2 Baseline Approach: Generalization of [81]*

The assumed baseline approach for inferring the microstructures from the properties desired, shown in Figure 18 and Figure 19, is based on [81]. The microstructure design of poly-crystals can be performed by tailoring the distribution of various crystal orientations [the oriental distribution function (ODF)] in the microstructure [81]. Structural optimization is carried out along different crystallographic directions to attain favorable properties. The multiple crystallographic directions embedded in the multi-dimensional ODF are used as control variables and the theoretical functions for properties are the objectives [81]. ML provides the opportunity to explore multiple design solutions and diminish search time in the high dimensional space of microstructure design problems, where the number of distinct design candidates can be infinite [61].

Two crucial ML steps, namely, search path refinement and search space reduction, are designed to develop heuristics that tour the search force to a much smaller preferable space [81]. A ML-based preprocessing is designed to locate critical region of a search space with a small overhead, so that the search force can be consciously concentrated [61].

*2.3 Combining Metrics: Method to Address Non-Uniqueness*

When "backward" predicting from a single output metric, such as strength, many input combinations may correspond to the same output. But one can address the non-uniqueness of the "backward" prediction by combining metrics. As more output metrics are combined, the set of corresponding input candidates will shrink, even drastically.

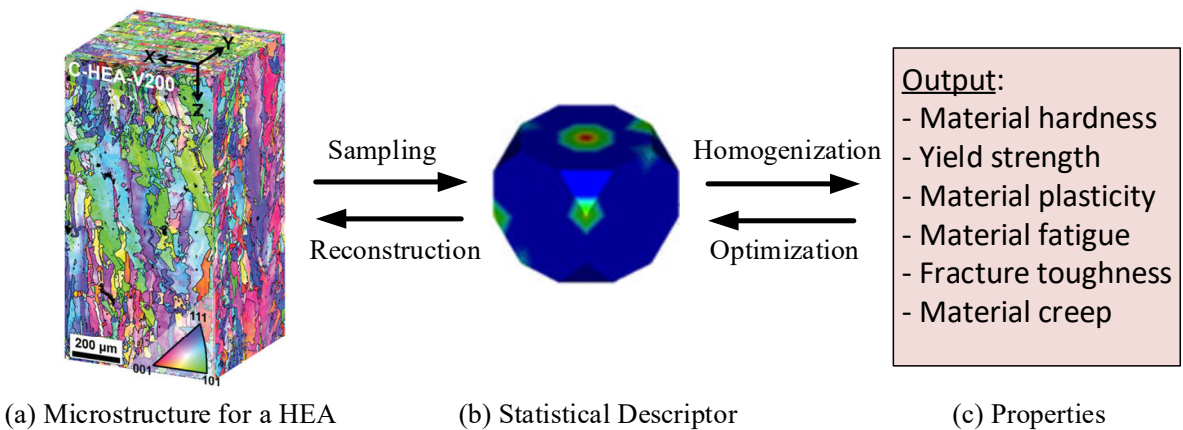

(a) Microstructure for a HEA          (b) Statistical Descriptor          (c) Properties

**Figure 18**: A baseline approach for inferring the microstructure from desired material properties [61].

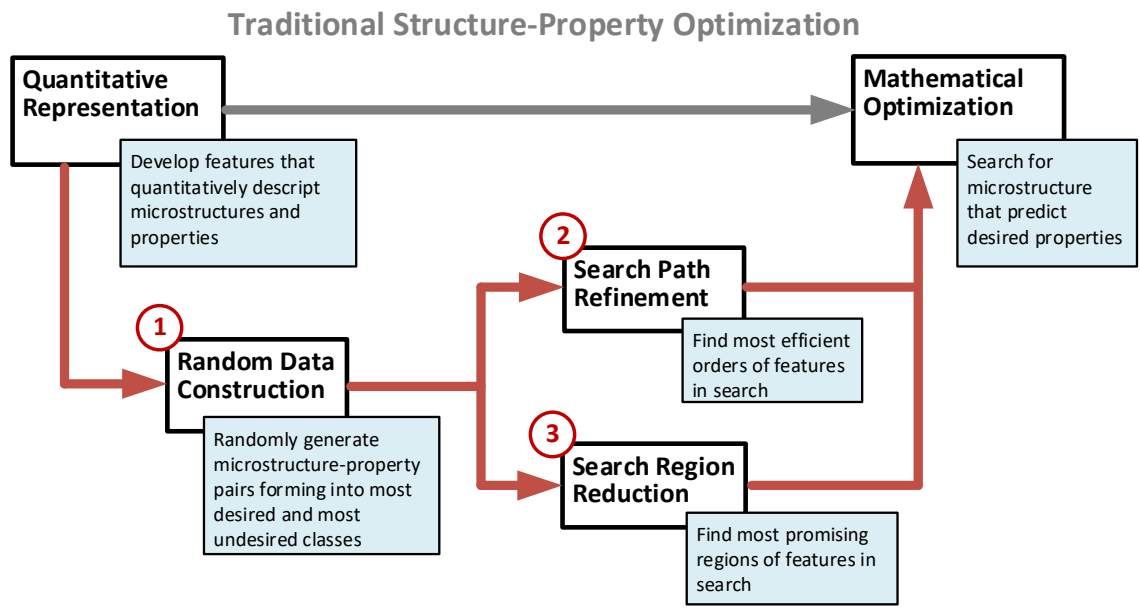

**Figure 19**: The Figure summarizes material structure optimization through microstructure-property pairs [61, 81].

Towards Simplification – Suitability of ML vs. Polynomial Fit in Absence of Discrete Jumps in Data

One can study if the alloy candidates of interest exhibit discrete jumps in the material properties observed. With increasing $x$, the AlxCoCrFeNi alloys change from the single face-centered-cubic (FCC) phase to single body-centered-cubic (BCC) phase with a transition duplex FCC / BCC region. If the data does involve continuous change (continuous 1st and 2nd derivatives), and there are not discrete jumps, similar to the on-set of superconductivity (where the conductivity suddenly exhibits a discrete jump to infinity), one can employ polynomial fit (of much lower complexity), in conjunction with ML, or at least as a part of a hybrid solution [61].

6. More on A Custom, Hybrid Solution for "Backward" Prediction: Polynomial Fit Employed for Complexity Reduction

One can employ ML to predict if one is close to a state transition. If not, then polynomial fit may suffice. In case of close proximity to a state transition, a ML approach may be necessitated [61].

Similar to [81] and [87], one can employ ensample prediction, i.e., build many models (often hundreds) in order to predict a given quantity of interest. Each model will make a prediction for a given new test point, and the final ensemble prediction will be given by the average value of all the individual model predictions. A polynomial function with numerical representation can be inverted. If the polynomial is monotonically increasing or decreasing, the inverse is unique. Otherwise, more than one inverse may exist, and the designer is at liberty of picking the best one [61].

7. Prediction of Distributions (Mean and Variance) – Stochastic Prediction

It makes more sense to develop a stochastic, as opposed to deterministic, predictors. The outputs of the stochastic predictors involve mean and variance, not a single scalar quantity. Through stochastic prediction, one can determine what % of the data falls within a single std. dev ($\sigma$), what % of the data falls within $2\sigma$'s, etc. [61].

## 4. Physics-Based Models

### 4.1. Key Take-Aways

1. This section presents methods for employing physics-based models, i.e., models that account for physical dependencies, and factor in the underlying physics as *a priori* information, during the prediction process.

2. For cases where an ANN is deemed suitable for the application at hand, this chapter introduces custom kernel functions consistent with the underlying physics, for the purpose of attaining tighter coupling, better prediction, and extracting the most out of the - usually limited - input data available.

3. There can be great benefits derived from combining ML with physics-based modeling approaches for alloys and composites, for improved prediction accuracy. Such modeling approaches can offer physical insight as unexplored regions of the composition space are investigated. These approaches may include thermo-dynamics, first-principle effects, empirical rules, mesoscale theory, models for dislocation dynamics and slip bands to help with accurate prediction of stress/life curves, or feature representations for distinguishing between hot corrosion attacks.

4. In case of the identification of HEA compositions yielding the high tensile strength, we present prediction methods capable of yielding *consistency* amongst (a) prediction of HEA compositions with attractive tensile strength, (b) empirical rules of thermodynamics [105, 106], and (c) experimental results, despite relatively limited data being available, and the corresponding selection of a simple prediction algorithm (multi-variate regression).

### 4.2. Generic Approach to Incorporating Physics-Based Models

For incorporating a physics-based model into a prediction mechanism, we recommend first developing qualitative understanding of the physics and the dependencies underlying the data available as well as understanding of the underlying the sources of variation. We then recommend crafting a generic mathematical model describing the data. In case of limited data, a simple, linear model may be a suitable starting point. But if supported by sufficient data for training, one can employ a more sophisticated model. Next, we recommend introducing non-linearity into the model, based on the underlying physics. The kernel functions of the non-linear models may utilize $\tanh(\cdot)$, $\log(\cdot)$, or $\exp(\cdot)$ functions, based on applications. In the case of reliability analysis, one can choose exponential functions. By carefully formulating the structure of the models, such as to capture underlying dependencies, together with *a priori* knowledge, derived from the underlying physics, one can expect to infer more from the – usually limited - data available than when directly using the same data to train generic, out-of-the-box models (with no *a priori* knowledge incorporated) [61].

### 4.3. Overview over the Primary Physics-Based Models Considered

There can be great benefits derived from combining ML with physics-based modeling approaches for design of alloys or composites (in particular HEMs), for improved prediction accuracy. These modeling approaches may include [61]

1. Thermo-dynamics (CALPHAD) [3, 39, 67, 68].

2. First principle calculations of bond energies and phase stability [55, 67, 106, 107].

3. Empirical rules [105-108].

4. Mesoscale models to predict distribution of phases in the microstructure and their morphologies, as influenced by thermal history and alloy chemistry [109, 110].

5. Models involving dislocation dynamics, solid solution strengthening or slip band information.

6. Feature representations useful for characterizing, and distinguishing between, chemical reactions associated with specific corrosion attacks, such as CMAS or calcium sulfate attacks.

Table 8 lists leads towards incorporating physics-based intuition into machine learning algorithms for predicting the material properties of alloys. In reference to Table 8, mechanical nano-twins at low temperature can result in continuous steady strain hardening, which improves both fracture toughness and strength, according to Gludovatz et al. [55]. By incorporating physics-based models, one can help the ML algorithms to avoid fitting to the input data. The physics-based models can also help with extrapolation into uncharted territories (into subspaces of the parameter space for which there are few or no experimentally obtained data points). The physics-based models can help with developing understanding into complex combinations of material and process-induced imperfections [61].

| ID | Quantity Predicted | Sources of the Input Data | Leads towards Incorporating Physics-Based Intuition |
|----|--------------------|---------------------------|-----------------------------------------------------|
| 1 | Hardness | [3-5, 111-113] | X. Q. Chen et al. can model hardness of polycrystalline materials [114] |
| 2 | Yield Strength | [1, 3-5, 55, 112, 115-117] | W.A. Curtin et al. can predict yield strength, only based on edge or screw dislocation. Can predict yield strength of high-entropy alloys [118-123]. W. Jiang et al. have theory to predict strength and explain why strength higher than conventional metal [124]. |
| 3 | Plasticity/Ductility | [3-5, 22, 55, 115] | [125-129] |
| 4 | Fracture Toughness | [3, 4, 55, 130-133] | [134, 135] |
| 5 | Fatigue Resistance | [24, 132, 133, 136-141] | For prediction of the S/N curve, refer to Example 6 in Section 5.6; but also [24, 142] |
| 6 | Creep Resistance | [143-145] | (a) Dislocation theory to predict regular (smooth) creep behavior [145] (b) See how stress changes; based on nano-indentation [146-148] |

**Table 8**: Further specifics on application of machine learning to prediction of material properties [61].

1. Thermodynamics: Interaction of ML with Phase Stability Models from CALPHAD

The CALPHAD methodology employs a phenomenological approach to calculate multi-component phase diagrams, based on binary phase information, as shown in Figure 6, Figure 20 and Figure 21. Per Figure 6 and Figure 22. CALPHAD enables material designers to estimate the thermodynamic properties of ternary systems (A-B-C combinations), based on (a) thermodynamic database and (b) Gibbs free energy. CALPHAD can also help material designers analyze the phase stability of quad systems (A-B-C-D combinations), based on properties of the binary and ternary systems [61, 73].

Applications, based on the CALPHAD methodology, such as Thermo-Calc, can tell material designers how stable, or how meta-stable, each phase in the ternary or quad system is. CALPHAD can provide estimates for stability of specific phases in a multi-component systems consisting of up to 20 compositions [61].

CALPHAD can develop the phase diagrams to determine/confirm the stability and volume fraction of various phases at different temperatures, in collaboration with first-principle modeling and molecular dynamics calculations. Moreover, CALPHAD can predict the compositions of different phases, which can be compared with atom-probe tomography results and first-principle predictions [61].

CALPHAD relies on lots of experimental data, much of which was collected in the fifties or sixties. *There are gaps in the CALPHAD database*, such as in regards to B2 phase in multi-component systems (the B2 phase that does not appear in binary or ternary systems). As a work-around, the B2 phase may be modeled into binary or ternary systems, even though it is not stable there (only meta-stable) [61].

The methodology of accelerated design and qualification of new alloys can be extended such as to efficiently exploit the configurational entropy in high-entropy alloy systems. The solution thermodynamics of solvent-rich systems is well described by binary interactions. But the HEA systems involve significant contributions from ternary interactions that also need to be quantified for accurate prediction of phase stability. Extension of CALPHAD databases for this purpose can be facilitated by DFT calculations of mixing enthalpies in equi-atomic ternary systems [61].

The determination of the most appropriate heat-treatment process may also rely on thermodynamics calculations (formation of a $2^{nd}$ phase) [61].

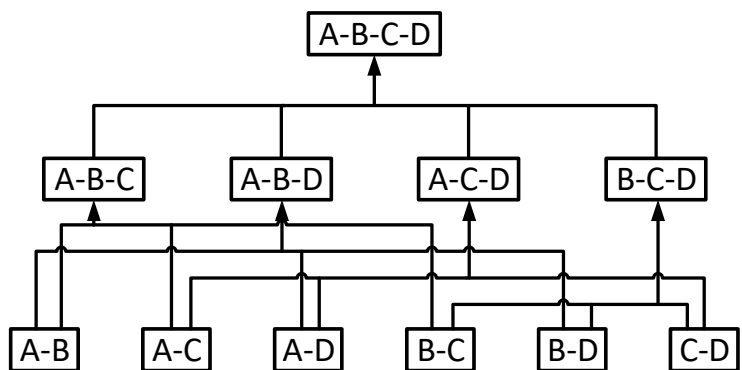

**Figure 20**: A schematic illustrates how properties of a quaternary system (A-B-C-D) are associated, at a high level, with the properties of the constituent binary systems (A-B, A-C and B-C).

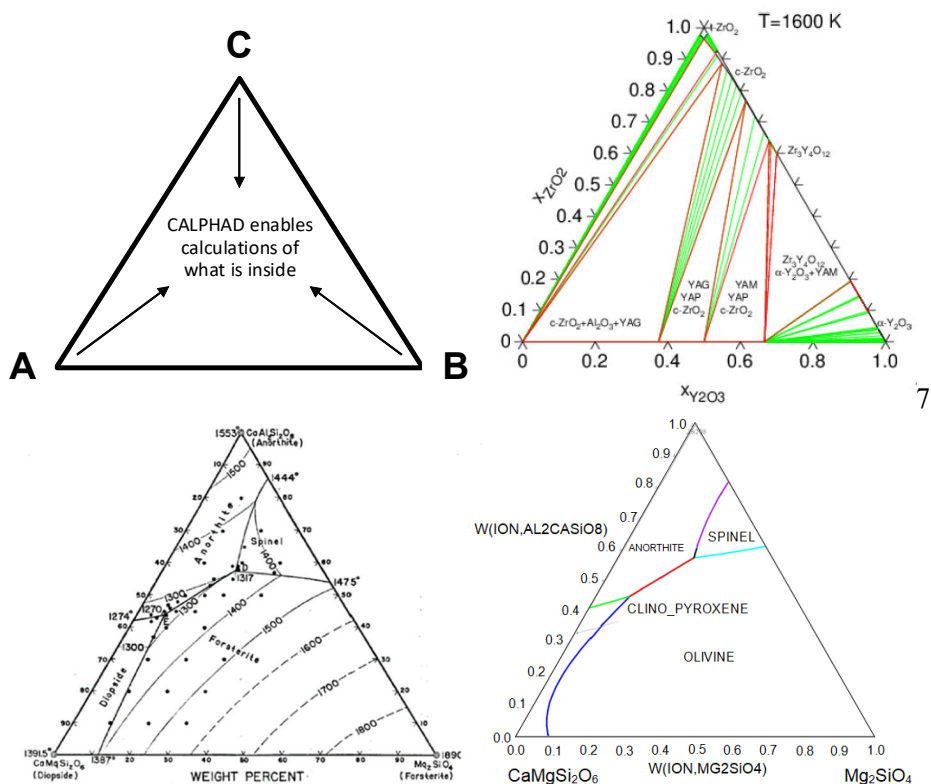

**Figure 21**: The Figure presents an illustration of analysis of properties of ternary material systems in CALPHAD [61].

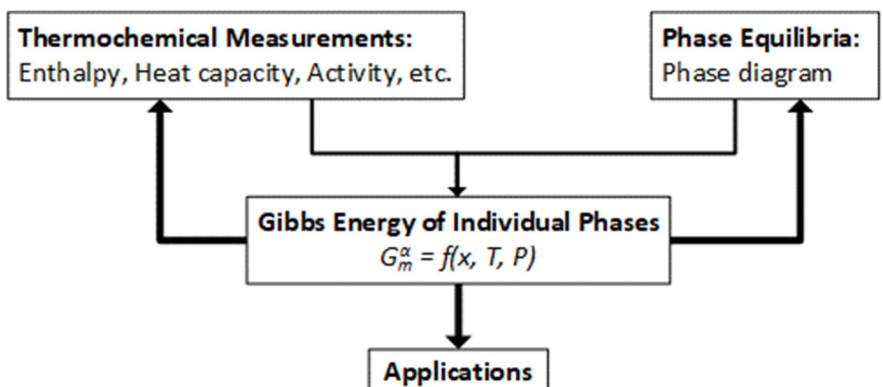

**Figure 22**: The Figure illustrates, at a high level, how CALPHAD computes the Gibbs free energy of individual phases from thermochemical measurements and phase equilibria information [61, 73].

## 2. First-Principle Effects

Investigations of alloys from first-principle perspective involve computational modeling at atomic scale, for example of bond energies or phase stability. Here, quantum and statistical mechanical first-principles are employed for modeling of molecular dynamics The first-principle studies leverage quantum mechanics calculations, with density functional theory (DFT) approximating the Schrodinger equation, to simulate the electronic properties and stability of material candidate from which the most promising leads are confirmed experimentally. Even so, DFT calculations tend to be computationally expensive, often taking hours to days for a single molecular structure, and more accurate results are often associated with lengthier calculations performed at higher levels of theory. On the other hand, properly trained neural networks can (in theory) yield highly accurate predictions with relatively low computational cost [61].

As suggested by Figure 6, applications, based on the CALPHAD methodology, such as Thermo-Calc, do not carry out first-principle (DFT) calculations explicitly. However, applications based on CALPHAD can accept first-principle data from applications such as the Vienna Ab initio Simulation Package (VASP). VASP is a computer program for atomic scale materials modelling, e.g. electronic structure calculations and quantum-mechanical molecular dynamics, from first principles. VASP computes an approximate solution to the many-body Schrödinger equation, either within DFT, by solving a Kohn-Sham equations, or within a Hartree-Fock (HF) approximation, by solving Roothaan equations [149]. Hybrid functionals that mix the Hartree-Fock approach with density functional theory are implemented in VASP as well. Furthermore, Green's functions methods and many-body perturbation theory (2nd-order Møller-Plesset) are available in VASP as well [61].

First-principle (DFT) calculations can be used to validate (sanity check) the prediction outcomes of traditional ML systems, i.e., as a part of a hybrid computational system [61].

## 3. Empirical Rules

The empirical rules of [105] predict the formation of solid solution phases, based on *Delta*, which describes the comprehensive effect of the atomic-size difference in multi-component alloy systems, and the mixing enthalpy of a solid solution, $\Delta H_{mix}$. The empirical rules specify which combinations of ($\Delta H_{mix}$, *Delta*) result in formation of a $S$ zone (a zone where only solid solution will form), which combinations result in formation of a $S'$ zone (a zone where a solid solution represents a main phase), which combinations result in a $B_1$ zone, which in a $B_2$ zone and which in a $C$ zone. The $B_2$ zone contains Mg and Cu based bulk metallic glasses, while the $B_1$ zone contains other kinds of bulk metallic glasses, such as Zr. In the $C$ zone, many intermediate phases are expected to form [61].

In Table 16, we apply the empirical rules of [105] to verify the sanity (phase stability) of compositions predicted to yield high tensile strength, on basis of machine learning or regression analysis. Similar to the first-principle (DFT) and thermodynamics (CALPHAD) calculations, the empirical rules can be incorporated into a hybrid computational paradigm, and used to validate (sanity check) the prediction outcomes of traditional ML [61].

One can formulate additional empirical rules, which are based on physics, and incorporate into ML prediction algorithms. One type of empirical rules may be based on the Poisson ratio, i.e., the elastic displacement/strain on different directions, in a lattice. Another category of empirical rules may address the short-range order (SRO) and shear transformation zone (STZ) expected, assuming one were to insert specific elements into a BCC lattice [61].

## 4. Mesoscale Models

The mesoscale models can be used to predict distribution of phases in the microstructure and their morphologies, as influenced by thermal history and alloy chemistry [109, 110]. Similar to the first-principle calculations, and the empirical rules, this phase information can be used to complement (validate) the prediction outcomes of traditional ML, i.e., as a part of a hybrid computational system [61].

## 5. Models Involving Dislocation Dynamics, Solid Solution Strengthening or Slip Band Information

Models involving dislocation dynamics or slip band information may be incorporated into prediction models for fatigue life. Such models may help with accurate prediction of stress/life (S/N) curves. The stress/life curves tend to be related to crack initiation, which can be associated with dislocation dynamics and slip band information [61]. Further, solid solution strengthening can be related to lattice distortion, which can be used to enhance the strength of HEAs, as shown in [150].

## 6 Environmental Resistance (Oxidation, Corrosion or Radiation)

Depending on the chemical reactions involved, and the temperatures at which they occur, one can derive a list of features that properly describe the data, e.g., using canonical component analysis. This can help in terms of developing distinguishing characteristics between CMAS and calcium sulfate ($CaSO_4$) hot corrosion, with and without the influence of sea salt, and with developing coatings resistant to CMAS and calcium sulfate hot corrosion [61].

## 4.4. General Approach to Construction of a Physics-Based Model - Application to Prediction of Ultimate Tensile Strength

1. Approach

First, one can start out by capturing the physics-based dependencies, per Figure 23. Assuming a linear model works, each arrow in Figure 23 represents a coefficient in a matrix. In this case, the parameters at the $1^{st}$ level consist of manufacturing, heat treatment, and processing. The parameters at the $2^{nd}$ level consist of microstructure, grain size, and defect levels. As depicted in Figure 23 the dependence relationships still represent a linear model [61].

Second, the approach assumes constructing an initial, linear regression model with input parameters from a continuous range. The first level of the model can be represented as follows [61]:

$$\mathbf{z} = [\text{microstructure, grain size, defect level}], \tag{51}$$

$$\mathbf{x} = [\text{manufacturing, heat treatment, processing}], \tag{52}$$

$$\mathbf{z} = \mathbf{A}\,\mathbf{x} + \mathbf{c}. \tag{53}$$

Now, the second level in Figure 23 can be written as shown below [61]:

$$\text{UTS} = y = \mathbf{b}^T \mathbf{z} + d. \tag{54}$$

This equation could give rise to an overall linear model of the following type [61]:

$$\text{UTS} = y = \mathbf{b}^T (\mathbf{A}\,\mathbf{x} + \mathbf{c}) + d = \mathbf{b}^T \mathbf{A}\,\mathbf{x} + (\mathbf{b}^T \mathbf{c} + d) = \tilde{\mathbf{a}}^T \mathbf{x} + e. \tag{55}$$

Note if, say, microstructure is a function of other input parameters, then these other input parameters should not be present at the current level, but should be accounted for at a preceding level [61].

Third, alternatively, one can construct an initial, linear model as follows [61]:

$$\text{UTS} = y = \tilde{\mathbf{a}}^T \mathbf{x}_B + e, \tag{56}$$

where, in the case of prediction of the UTS, $\mathbf{x}_B$ can be defined as [61]:

$$\mathbf{x}_B = [ \ \%\text{Al}, \%\text{Mo}, \%\text{Nb}, \%\text{Ti}, \%\text{V}, \%\text{Ta}, \%\text{Zr}, \%\text{Hf}, \%\text{Cr} \ ]. \tag{57}$$

Fourth, the set of input parameters ($\mathbf{x}$) can be extended, by adding other continuous-valued input parameters, such as temperature. At this point, we have accounted both for the compositions, the temperature, and the continuous-valued input parameters. This arrangement involves a relatively straight forward extension [61].

Fifth, one can further extend the set of input parameters ($\mathbf{x}$), by adding categorical inputs, such as for the manufacturing process used. At this point, the problem becomes a mixed-integer optimization problem. Our approach assumes gradual introduction of complexity into the model [61].

Sixth, one can introduce nonlinearities (non-linear kernel functions), based on the underlying physics. In case of Figure 23 [Eq. (55)], this formula can be rewritten as [61]:

$$\text{UTS} = y = \tilde{\mathbf{a}}^T f ([\text{microstructure, grain size, defect level}]) + e = \tilde{\mathbf{a}}^T f (\mathbf{x}) + e. \tag{58}$$

Here, $f(\cdot)$ may represent the sigmoid function, a $\log(\cdot)$ function, or an $\exp(\cdot)$. In this chapter, we assume that $f(\cdot)$ is selected such as to suit the application at hand [61].

Seventh, in case of the prediction presented in Figure 30 and Figure 31, the prediction model can be represented as [61]:

$$\text{UTS} = y = h(\mathbf{x}_B). \tag{59}$$

Some of the underlying physics may be common for regular alloys and HEAs. Hence, one may develop certain aspects of the model, through analysis of data available for regular alloys. Figure 24 and Figure 25 are presented with this in mind [61].

Eighth, one can systematically expand the model, such as to introduce additional complexity [61].

Ninth, one can systematically retrain the model, as additional complexity is introduced. If the addition of a non-linearity or an input parameter (complexity) does not improve the prediction accuracy of the model, one can analyze why the observed prediction accuracy has not conformed with expectations [61]?

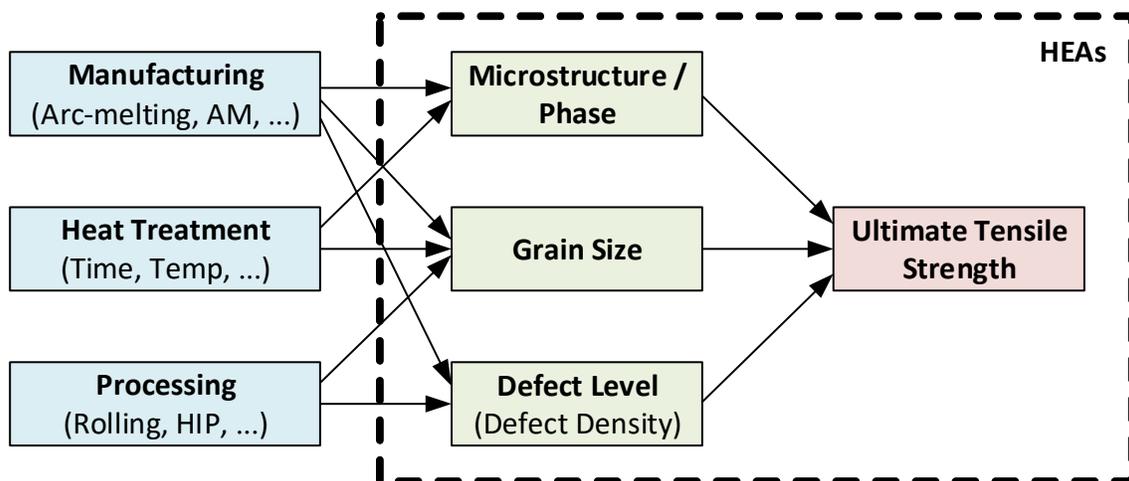

**Figure 23**: The Figure presents underlying physical dependencies structured in a fashion resembling a neural network. Our intent is to construct models capturing the underlying physics. The model shows that microstructure formed depends on the heat treatment process applied [61].

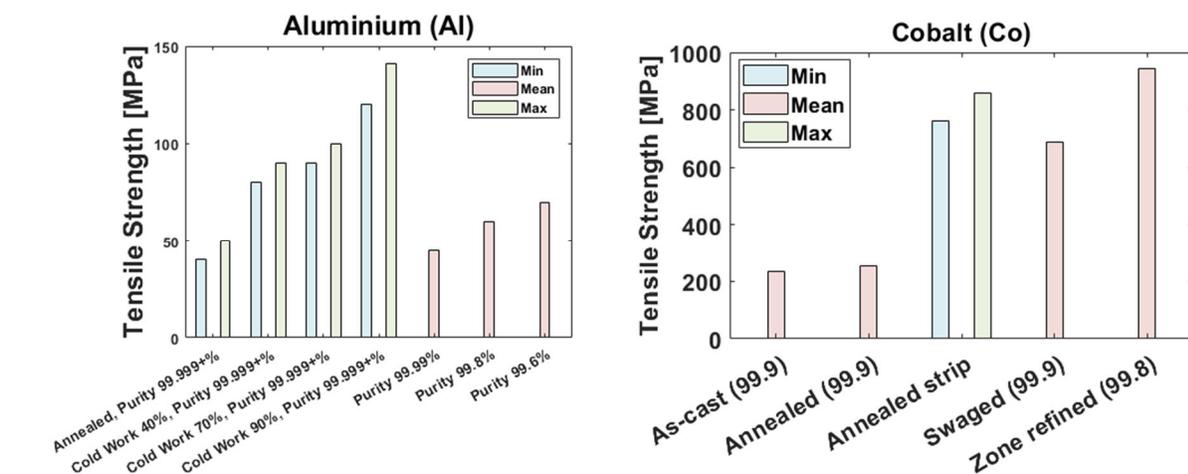

**Figure 24**: The Figure presents variations in tensile strength of Aluminum and Cobalt, based on purity and processing conditions [61].

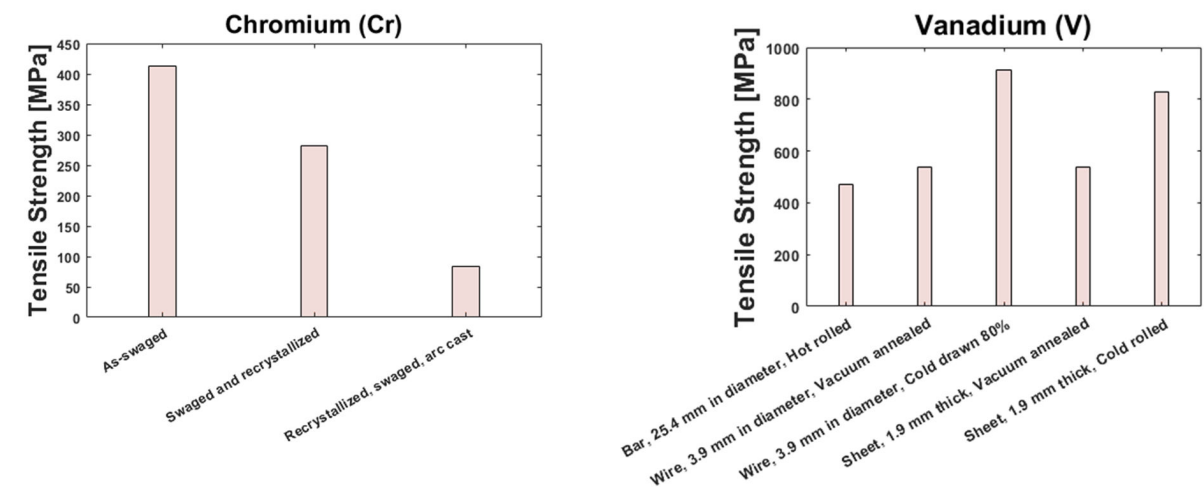

**Figure 25**: The Figure presents variations in tensile strength of Chromium and Vanadium, based on processing conditions [61].

2. Further Theoretical Considerations (Justifications)

Broadly speaking, for the selection of kernel functions based on underlying physics, one should first look at the general shape of the data. Sometimes, the selection may be based on qualitative physics-based insights of an expert. The expert may know what to expect, or if there appears to be too much or too little of a given concentration [61].

A parametrized description related to the microstructure may be based on the following parameters: Radius asymmetry, difference in atomic radii, number of valence electrons, enthalpy of mixing, ideal entropy of mixing, mean melting temperature, difference in Pauling electro-negativity, electro negativity difference, cohesive energy, first ionization energy, covalence radius, electron affinity, molar volume, etc. [61].

The parametrized description of the defect levels may involve a significant undertaking, as noted above, since there are so many different types to consider. A systematic approach to the parametrized description of defect levels may start with a single case (e.g., powder-bed AM) and a single category of defects (macro-scale, micro-scale or nano-scale defects) [61].

A parametrized description of the grain size may similarly involve a significant undertaking. A systematic approach  may  exclude metallic glasses (amorphous metals), but instead focus on characterizing the distribution in the size and shape of polycrystalline grains [61].

The simplest description of the heat treatment process may involve a simple listing by categories. However, such categorical designation may be subject to a level of arbitrariness with regards to how the data scatters along the axis listing the categories. In case of hot isostatic pressing, one would need to account for the temperature, pressure and duration of the HIP.  Similarly, in case of annealing, one would need to account for the temperature and duration of homogenization [61].

## 4.5. Necessary Steps towards Accurate Prediction: Characterization of Expected Sources of Variations – Application to Prediction of Ultimate Tensile Strength

In order to yield highly accurate predictions, one needs to understand the sources causing variations in the property predicted, and properly account for these sources [61].

1. Expected Dependence of Tensile Strength on Alloy Type

As an example, we expect the mechanical properties of refractory HEAs to differ from those of traditional alloys. For refractory HEAs and transition metals, we expect higher yield strength and lower ductility, compared to other alloys [61]. By increasing the concentration of aluminum (Al) in transition metal type materials, one has a reason to believe the strength will likely improve [151].

2. Expected Dependence of Tensile Strength on Temperature

Usually, when the temperature increases, the ultimate tensile strength will tend to decrease. However, we are unaware of a theoretical model describing this relationship. However, with that said, Figure 26 captures empirical results supporting this observation [61].

3. Expected Dependence of Tensile Strength on Manufacturing Technique

Dependence of the tensile strength on the manufacturing technique employed may depend on specifics of the implementation. Traditionally, arc-melting and spark plasma sintering have been the two main processing techniques employed to fabricate bulk HEAs [152]. To successfully produce homogeneous bulk HEAs by arc-melting, extensive re-melting and intermittent ingot inversions are required, and powder alloying and refinement (typically via balling milling) is necessary when processing via the SPS route [152]. The main undesirable feature of the metal additive manufacturing process are the non-equilibrium thermal cycles, consisting of the solid-melting crystallization and solid-remelting recrystallization under fast heating and cooling conditions, which can generate anisotropic microstructures and defects [61].

4. Expected Impact of Grain Size on Yield or Tensile Strength

Grain-boundary strengthening (or Hall–Petch strengthening) is a method of strengthening materials by changing their average grain size [153]. It is based on the observation that grain boundaries are insurmountable borders for dislocations and that the number of dislocations within a grain have an effect on how stress builds up in the adjacent grains, which will eventually activate dislocation sources and thus enable deformation in the neighboring grains. Therefore, by changing grain size, one can influence the number of dislocations piled up at the grain boundary, which impacts the yield and tensile strengths. The Hall-Petch relation models the relationship between the yield stress and the grain size as [61]:

$$\sigma_y = \sigma_0 + \frac{k_y}{\sqrt{d}}. \tag{60}$$

Here, $\sigma_y$ is the yield stress, $\sigma_0$ is a materials constant for the starting stress for dislocation movement (or the resistance of the lattice to dislocation motion), $k_y$ is the strengthening coefficient (a constant specific to each material), and $d$ is the average grain diameter [153].

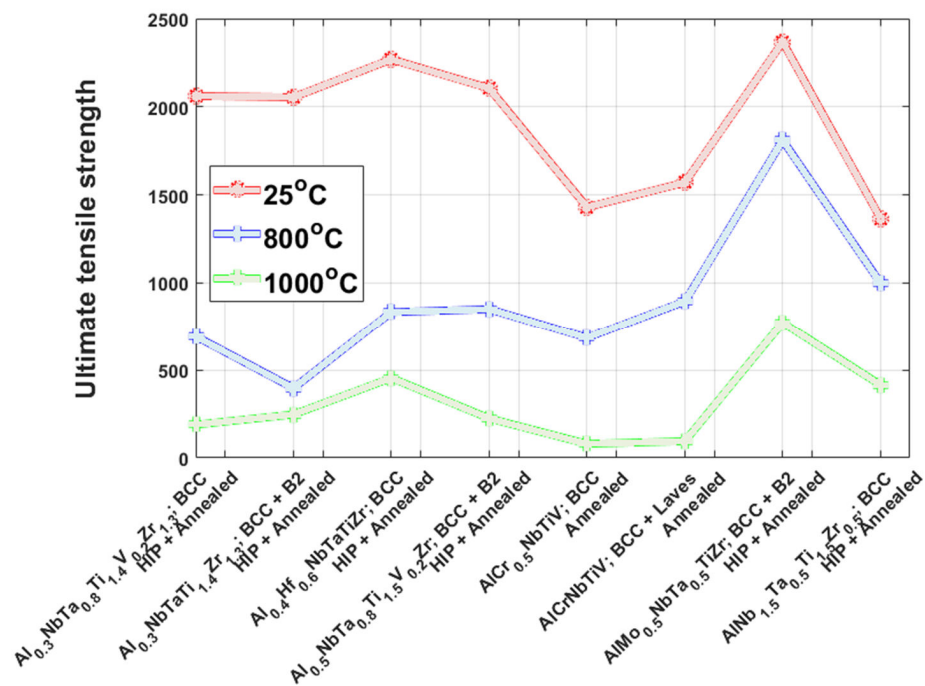

**Figure 26**: Towards the characterization of temperature dependence of the ultimate tensile strength of HEAs. For a given temperature, the data points listed correspond to the same heat-treatment process [61].

5. Expected Dependence of Endurance Limit on UTS, for Fixed Defect Level and Heat Treatment Process
Upon studying stress states, fracture surfaces, and tensile stress at the fracture-initiation site, one can expect the fatigue-endurance limit to scale in proportion with the UTS, for a fixed defect level and heat-treatment process [61]. Reference [154] cites a study of the stress-life fatigue behavior of a $Zr_{41.25}Ti_{13.75}Ni_{10}Cu_{12.5}Be_{22.5}$ bulk metallic glass using notched cylindrical bars, where the fatigue endurance limit of ~1/2 of the UTS was reported. This result was significantly higher than the value of ~1/10 of the fatigue endurance limit previously reported using four-point bend specimens [154].

6. Expected Dependence of Endurance Limit on Defect Levels, for Fixed UTS and Heat Treatment Process
Defect levels are here taken to broadly represent microstructural effects. While one expects the increased defect level to exert adversarial impact on the endurance limit, complete characterization of microstructural aspects may involve a significant undertaking. Reference [155] reports on the intrinsic role of microstructure on persistent slip bands. Reference [155] notes that although the nano-sized $L1_2$ precipitates enhance tensile strength, no improvement in fatigue properties have been observed [61].

### 4.6. Necessary Steps towards Accurate Prediction: Characterization of Sources of Observed Variations – Application to Prediction of Ultimate Tensile Strength

Table 9 *shows that one can expect ~ 2x variations in the endurance limit, based on defect levels (defect size, density, and type) and raw material purity*, for a fixed UTS. This trend suggests that the explicit access to the information on the defect level may be needed in order to accurately predict the endurance limit. The samples in Table 9 were homogenized at 1,000 °C for 6 hour, water quenched, and then cold rolled. For Condition 1, shrinkage pores and macro-segregation remained in some portions. For Conditions 2 and 3, shrinkage pores and macro-segregation were removed before cold rolling [61] .

*Table 10 similarly illustrates that one can expect ~ 2x variations in the endurance limit, likely caused by variations in the grain size*, even for the same microstructure (FCC) and similar process (hot-rolled and heat-treated). Together, Table 9 and Table 10 illustrate that accurate prediction of the endurance limit is not possible, based on the UTS alone. One also needs to know the defect levels (the defect size, density, and type), the grain size, and even

parameters of the heat-treatment process [61]. These observations are consistent with those of [24] as well as with those of [141].

1. Comparison across Compositions, Process Parameters, Defect Levels, and Grain Sizes for a Given UTS – Further Explanations of the Scatter

1. UTS ≈ 1,100 MPa

Variations in the defect level alone can result in ~2x variations in the endurance limit for the multi-variate data point, as noted in Table 9. At UTS ≈ 1,100 MPa, the variations observed can additionally be explained in terms of variations in the grain size, microstructure and composition, per Table 10 [61].

2. UTS ≈ 1,340 MPa

At UTS ≈ 1,340 MPa, the variations observed can similarly be explained in terms of variations in the microstructure, grain size, and processing parameters, per Table 9 and Table 10. One cannot expect the accurate prediction of the fatigue resistance, unless knowing parameters of the heat treatment process and the defect levels, in addition to the UTS [61].

| Composition | Micro structure | Grain Size (um) | Process | Defects Reported | Tensile Strength (MPa) | Endurance Limit (MPa) |
|---|---|---|---|---|---|---|
| Al0.5CoCrCuFeNi Condition-1 | 2 FCC | 2 and 1 mm (matrix phase and Cu-rich) | annealed + cold-rolled | High defect level | 1,344 | 270 |
| Al0.5CoCrCuFeNi Condition-3 | 2 FCC | 2 and 1 mm (matrix phase and Cu-rich) | annealed + cold-rolled | high-purity raw elements | 1,344 | 360 |
| Al0.5CoCrCuFeNi Condition-2 | 2 FCC | 2 and 1 mm (matrix phase and Cu-rich) | annealed + cold-rolled | commercial-purity raw elements | 1,344 | 382 |
| Al0.5CoCrCuFeNi Condition-1 | 2 FCC | 2 or 1 mm (matrix phase or Cu-rich) | annealed + cold-rolled | Few defects- | 1,344 | 472 |

**Table 9**: For a given composition (Al$_{0.5}$CoCrCuFeNi), microstructure, grain size, and process, the endurance limit exhibits a high degree of correlation with the defects reported [24, 61, 141].

| Composition | Micro-structure | Grain Size (um) | Manufact-uring Technique | Process | Defects Reported | Load Ratio (R) | Tensile Strength (MPa) | Endurance Limit (MPa) |
|---|---|---|---|---|---|---|---|---|
| CoCrFeNiMn | FCC (random solid solution) | 245.48 (avg.) | Vacuum induction melting | hot-rolled + heat-treated | Not Specified | 0.1 | 625.6 | 126 |
| CoCrFeNiMn | FCC (no sign of martensitic transform.) | 41 | Vacuum induction melting | hot-rolled at 1373K + solution-treated at 1073K + water-quenched | | -1 | 585 | 250 |
| CoCrFeNiMn | FCC (single phase, coarse grained) | 250 – 500 | Thermite-type self-propagating high-temperature synthesis (centrifugal set-up) | as-sintered | | 0.1 | 362 | 90 |
| CoCrFeNiMn | | 100 – 300 | | laser beam welded | | 0.1 | 349 | 90 |

| CoCrFeNiMn | Only micro-structural characteristics Specified | 0.407 (median) | Powder metallurgy (ball milling process) | Spark plasma sintering (SPS) at 1150 °C for 5 min | 0.1 | N/A (Bending test) | 495 |
| CoCrFeNiMn | | 0.628 (median) | | | 0.1 | | 450 |

**Table 10**: Variations in endurance limits for CoCrFeNiMn (the Cantor alloy) [61, 156-159].

2. <u>Comparison across Process Parameters, Defect Levels, Grain Sizes, and UTS for a Given Composition</u>

1. Comparison for AlCoCrFeNi$_{2.1}$

We believe that the increments of the UTS and endurance limit, shown in Table 11, are – at least in part - caused by the lower defect level for the AlCoCrFeNi$_{2.1}$ cold-rolled and heat-treated eutectic HEA (EHEAw) composition. Since the EHEAw samples were annealed after cold rolling, the grain sizes are likely similar [61]. In addition to defect structures and grain sizes, the variations may be impacted by persistent slip bands in the micro-structure [106, 160].

| Composition | Micro-Structure | Grain Size (um) | Process | Defects Reported | Tensile Strength (MPa) | Endurance Limit (MPa) |
|---|---|---|---|---|---|---|
| AlCoCrFeNi2.1 EHEAw | FCC+BCC | Not Specified | cold-rolled + heat-treated | Not Specified | 1,340 | 466 |
| AlCoCrFeNi2.1 EHEAc | | | as-cast | | 1,057 | 374 |

**Table 11**: Variations in endurance limits for the as-cast eutectic HEA (EHEAc) and cold-rolled and heat-treated eutectic HEAs [61, 137].

2. Further Comparison for CoCrFeNiMn

Similar to the case for AlCoCrFeNi$_{2.1}$, we believe that the hot-rolled and heat-treated process may have contributed to a little higher UTS and endurance limit for CoCrFeNiMn listed in Table 10 and Figure 27, compared to the cases of as-sintered or laser-beam welded. But more importantly, we believe that the smaller grain size is resulting in a higher strength and endurance limit, per the Hall-Petch relation (Eq. (61)). The hot rolling and heat treatment may have given rise to smaller grain sizes, compared to sintering or laser beam welding. Figure 27 demonstrates clear inverse correlation between the endurance limit and the grain size, but positive correlation between the endurance limit and the UTS [61]. For information on the fatigue-crack growth behavior of CoCrFeNiMn, refer to [133].

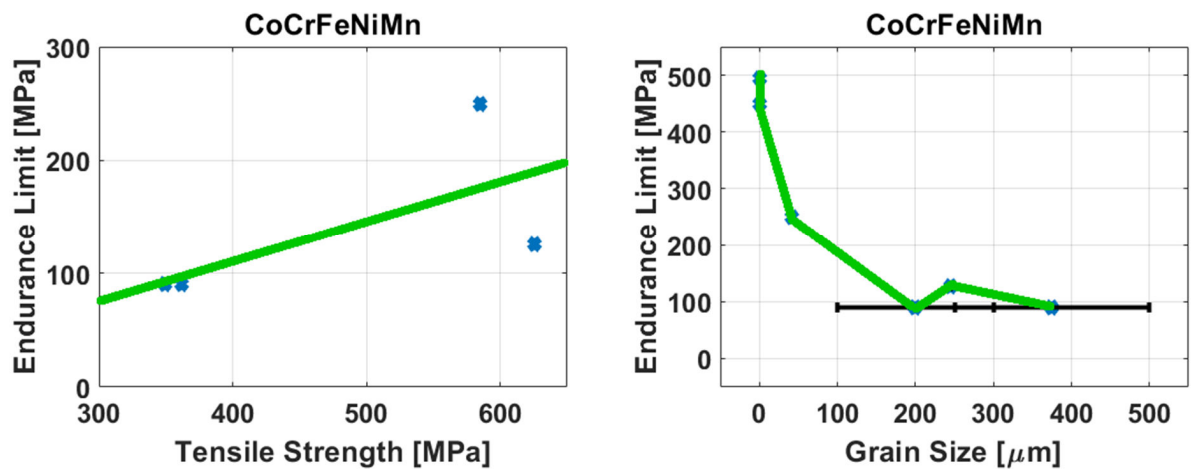

**Figure 27**: The Figure captures analysis of the dependence of the endurance limit of CoCrFeNiMn (the Cantor alloy) on gain size (right) and tensile strength (left) [61].

3. Comparison for 4340 Steel

Since both 4340 steel samples were heat treated, we expect similar defect levels. Judging from Table 9 - Table 11, the difference in the tensile strength and endurance limit, listed in Table 12, likely is caused by the difference in the grain size, and possibly – depending on the annealing temperature (not specified in [152]) – microstructure (phases) [61].

| Composition | Micro-structure | Grain Size (um) | Process | Defects Reported | Tensile Strength (MPa) | Endurance Limit (MPa) |
|---|---|---|---|---|---|---|
| 4340 Steel | Not Specified | Not Specified | Annealed | Not Specified | 745 | 170 |
| 4340 Steel | | | Quenched & tempered at 538°C | | 1,260 | 335 |

**Table 12**: Variations in endurance limits for 4340 steels [61, 161].

## 4.7. Example 1: Identification of Compositions Yielding High Tensile Strength

1. Review of the Original Data Set - Rational for Restricting Analysis to Room-Temperature Data

As illustrated in Figure 26, the ultimate tensile strength exhibits the significant dependence on the temperature. While all compositions in Figure 26 contained a BCC phase, and are subjected to some type of annealing, the tensile strength at 1,000°C can be ~ 1/8 (~ 12%) to ~ 1/3 (~ 33%) of the tensile strength at room temperature. With this in mind, and to maintain consistency across compositions, we elect only to apply our optimization framework to tensile strengths at room temperature. Our original data set, listed in Table 13, contains some 24 compositions that yield relatively-high UTS at room temperature. We derive two feature sets, hereafter referred to as A and B from the original data set in Table 13 [61]:

$$\text{Feature Vector A} = \mathbf{x}_A = [\ \%Al, \%Mo, \%Nb, \%Ti, \%V, \%Ta, \%Zr, \%Hf\ ] \quad (61)$$

$$\text{Feature Vector B} = \mathbf{x}_B = [\ \%Al, \%Mo, \%Nb, \%Ti, \%V, \%Ta, \%Zr, \%Hf, \%Cr\ ] \quad (62)$$

We have available nineteen (19) instances of the feature vector, A, and twenty two (22), of the feature vector, B. While the set of input data may seem small, we will show that it suffices for the meaningful prediction, provided that a suitable optimization technique is selected [61].

2. Analysis of Variations in UTS for the Pure Elements – Selection of a Suitable Prediction Model

In order to develop insight into the causes of variations in tensile strengths for the pure elements comprising feature vectors A and B, and for the identification of a model for predicting compositions yielding high tensile strengths, and presumably attractive fatigue resistance, we present Figure 24 and Figure 25. Figure 24 shows that processing conditions and purity can contribute to variations in tensile strengths of Al of ~3x and of ~4x in the tensile strength of Co. Figure 25 similarly illustrates that processing methods have significant influence on the tensile strength of V and Cr. For the Vanadium, the variations in the tensile strength are ~2x, and for the Chromium, we are looking at variations of up to ~5x (!) This trend suggests that an accurate model for predicting the tensile strength may indeed have the form of Eq. (50). But for relative comparison of tensile strengths across compositions, for the same heat-treatment process and defect levels, and at fixed (room) temperature, the model [61]

$$\text{UTS} = h(\text{composition}) \quad (63)$$

may suffice. For the prediction presented in Figure 28 and the experimental verification outlined in Figure 32, we employ the prediction model of Eq. (63) [61].

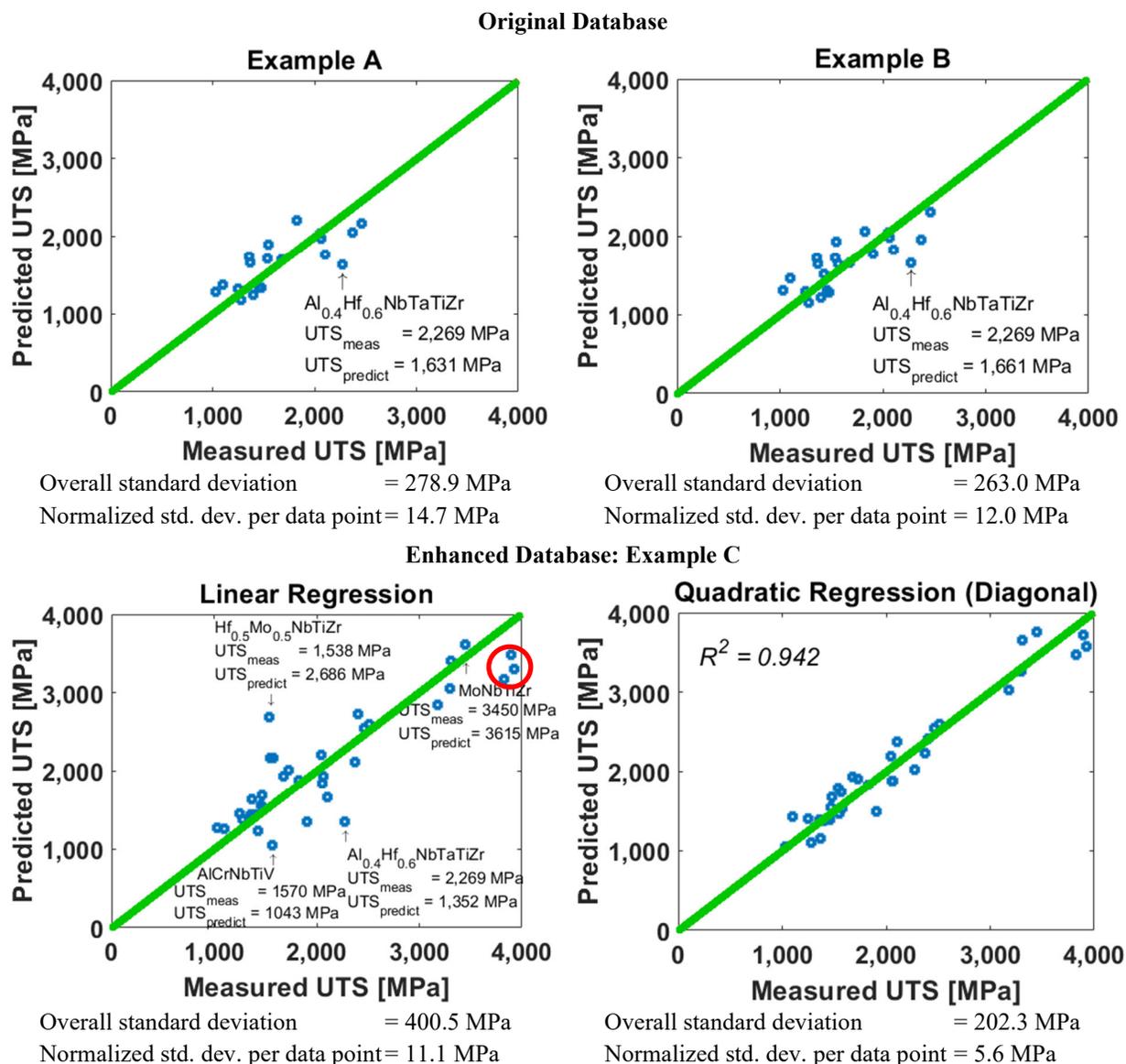

**Figure 28**: The Figure presents regression analysis applied to the Data Sets from Examples A, B and C [61].

### 3. Selection of a Suitable Optimization Technique

Given the small size of the data set in Table 13, it suffices to say that we are not ready for traditional ML models. Models, such as artificial neural networks, decision trees, support vector machines, Bayesian networks, or genetic algorithms, tend to be effective in organizing and extracting complex patterns from large sets of data, as noted above. *But for the application and limited data set at hand, it makes sense to select a simple linear-prediction model, multivariate linear regression, to begin with, and build from there.* As suggested by Agrawal et al. [82], changing the method may not change the results that much. According to Figure 5 and Table 2 in [82], the linear regression yields $R^2$ of 0.963, when predicting the fatigue strength of the stainless steel, compared to $R^2$ of 0.972 for the artificial neural networks [61].

Our intent is to start out with the statistical (linear) regression analysis, and account for the underlying sources of (input) variations. We intend to then expand the model, and add non-linearity, based on the underlying physics, and as necessitated by the application at hand and the data available [61].

| | No. | Composition | Micro-structure | Process | Select Process Specifics | UTS (MPa) | Example (Data Set) |
|---|---|---|---|---|---|---|---|
| **Original Database** | 1 | $Al_{0.25}NbTaTiZr$ | BCC + B2 | HIP + anneal | HIP: 2 hr., 1400°C | 1,830 | A, B, C |
| | 2 | $Al_{0.2}MoTaTiV$ | BCC | As-cast | N.A. Remelted a few $x$ | 1,249 | A, B, C |
| | 3 | $Al_{0.3}NbTaTi_{1.4}Zr_{1.3}$ | 2 BCC | HIP + anneal | HIP: 2 hr.,1200°C | 2,054 | A, B, C |
| | 4 | $Al_{0.3}NbTa_{0.8}Ti_{1.4}V_{0.2}Zr_{1.3}$ | BCC | HIP + anneal | HIP: 2 hr.,1200°C | 2,061 | A, B, C |
| | 5 | $Al_{0.4}Hf_{0.6}NbTaTiZr$ | BCC | HIP + anneal | HIP: 2 hr.,1200°C | 2,269 | A, B, C |
| | 6 | $Al_{0.5}Mo_{0.5}NbTa_{0.5}TiZr$ | BCC + B2 | HIP + anneal | HIP: 2 hr., 1400°C | 2,460 | A, B, C |
| | 7 | $Al_{0.5}NbTa_{0.8}Ti_{1.5}V_{0.2}Zr$ | 2 BCC | HIP + anneal | HIP: 2 hr.,1200°C | 2,105 | A, B, C |
| | 8 | $Al_{0.6}MoTaTiV$ | BCC | As-cast | N.A. Remelted a few $x$ | 1,033 | A, B, C |
| | 9 | $AlCr_{0.5}NbTiV$ | BCC | Annealed | Homog: 24 hr., 1200C | 1,430 | B, C |
| | 10 | $AlCrNbTiV$ | BCC+Laves | Annealed | Homog: 24 hr., 1200C | 1,570 | B, C |
| | 11 | $AlMo_{0.5}NbTa_{0.5}TiZr$ | BCC + B2 | HIP + anneal | HIP: 2 hr., 1400°C | 2,370 | A, B, C |
| | 12 | $AlNb_{1.5}Ta_{0.5}Ti_{1.5}Zr_{0.5}$ | BCC | HIP + anneal | HIP: 2 hr., 1400°C | 1,367 | A, B, C |
| | 13 | $AlNbTa_{0.5}TiZr_{0.5}$ | B2 | HIP + anneal | HIP: 2 hr., 1400°C | 1,357 | A, B, C |
| | 14 | $AlNbTiV$ | BCC | Annealed | Homog: 24 hr., 1200C | 1,280 | A, B, C |
| | 15 | $AlNbTiVZr$ | B2+Al3Zr5 +Laves | Annealed | Homog: 24 hr., 1200°C 99.9+% purities; Prior to annealing, the samples were encapsulated in vacuumed ($10^{-2}$ Torr) quartz tubes. | 1,675 | A, B, C |
| | 16 | $AlNbTiVZr_{0.1}$ | B2+Al3Zr5 | Annealed | | 1,395 | A, B, C |
| | 17 | $AlNbTiVZr_{0.25}$ | B2+Al3Zr5 | Annealed | | 1,480 | A, B, C |
| | 18 | $AlNbTiVZr_{1.5}$ | B2+Al3Zr5 +Laves | Annealed | | 1,550 | A, B, C |
| | 19 | $CrHfNbTiZr$ | BCC+Laves | Annealed | 973K for 600 sec | 1,908 | B, C |
| | 20 | $Hf_{0.5}Mo_{0.5}NbTiZr$ | BCC | As-cast | N.A. Melt 5 times | 1,538 | A, B, C |
| | 21 | $MoTaTiV$ | BCC | As-cast | N.A. Remelted a few $x$ | 1,454 | A, B, C |
| | 22 | $HfNbTaTiZr$ | BCC | Cold roll + anneal | 1373K anneal in He atmos. for 5 hr | 1,095 | A, B, C |
| **Enhanced Database** | 23 | $CrMo_{0.5}NbTa_{0.5}TiZr$ | 2BCC+FCC | HIP + anneal | 1723K/207MN/m²/3hr | 2,046 | C |
| | 24 | $MoNbTaV$ | BCC | As-cast | N.A. Remelted a few $x$ | 2,400 | C |
| | 25 | $CrNbTiZr$ | BCC+Laves | HIP + anneal | HIP at 1473K & 207MPa for 2 hr. | 1,575 | C |
| | 26 | $CrNbTiVZr$ | BCC+Laves | HIP + anneal | | 1,725 | C |
| | 27 | $HfNbTiVZr$ | BCC+Unknown | As-cast | N.A. Remelted a few $x$ | 1,463 | C |
| | 28 | $MoNbTiV_{0.25}Zr$ | BCC | As-cast | N.A. Remelted a few $x$ | 3,893 | C |
| | 29 | $MoNbTiV_{0.5}Zr$ | BCC | As-cast | N.A. Remelted a few $x$ | 3,307 | C |
| | 30 | $MoNbTiV_{0.75}Zr$ | BCC | As-cast | N.A. Remelted a few $x$ | 3,929 | C |
| | 31 | $MoNbTiV_{1.5}Zr$ | 2 BCC | As-cast | N.A. Remelted a few $x$ | 3,300 | C |

| 32 | MoNbTiV$_2$Zr | 2 BCC | As-cast | N/A. Remelted a few $x$ | 3,176 | C |
| 33 | MoNbTiV$_3$Zr | 2 BCC | As-cast | N/A. Remelted a few $x$ | 2,508 | C |
| 34 | MoNbTiVZr | BCC | As-cast | N/A. Remelted a few $x$ | 3,828 | C |
| 35 | MoNbTiZr | BCC | As-cast | N/A. Remelted a few $x$ | 3,450 | C |
| 36 | MoTaTiV | BCC | As-cast | N/A. Remelted a few $x$ | 1,454 | C |
| 37 | Al | Purity: 99.99% | | | 45 | C |
| 38 | Mo | Annealed | | | 324 | C |
| 39 | Nb | Annealed | | | 275 | C |
| 41 | V | Cold rolled | | | 828 | C |
| 40 | Ti | Purity 99.9% | | | 235 | C |
| 42 | Ta | Cold worked | | | 900 | C |
| 43 | Zr | Typical | | | 330 | C |
| 44 | Hf | Typical | | | 485 | C |
| 45 | Cr | As-swaged | | | 413 | C |

**Table 13:** Compositions from the original and enhanced databases yielding the high UTS at room temperature (25 °C). Compositions No. 1 –36 are all fabricated using arc melting [61].

4. Setting up the Optimization Problem

1. Multi-Variate Linear Regression

When applying the linear regression, we solve a constrained optimization problem of the form [61]

$$\min_{\mathbf{x}} \|\mathbf{B}\,\mathbf{x} - \mathbf{y}\|_2^2 \ \text{ such that } \ \begin{matrix} 0 \leq x_i \leq 100 \\ \sum_i x_i = 100 \end{matrix}. \tag{64}$$

Here, $\mathbf{y}$ represents a vector of tensile-strength values, but $\mathbf{B}$ the training set of compositions [a stacked version of $\mathbf{x}$ vectors, derived from Table 13]. To solve this constrained optimization problem, one can use a function from Matlab® or Octave called lsqlin [61].

2. Quadratic Regression with Diagonal Matrix (for Comparison)

When applying the quadratic regression, we model the UTS ($y$) as [61]:

$$y = \mathbf{x}^\mathsf{T}\,\mathbf{A}\,\mathbf{x} + \mathbf{b}^\mathsf{T}\,\mathbf{x} + c. \tag{65}$$

Assuming a general $\mathbf{A}$ matrix and a 9-element feature vector ($\mathbf{x}_B$), this model consists of [61]

$$9 + 9 \times 9 + 1 = 91 \text{ parameters.} \tag{66}$$

Using an unconstrained model with 91 parameters to fit to the data sets in Table 13 does not make sense, since the number of model parameters greatly exceeds the number of data points. Hence, we should be able to fit the model perfectly to the data. In order to reign in the model complexity, we restrict the $\mathbf{A}$ matrix to a diagonal form. In this case, the model consists of [61]

$$9 + 9 + 1 = 19 \text{ parameters,} \tag{67}$$

i.e., fewer parameters than listed for Data Set C in Table 13 [61].

In the case of a diagonal $\mathbf{A}$ matrix [61],

$$\mathbf{A} = \begin{bmatrix} a_{11} & 0 & \cdots & 0 \\ 0 & a_{22} & \cdots & 0 \\ 0 & 0 & \cdots & 0 \\ \vdots & & \ddots & \vdots \\ 0 & 0 & \cdots & a_{mm} \end{bmatrix}, \tag{68}$$

the quadratic regression can be cast as an enhanced version of standard linear regression. To accomplish this process, we write the training set, as shown below [61]:

$$\begin{aligned} y_1 &= a_{11}\,x_1(1)\,x_1(1) + a_{22}\,x_1(2)\,x_1(2) + a_{33}\,x_1(3)\,x_1(3) + \ldots + a_{99}\,x_1(9)\,x_1(9) + \mathbf{b}^\mathsf{T}\,\mathbf{x}_1 + c \\ y_2 &= a_{11}\,x_2(1)\,x_2(1) + a_{22}\,x_2(2)\,x_2(2) + a_{33}\,x_2(3)\,x_2(3) + \ldots + a_{99}\,x_2(9)\,x_2(9) + \mathbf{b}^\mathsf{T}\,\mathbf{x}_2 + c \\ y_3 &= a_{11}\,x_3(1)\,x_3(1) + a_{22}\,x_3(2)\,x_3(2) + a_{33}\,x_3(3)\,x_3(3) + \ldots + a_{99}\,x_3(9)\,x_3(9) + \mathbf{b}^\mathsf{T}\,\mathbf{x}_3 + c \\ &\cdots \\ y_N &= a_{11}\,x_N(1)\,x_N(1) + a_{22}\,x_N(2)\,x_N(2) + a_{33}\,x_N(3)\,x_N(3) + \ldots + a_{99}\,x_N(9)\,x_N(9) + \mathbf{b}^\mathsf{T}\,\mathbf{x}_N + c \end{aligned} \tag{69}$$

We then rearrange the terms as follows [61]:

$$\begin{aligned} y_1 &= c + x_1(1)\,x_1(1)\,a_{11} + x_1(2)\,x_1(2)\,a_{22} + x_1(3)\,x_1(3)\,a_{33} + \ldots + x_1(9)\,x_1(9)\,a_{99} + \mathbf{x}_1^\mathsf{T}\,\mathbf{b} \\ y_2 &= c + x_2(1)\,x_2(1)\,a_{11} + x_2(2)\,x_2(2)\,a_{22} + x_2(3)\,x_2(3)\,a_{33} + \ldots + x_2(9)\,x_2(9)\,a_{99} + \mathbf{x}_2^\mathsf{T}\,\mathbf{b} \\ y_3 &= c + x_3(1)\,x_3(1)\,a_{11} + x_3(2)\,x_3(2)\,a_{22} + x_3(3)\,x_3(3)\,a_{33} + \ldots + x_3(9)\,x_3(9)\,a_{99} + \mathbf{x}_3^\mathsf{T}\,\mathbf{b} \end{aligned} \tag{70}$$

...

$$y_N = c + x_N(1)\,x_N(1)\,a_{11} + x_N(2)\,x_N(2)\,a_{22} + x_N(3)\,x_N(3)\,a_{33} + \ldots + x_N(9)\,x_N(9)\,a_{99} + \mathbf{x}_N^\top\,\mathbf{b}$$

This process results in the linear system [61]

$$\begin{bmatrix} y_1 \\ y_2 \\ y_3 \\ \vdots \\ y_N \end{bmatrix} = \begin{bmatrix} 1 & x_1(1)x_1(1) & x_1(2)x_1(2) & \cdots & x_1(9)x_1(9) & \mathbf{x}_1^T \\ 1 & x_2(1)x_2(1) & x_2(2)x_2(2) & \cdots & x_2(9)x_2(9) & \mathbf{x}_2^T \\ 1 & x_3(1)x_3(1) & x_3(2)x_3(2) & \cdots & x_3(9)x_3(9) & \mathbf{x}_3^T \\ 1 & \vdots & \vdots & \ddots & \vdots & \vdots \\ 1 & x_N(1)x_N(1) & x_N(2)x_N(2) & \cdots & x_N(9)x_N(9) & \mathbf{x}_N^T \end{bmatrix} \begin{bmatrix} c \\ a_{11} \\ a_{22} \\ a_{33} \\ \vdots \\ a_{99} \\ \mathbf{b} \end{bmatrix} \tag{71}$$

$$\breve{\mathbf{y}} = \breve{\mathbf{X}}\,\breve{\mathbf{b}} \tag{72}$$

The least-squared solution of Eq. (71) can now be obtained in closed form as [61]:

$$\hat{\breve{\mathbf{b}}} = \left(\breve{\mathbf{X}}^T\,\breve{\mathbf{X}}\right)^{-1}\breve{\mathbf{X}}^T\,\breve{\mathbf{y}} . \tag{73}$$

5. <u>Prediction of Composition Yielding Higher UTS, and Presumably More Attractive Fatigue Resistance, Than Previously Observed – Based on Data Sets A and B</u>

Figure 29 and Figure 30 illustrate the process of predicting compositions yielding higher UTS, and presumably higher fatigue resistance, based on Data Sets A and B. The prediction process consists of three main steps [61]:

First, we identify the composition with the highest measured UTS, which in the case of Data Sets A and B is $Al_{0.5}Mo_{0.5}NbTa_{0.5}TiZr$, with the measured UTS of 2,460 MPa [61].

Second, we decrease the concentrations corresponding to negative (or small, positive) values of the **a** vector. In case of Figure 29 and Figure 30 this results in decreasing the concentrations of Ti and Hf. But the concentration of Hf in $Al_{0.5}Mo_{0.5}NbTa_{0.5}TiZr$ is already 0.0%, so the Hf cannot be decreased further [61].

Third, we increase the concentration of elements corresponding to the largest values of the weighting vector, **a**. These elements exhibit the largest correlation with (or contributions to) the UTS observed. Hence, by increasing these elements, one can expect the largest relative increase in the UTS. In an effort to maximize the UTS, this process results in increasing the concentrations of Nb and Zr (and for Data Set B, the concentration of Cr) [61].

*Both Data Sets A and B give rise to the same, predicted composition ($Al_{0.5}Mo_{0.5}Nb_{1.5}Ta_{0.5}Zr_{1.5}$). This consistency suggests that the prediction algorithm may be somewhat immune to minor variations or redundancy (or even discrepancy) in the input data* [61].

Next, we assess the predictive capability of the regression model. The top two figures in Figure 28 show the predicted UTS as a function of the measured UTS. Here we are applying the same data points from Table 13 (19 and 22, respectively) for training and testing. Even so, we are looking at $R^2 = 0.592$ and normalized standard deviation per data point of 15.0 MPa for Data Set A. Similarly, for Data Set B, we are looking at $R^2 = 0.591$ and normalized standard deviation of 12.2 MPa. We attribute the spread observed to the fact that we are obtaining the data from the open literature, and that variations primarily in the process, but also the microstructure, shown in Table 13, are not accounted for in the prediction model in Eq. (63). Figure 24 and Figure 25 clearly indicate that *the heat-treatment process applied can significantly impact the UTS observed* [61].

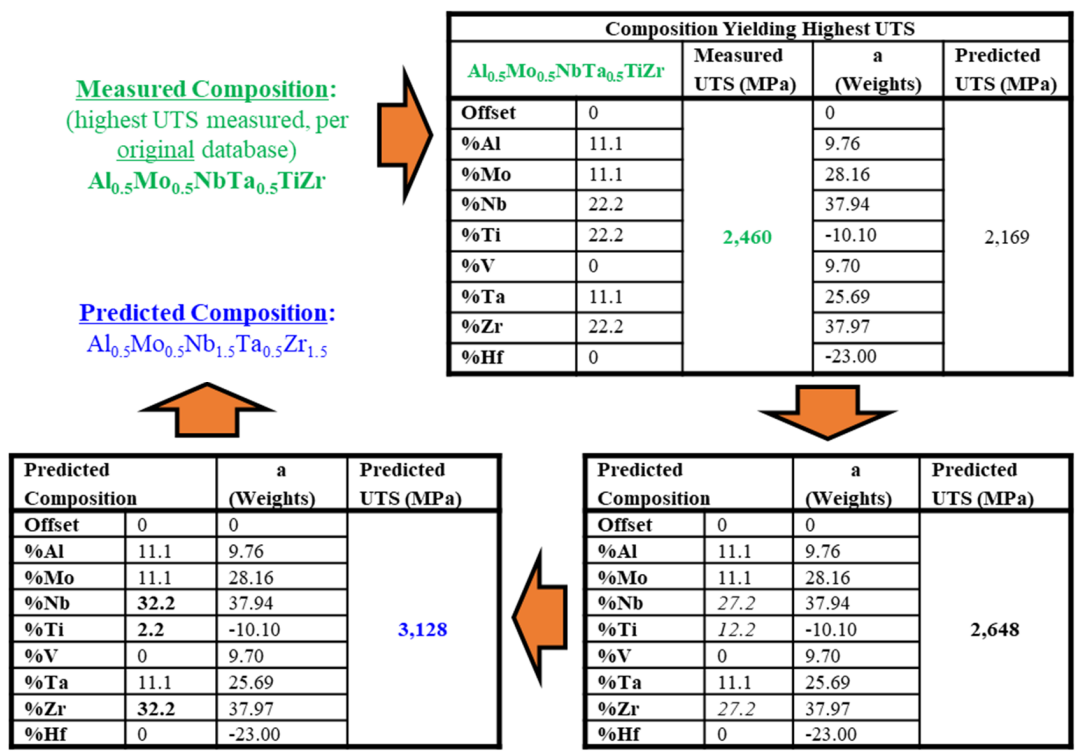

**Figure 29**: The Figure presents an example, where a linear prediction model is derived from Data Set A in Table 13 through linear regression analysis [61].

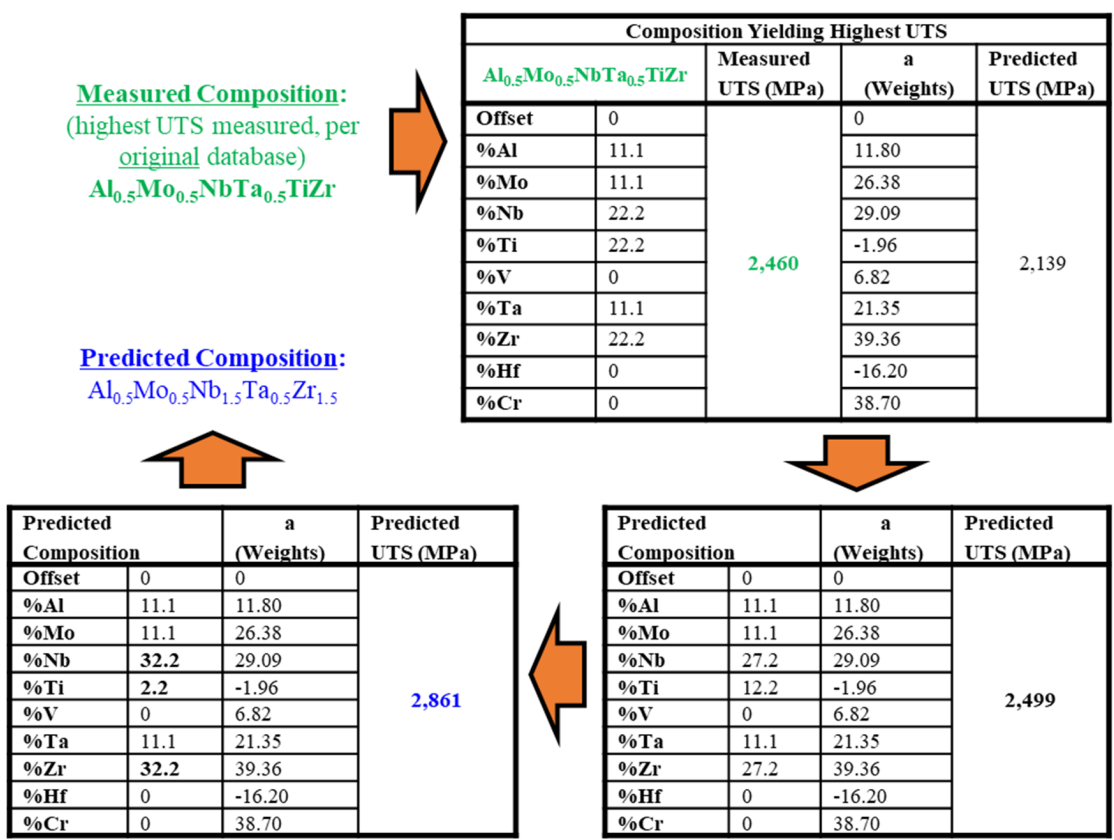

**Figure 30**: The Figure presents an example, where a linear prediction model is derived from Data Set B in Table 13 through linear regression analysis [61].

6. Towards Understanding What Is Causing Limitations of the Model –Analysis of Variance (Outliers)

For the purpose of confirming the conjecture about variance in the post-processing (heat-treatment process) applied being a major cause of the variance observed in Figure 28, we pick a few outliers for further analysis. We are in particular interested in identifying (analyzing) the processing applied to the outliers. Table 14 and Table 15 summarize our analysis of the annotated outliers from Figure 28. Our anticipated outcome can be characterized as follows [61]:

For outliers above the red lines in Figure 28 corresponding to [61]

$$\text{Measured(UTS)} < \text{Predicted(UTS)},\tag{74}$$

we expected poor processing (no heat-treatment process or a cheap process) to be applied. Here, the other compositions, esp., the ones with good heat treatments, impact the overall weighting (the **a** vector) such as to improve the overall prediction of the UTS for this data point [61].

But for outliers below the red lines in Figure 28, corresponding to [61]

$$\text{Measured(UTS)} > \text{Predicted(UTS)},\tag{75}$$

we think good processing may have been applied. Here, the other compositions, esp. the ones with poor heat treatment, impact the overall weighting (the **a** vector) such as to degrade the overall prediction of the UTS for this data point.

*The results from Table 14 indeed serve to confirm our conjecture*: The outlier, $Al_{0.4}Hf_{0.6}NbTaTiZr$, from Figure 28 falls below the red line, and has a reasonably-good heat treatment applied (HIP for 2 hours at.1,200 °C). The other outlier from Figure 28, $Hf_{0.5}Mo_{0.5}NbTiZr$, shows up quite a bit above the red line, and has no heat-treatment process applied [61].

*The results from Table 15 further serve to confirm our conjecture*: The outlier, AlCrNbTiV, from Figure 28 falls below the red line, and has a reasonably-good heat treatment applied (annealed for 24 hours at 1,200 °C). The other outlier from Figure 28, MoNbTiZr, shows up somewhat above the red line, and has no heat-treatment process applied. [For full disclosure, the compositions within the green circle had no heat-treatment applied, but appeared beneath the red line]. This trend may have had to do with the fact that none of the compositions with highest UTS had heat-treatments applied [61].

Overall, *these observations strengthen our belief in that the prediction accuracy, measured in terms of $R^2$ and the standard deviation normalized per data point, is primarily limited by the quality of (variance in) the input data.* These limitations in the prediction accuracy are consistent with the variations observed in Table 13, Figure 24 and Figure 25. The observations further underline the importance of selecting an optimization technique suitable for the application at hand and data available, and suggest that multi-variate regression is indeed suitable for analysis of the tensile strength [61].

The methodology presented here is not specific to the tensile strength. Comparison, such as Eqs. (75) and (76), can be presented as a part of outlier analysis for other quantities of interest, as long as combinations of predicted and measured values are available [61].

| Outlier | $Al_{0.4}Hf_{0.6}NbTaTiZr$ | $Hf_{0.5}Mo_{0.5}NbTiZr$ |
|---|---|---|
| **Process** | HIP + anneal | As-cast |
| **Process Specific** | HIP: 2 hr.,1,200 °C | N/A. Re-melt 5 times |
| **UTS$_{predict}$** | 1,657 MPa | 2,700 MPa |
| **UTS$_{meas}$** | 2,269 MPa | 1,538 MPa |
| **Above or Below Red Line in Figure 28?** | Below | Above |
| **Expectation** | Since UTS$_{meas}$ > UTS$_{predict}$, we expect good processing applied | Since UTS$_{meas}$ < UTS$_{predict}$, we expect poor processing applied |
| **Observation** | Indeed, here a reasonably good heat treatment process has been applied | Indeed, here no heat treatment process was applied |

**Table 14**: Analysis of .properties, in particular heat-treatment properties, corresponding to two of the annotated outliers in Figure 28 [61].

| Outlier | AlCrNbTiV | MoNbTiZr |
|---|---|---|
| **Process** | Annealed | As-cast |
| **Process Specific** | Homogenized for 24 hr. at 1,200 °C | N/A. Remelted a few times. |
| **UTS$_{predict}$ (MPa)** | 1,043 | 3,615 |

| UTS$_{meas}$ (MPa) | 1,570 | 3,450 |
|---|---|---|
| **Above or Below Red Line in Figure 28?** | Below | Above |
| **Expectation** | Since UTS$_{meas}$ > UTS$_{predict}$, we expect good processing applied | Since UTS$_{meas}$ < UTS$_{predict}$, we expect poor processing applied |
| **Observation** | Indeed, here a reasonably good heat treatment process has been applied | Indeed, here no heat treatment process was applied |

**Table 15**: Analysis of .properties, in particular heat treatment properties, corresponding to other two of the annotated outliers in Figure 28 [61].

7. <u>Assessing the Need for a More Sophisticated Prediction Model – Further Analysis of Data Set C</u>

1. Criterion for Assessment of Suitability of the Linear Regression, upon Addition of New Data
The results for Data Sets A and B in Figure 28 show that one can predict the UTS, using multi-variate linear regression, but that the variance is somewhat large, for reasons stated above [61].
Further assessment of the suitability of the linear regression can be obtained, by looking at the error margins, upon the addition of new data: If the variances decrease (improve), upon the addition of new data, then linear regression is a good model. If the variance stays approximately the same, then the linear regression is a questionable technique. But if the variance increases (deteriorates), upon the addition of new data, then we may need to look for another method [61].

2. Prediction of Compositions with Higher UTS Than Previously Observed, Based on Data Set C
Figure 31 illustrates the process of predicting compositions yielding higher UTS, and presumably higher fatigue resistance, than previously observed, based on Data Set C from Table 13. The prediction process consists of the same three, main steps, as the prediction in Figure 29 and Figure 30. The 4,847.1 MPa tensile strength for the predicted composition, MoNbZr, is 918 MPa higher than the highest UTS that has been previously observed (which is 3,929 MPa, observed for MoNbTiV$_{0.75}$Zr) [61].

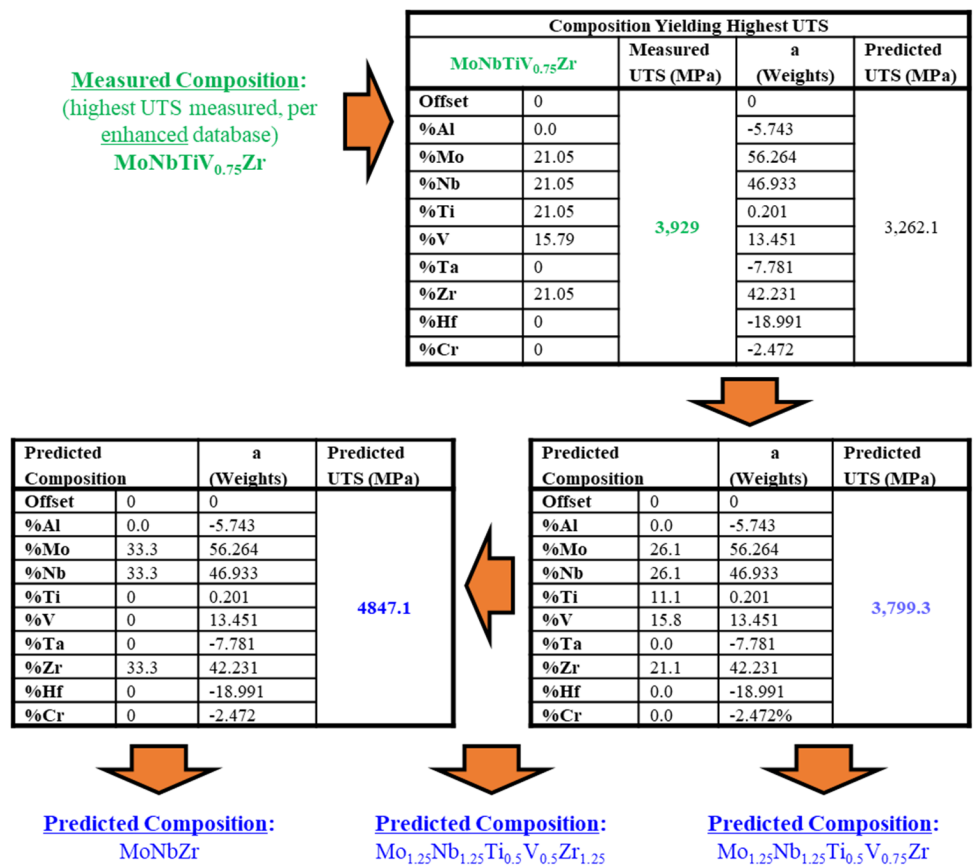

**Figure 31**: The Figure presents an example, where a linear prediction model is derived from Data Set C in Table 13 through linear regression analysis [61].

If desired, one can include the tensile strength of the pure elements comprising Feature Vector B, such as to introduce natural "barriers", which - together with the measured compositions - can limit the extent of the extrapolation. The nine (9) pure elements in Feature Vector B provide references, which one can compare against, during the extrapolation [61].

When expanding the data set, we have considered also expanding the feature vector, by adding Co, Fe, and Ni. But eventually, we have decided against it. While expanding the feature vector has allowed us to introduce at least forty (40) additional compositions, and the addition of Cu will allow us to introduce several new compositions on top of that, all of these compositions will correspond to lower UTS than the highest UTS observed in Table 13. Out of these additional compositions considered, the UTS, which has come closest to the composition in Table 13 with the highest UTS ($MoNbTiV_{0.25}Zr$ with UTS of 3,893 MPa), is AlCoCrFeNi, which exhibits UTS of 3,531 MPa. Out of the additional compositions considered, a dozen or so exhibit UTS in the range of 2,500 – 3,200 MPa. However, very few measure at higher UTS [61].

3. Suitability of Linear Regression for Data Set C

To the bottom left, Figure 28 presents results from applying the multi-variate linear regression to Data Set C. Based on the figure, and $R^2 = 0.77$, it does seem that the data points are more or less following linear regression. But the results also may not be conclusive; still more data may be necessary. Although quadratic dependence is not apparent in the figure, such (weak?) dependence may still be present in the data [61].

When comparing the accuracy of the fit for the linear regression for Data Sets B and C, we notice that the normalized standard deviation per data point has decreased somewhat (from 12.0 MPa to 11.1 MPa). Applying the criterion above, this trend suggests that *linear regression is a reasonably good technique*. Linear regression seems to provide reasonably-good description of the data [61].

4. Suitability of Quadratic Regression for Data Set C

To the bottom right, Figure 28 presents results from applying quadratic regression, with the diagonal **A** matrix per Eqs. (69) - (73), to Data Set C. Judging from the figure, $R^2 = 0.94$, and the normalized standard deviation of only 5.6 MPa per data point, it may seem that the prediction accuracy is limited by the prediction method more so than the variance in the inputs. Note though that, given limitations in the availability of the input data, we are using Data Sets A, B, and C (the same data sets) both for training and testing. Note also that the linear regression involves 10 model parameters, but the quadratic regression 19, per Eq. (67). Hence, the improvement in the accuracy of the fit, in case of the quadratic regression, is accomplished in part by fitting data to the 9 new model parameters. While quadratic regression with a diagonal matrix may indeed be a good technique for the application at hand, proper characterization of the relative merits of linear vs. quadratic regression will require separate data sets for training and testing. Further, although we may have sufficient data for a technique more sophisticated than linear regression (e.g., quadratic regression), we still believe the overall accuracy is primarily limited by variations in the inputs. This assessment is motivated by Figure 24 and Figure 25 [61].

9. Verifying Feasibility of the Predicted Compositions – Empirical Rules

Table 16 captures the outcomes from applying the empirical rules of [105, 106] to the formation of the compositions predicted, $Al_{0.5}Mo_{0.5}Nb_{1.5}Ta_{0.5}Zr_{1.5}$ $Mo_{1.25}Nb_{1.25}Ti_{0.5}V_{0.75}Zr$, $Mo_{1.25}Nb_{1.25}Ti_{0.5}V_{0.5}Zr_{1.25}$ and MoNbZr [61].

1. Expected Properties of $Al_{0.5}Mo_{0.5}Nb_{1.5}Ta_{0.5}Zr_{1.5}$, Based on the Empirical Rules

By comparing the calculated parameters for the atomic difference, $\delta_r$, and the enthalpy of mixing, $\Delta H_{mix}$, from Table 16 to Figure 2 from [105], one can see that a solid solution will likely form both in the predicted composition, $Al_{0.5}Mo_{0.5}Nb_{1.5}Ta_{0.5}Zr_{1.5}$, and in the reference composition, $Al_{0.5}Mo_{0.5}NbTa_{0.5}TiZr$. However, the predicted and reference compositions fall near the boundary between $S$ and $S'$ regions in Figure 2 from [105]. This trend suggests a small amount of the ordered solid solution precipitates may also form as a minor phase. We expect *$Al_{0.5}Mo_{0.5}Nb_{1.5}Ta_{0.5}Zr_{1.5}$* to be a stable composition with two types of phases [61].

2. Expected Properties of $Mo_{1.25}Nb_{1.25}Ti_{0.5}V_{0.75}Zr$, $Mo_{1.25}Nb_{1.25}Ti_{0.5}V_{0.5}Zr_{1.25}$, and MoNbZr, Based on the Empirical Rules

Again, by comparing the calculated parameters for the atomic difference, $\delta_r$, and the enthalpy of mixing, $\Delta H_{mix}$, from Table 16 to Figure 2 from [105], one notices that the predicted compositions, $Mo_{1.25}Nb_{1.25}Ti_{0.5}V_{0.75}Zr$, $Mo_{1.25}Nb_{1.25}Ti_{0.5}V_{0.5}Zr_{1.25}$ and MoNbZr, sit in the middle of the $S'$ region in Figure 2, which suggests that the composition has high chance of forming a solid solution main phase with ordered solid solution precipitates [61].

| Alloy | $Al_{0.5}Mo_{0.5}Nb$ $Ta_{0.5}TiZr$ | $Al_{0.5}Mo_{0.5}Nb_{1.5}$ $Ta_{0.5}Zr_{1.5}$ | $MoNbTi$ $V_{0.75}Zr$ | $Mo_{1.25}Nb_{1.25}$ $Ti_{0.5}V_{0.75}Zr$ | $Mo_{1.25}Nb_{1.25}$ $Ti_{0.5}V_{0.5}Zr_{1.25}$ | $MoNbZr$ |
|---|---|---|---|---|---|---|
| $\delta_r$ (%) | 4.41 | 4.89 | 5.65 | 5.74 | 5.79 | 5.56 |
| $\Delta S_{mix}$ (J/K/mol) | 14.43 | 12.18 | 13.33 | 12.96 | 12.70 | 9.13 |
| $\Delta H_{mix}$ (kJ/mol) | -10.52 | -10.17 | -2.70 | -3.16 | -3.08 | -3.56 |
| $\Omega$ | 3.16 | 2.89 | 11.79 | 10.19 | 10.22 | 6.66 |

**Table 16:** Assessment of viability of the predicted compositions ($Al_{0.5}Mo_{0.5}Nb_{1.5}Ta_{0.5}Zr_{1.5}$, $Mo_{1.25}Nb_{1.25}Ti_{0.5}V_{0.75}Zr$, $Mo_{1.25}Nb_{1.25}Ti_{0.5}V_{0.5}Zr_{1.25}$, and $MoNbZr$) through application of empirical rules [61, 105].

8.   Initial Efforts towards Experimental Verification

Figure 32 summarizes results from the experimental verification of the strength of the predicted composition, $Al_{0.5}Mo_{0.5}Nb_{1.5}Ta_{0.5}Zr_{1.5}$, in comparison to the reference, $Al_{0.5}Mo_{0.5}NbTa_{0.5}TiZr$. *We conclude from the experimental results in Figure 32 that the candidate composition indeed exhibits higher UTS than the reference, hence confirming the outcome of our prediction.* Coupons for both the predicted and reference compositions were prepared, using arc melting and compression testing, for as-cast samples, and with no heat treatment applied. The casted samples had the initial diameter of a quarter inch (6.27 mm) based on the mold used. The diameter of the samples was then decreased by grinding the cylindrical surfaces to diameter of 4 – 5 mm, in order to obtain higher stress, and ultimately break the samples, using a mechanical compression testing machine with maximum force of 15,000 lbs. From the compression tests, it turned out that the predicted composition exhibited higher yield strength and fracture strength than the reference, as shown in Figure 32. But both compositions were very brittle [61].

For further information on the experimental verification, refer to [162].

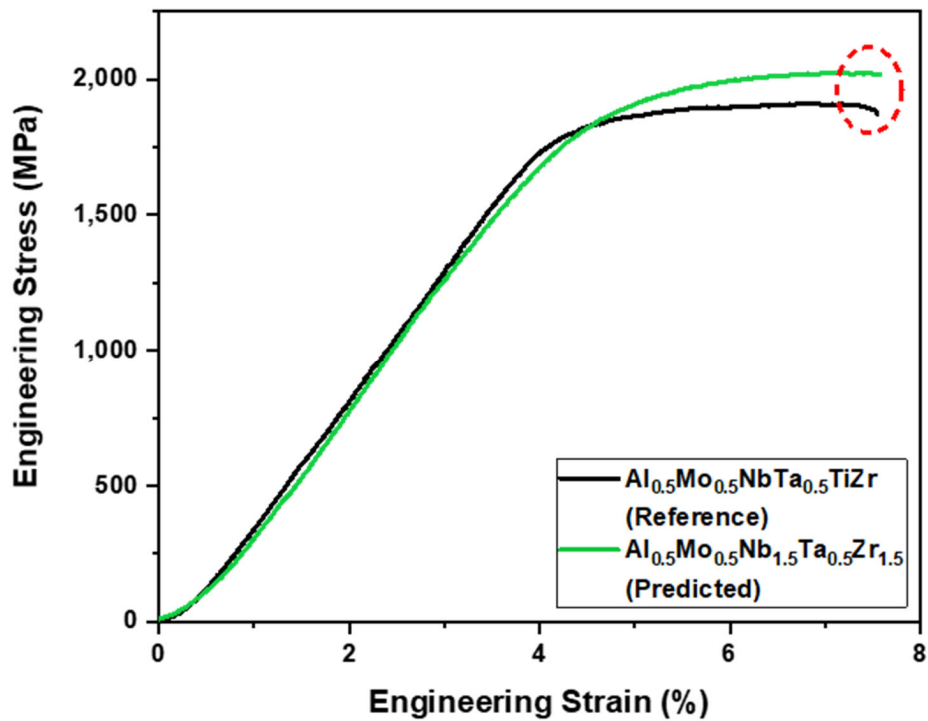

**Figure 32**: Experimental verification of superior compression strength of the predicted composition ($Al_{0.5}Mo_{0.5}Nb_{1.5}Ta_{0.5}Zr_{1.5}$) over the reference composition ($Al_{0.5}Mo_{0.5}NbTa_{0.5}TiZr$). Coupons for both compositions were prepared through arc melting and with no heat treatment applied [61].

### 4.8.   Example 2: Prediction of Fatigue Resistance (Endurance Limit)

In the case of prediction of fatigue resistance, this chapter presents an indirect approach for predicting HEA compositions yielding attractive fatigue resistance, despite limited fatigue data currently being available from the open literature. We present a method that harvests a correlation identified between fatigue resistance and the ultimate tensile strength.

The overall methodology for predicting the fatigue resistance is presented in Figure 33. This is a two-stage prediction process, where we first determine the tensile strength, given an input combination, and then determine the fatigue resistance, given the tensile strength and the inputs [61].

Figure 34 compares the high-cycle fatigue properties of HEAs to those of conventional alloys. The HEAs seem to generally result in higher UTS and endurance limits, compared to conventional alloys. Still, there is a lot of scatter in the data. But despite the scatter, the endurance limit seems primarily correlated with the UTS. Hence, by identifying compositions with larger UTS, one can expect greater fatigue resistance. A key to accurate prediction entails understanding, and properly accounting for (explaining), the sources of variations in the data [61].

*The scatter in the data is caused by input sources, such as defect levels, process parameters, or grain sizes, which are not accounted for in the prediction model, or accounted for in the prediction model, but not available at the time of prediction* [61].

The endurance limits in Figure 34 represent multi-dimensional data points, which in addition to the UTS exhibit dependence on the grain size, process parameters, and defect characteristics [61].

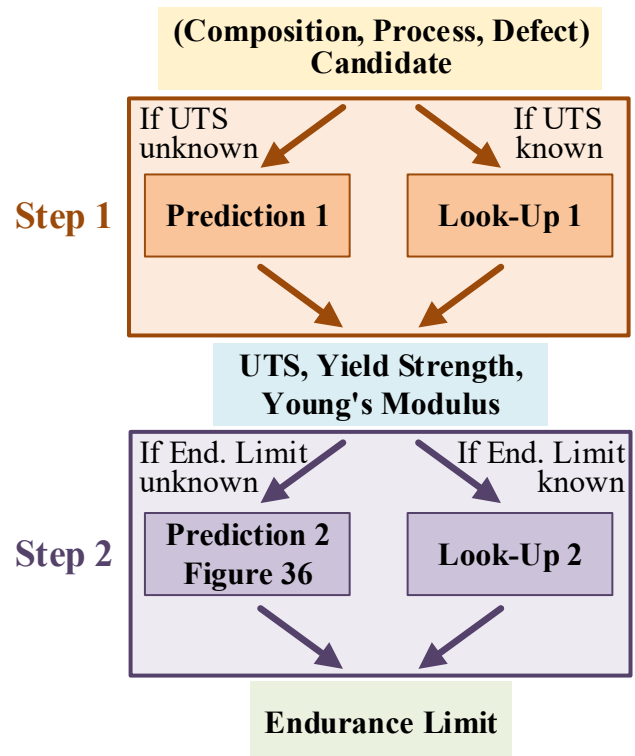

**Figure 33**: The Figure presents overall methodology for deriving the output quantity of interest (the endurance limit) from the input sources (compositions, processes and defects) [61].

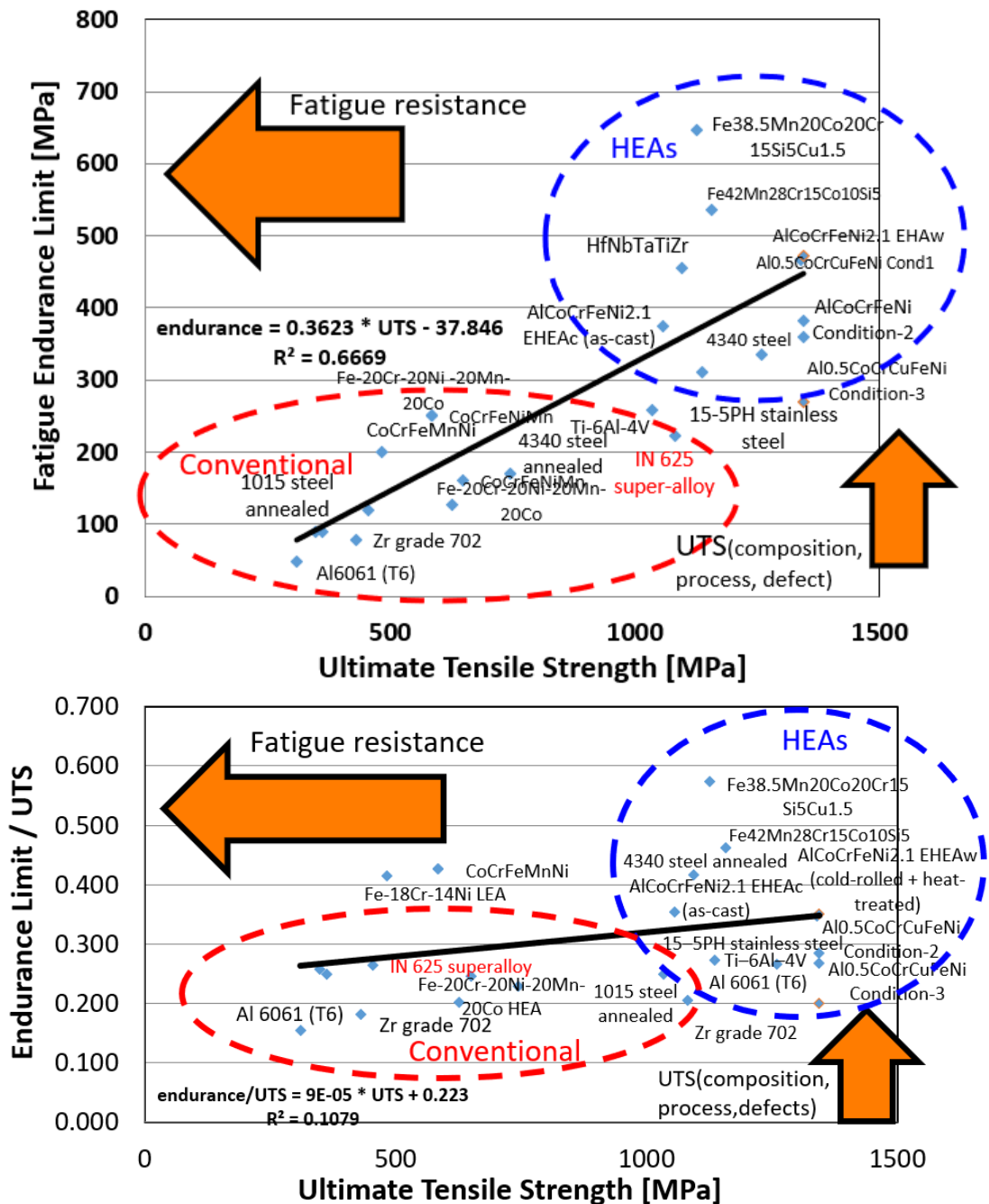

**Figure 34:** The Figure captures Step 2 in the prediction methodology outlined in Figure 33 [61].

## 1. Formulation of the Input Combinations

Our intent is to capture in the input sources that contribute to variations in the fatigue resistance (to variations in the output). In this chapter, we like to model the input combination as [61]:

$$\text{input combination} = (\text{composition}, \text{heat treatment process}, \text{defect level}). \tag{76}$$

Here defects are defined broadly such as to include inhomogeneities, impurities, and unwanted features [61].

While we are primarily looking for compositions yielding attractive fatigue resistance, we will see (from Table 9) that the defect level also significantly affects the fatigue resistance. Hence, the accurate prediction of the fatigue resistance (endurance limit) may be impossible, without knowing the defect information. Similarly, we preferably would like to determine the heat-treatment process that yields the least scatter in the fatigue resistance observed [61].

2. Estimating the Tensile Strength, Given an Input Combination

Step 1 in Figure 33 summarizes the overall approach. If the UTS corresponding to a given input combination is known, we can apply a simple look-up. If the UTS corresponding to a given input combination is not known, we can apply prediction (interpolation or extrapolation), based on the nearest neighbors [61].

Figure 23 captures a general model of physical dependencies for prediction of the UTS. This model is a generalization of the input sources modeled in Eq. (76). Capturing of the physical dependencies helps greatly in terms of the incorporation of *a priori* knowledge, derived from the underlying physics, and in terms of making the most of the - usually limited - input data available [61].

3. Arriving at the Fatigue Resistance, Given the Tensile Strength and the Remaining Inputs

Step 2 in Figure 33 summarizes one possible approach. Here, we model the fatigue resistance (the endurance limit) as [61]:

$$\text{endurance limit} = f(\text{UTS, heat treatment process, defect level, ...}). \tag{77}$$

If the endurance limit corresponding to a given input combination is known, we can apply a simple look-up. If the endurance limit corresponding to a given input combination is not known, we can predict the endurance limit, on basis of the estimated [61]

$$\text{UTS} = h(\text{composition, heat treatment process, defect level}), \tag{78}$$

as shown in Figure 34.

The significant variations observed in Figure 34 reinforce the need for access to the relevant input data, for accurate prediction. To obtain the prediction accuracy within the measurement error (10% - 20%), or within the accuracy limits on fatigue life expected for given parts, one may need to know quite a bit more about the input (configurational) parameters. An accurate prediction model seems to have the form [61]:

$$\text{endurance limit} = f(\text{UTS, process, defect(process), grain(process), microstructure (process)}, T, ...), \tag{79}$$

where [61]

$$\text{UTS} = h(\text{ composition, heat treatment process, defect level(process), grain size}, T). \tag{80}$$

Here, defect (process), grain (process), and microstructure (process) are taken to represent the defect level, grain size, and microstructure, resulting from the application of a given heat-treatment process, respectively. The term microstructure (process) may be interpreted broadly such as to include both microstructures and phase properties. In case of $TiAl_6V_4$, the fatigue properties may further depend on the application of hot iso-static pressing (HIP), which – again through a broad interpretation - may be captured as a part of the heat treatment process, and on the type of machining applied, which may be broadly combined with defect(process) [61].

## 4.9.  Example 3: Towards Coatings and Base Alloys Resistant to Hot Corrosion

1. Overall Goal

Our intent here is to provide insight into how models or algorithms for machine learning can be employed, for the purpose of detecting data patterns and characteristic trends, for learning from accumulated data, and for evolving distinguishing characteristics between CMAS attack and calcium sulfate ($CaSO_4$) hot corrosion, with or without the influence of sea salt, in order to develop coatings resistant to CMAS and calcium sulfate hot corrosion. In this context, proper characterization of the reaction space is essential. In principle, one may assume that the base alloy consists of a HEA, or even that the ceramic coating consists of high-entropy ceramics. A HEA can also be used for the metallic bond coating. Figure 35 presents data analytics for hot corrosion in context with the overall effort to identify high-risk areas for deposit build-up in gas turbines. Here, it may make sense to link structure and chemistry to observed reaction mechanisms to accelerate materials design for the hot corrosive environments, and develop sophisticated physics-based prediction models from experimental data [61].

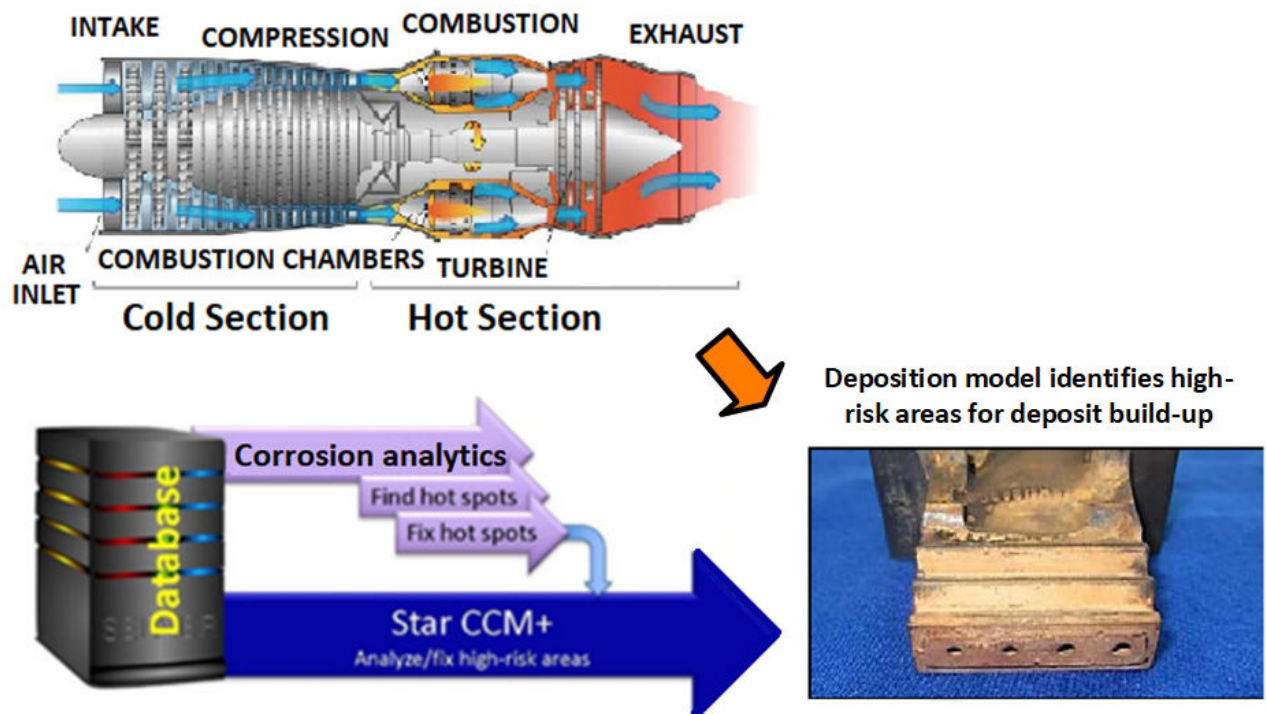

**Figure 35**: The Figure presents analytics for hot corrosion in context with the overall effort to identify high-risk areas for deposit build-up on actual components in gas turbines [61].

2. Brief Background on Thermal Barrier Coatings
Thermal barrier coatings (TBCs) are recognized as enabling materials for enhancing the performance and durability of gas turbine engines. A TBC system consists of a metallic bond coat together with a ceramic top coating (see Figure 36) [61].

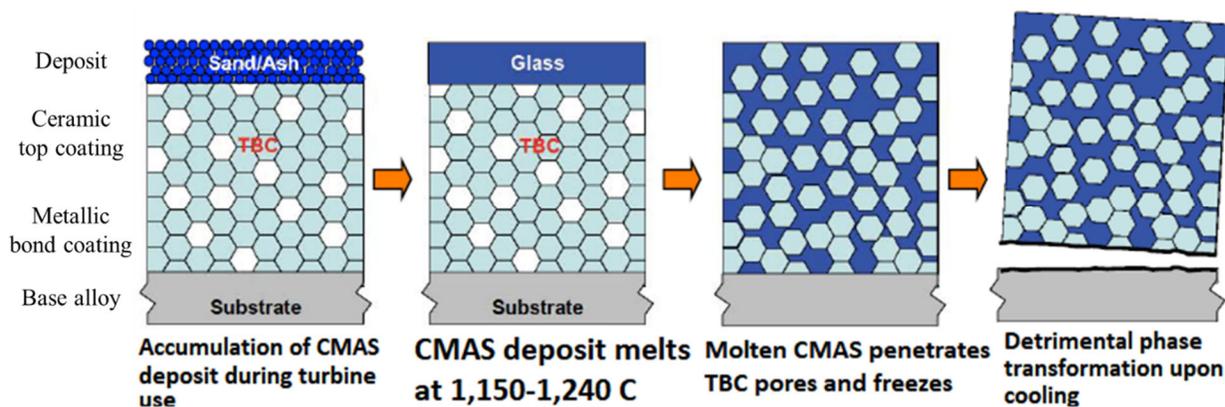

**Figure 36**: The Figure presents the essential mechanism behind corrosion of the thermal barrier coating instigated by a CMAS hot corrosion attack [61].

3. Brief Background on Deposition Models for Hot Corrosion
Hot corrosion has been studied for a number of years, although not through application of machine learning. Reference [163] presents a deposition model capable of identifying high-risk areas for deposit build-up for actual components in a gas turbine. The deposition model provides valuable information on where deposits are likely to take place inside a gas turbine. The model incorporates the particulate characteristics and component design conditions to identify high risk areas for deposit buildup for actual components. The deposition model covers the cases of hot corrosion for land-based, air-borne or sea water applications of gas turbines. In sea water, sodium chloride or sodium sulfate tend to be prevalent, and the corrosion reactions may occur at lower temperature. But the deposition model still can apply (with adjusted input values) [61].

4. Essence of CMAS and Calcium Sulfate Corrosion Attacks

The CMAS attacks the ceramic top coating first, and then attacks the metal side. For CMAS, reaction with the thermal barrier coating is the primary (and often the only) thing one needs to consider. The calcium sulfate ($CaSO_4$), on the other hand, soaks into the ceramics top coating first, and then attacks the bond coat surface and the base alloy (which in the past has usually consisted of a super-alloy). For the calcium sulfate, there is less interaction with the TBC, but more with the base alloy [61].

5. Specifics of Interaction of CMAS with the Thermal Barrier Coating

CMAS consists a combination of $SiO_2$, $CaO$, $Mg$, $Al_2O_3$, and $FeO$. For information on relative concentration of these constituents between different types of CMAS simulated sand, engine deposits, average earth's crust, Saudi sand, airport runway sand, Mt. St. Helen's volcanic ash, volcanic ash from Eyjafjallajokull glacier in Iceland, subbituminous fly ash or bituminous fly ash, we refer to [164].

CMAS degradation is both thermochemical and thermomechanical to the thermal barrier coatings, as shown in Figure 36. Molten CMAS (1,150 - 1,240°C) penetrates the TBC pores and freezes at a given depth within the TBC. Above 1250 °C, CMAS deposits can infiltrate and significantly dissolve YSZ top coatings via thermochemical interactions. Upon cooling, zirconia can reprecipitate with a spherical morphology and a composition that depended on the local melt chemistry. Upon cooling, the CMAS can destabilize the YSZ top coating through a detrimental phase transformation (t' → t → f + m) [61].

Calcium oxide ($CaO$) is known to react with chromium contained in MCrAlY (M = Ni, Co) alloys and nickel-based super-alloys to form a low-melting (1,100°C) calcium chromate. The reactivity of gamma-NiAl and gamma-Ni-based NiCoCrAlY alloys with $CaO$ at 1,100°C produced multi-layer scales of $Al_2O_3$ and calcium aluminates ($xCaO-yAl_2O_3$). Increasing the alloy chromium content only enhances corrosion severity. The reaction of two-phase beta-gamma MCrAlY alloys with $CaO$ progressed according to two distinct mechanisms [61]:

1. During the initial stage, formation of a liquid calcium chromate led to the rapid consumption of the Cr-rich gamma-phase. The extent of degradation was particularly important for a single-phase gamma-composition, and was significantly reduced by increasing the alloy beta fraction [61].

2. In the subsequent stage, a continuous $Al_2O_3$ layer was established at the base of the scale, which led to a much lower oxidation rate. Additions of $Al_2O_3$ or $SiO_2$ decreased the $CaO$ reactivity due to the formation of aluminates or silicates [61].

Upon cooling, the glass and reaction product phases solidify and the void structure, that is utilized to reduce thermal conductivity and provide the strain compliance, is lost leading to TBC delamination, as shown in Figure 36 [61].

6. Specifics of Interaction of Calcium Sulfate with the Thermal Barrier Coating and Base Alloy

Due to relatively short history, modeling of $CaSO_4$ is in part based on analogy with sodium sulfate ($Na_2SO_4$). Early research has shown that $CaSO_4$ tends to attack yttria, destabilizing zirconia-based TBCs. Both in as-sprayed and preoxidized condition, CoNiCrAlY suffered a significant damage by $CaSO_4$ deposits via a basic fluxing mechanism that yielded $CaCrO_4$ and $CaAl_2O_4$ [61].

7. Overview of the Reaction Space

The reaction space for CMAS, calcium sulfate and sodium sulfate ($Na_2SO_4$) deposition is summarized in Table 17. We are including sodium sulfate for historic reference. Research into hot corrosion and its preventive measures in gas turbine engines has mostly focused on sodium sulfate since the early 1950s [61].

| Attack / Deposit | Environment | Reactions | Temperature |
|---|---|---|---|
| Sodium sulfate | Natural / Air | $2\ NaCl + SO_2 + O_2 = Na_2SO_4 + Cl_2$ | ~ 900 °C |
| Calcium Sulfate | Natural / Air | $CaSO_4 \cdot 2H_2O = CaO + SO_3 + 2\ H_2O$ | ~ 1220 °C |
| CMAS (Calcium-magnesium-alumino-silicate; $CaO-MgO-Al_2O_3-SiO_2$) | Natural / Air (without sea salt) | $CaSO_4 + 2H_2O = CaSO_4$ | Dehydration at 150 °C |
| | | $CaCO_3 = CaO + CO_2$ | ~ 800 °C |
| | | $CaSO_4 = CaSO_3 + 1/2O_2$ $CaSO_3 = CaO + SO_2$ | ~ 1200 °C |
| | Reaction with TBC (7YSZ) | $CaO-MgO-Al2O3-SiO2-Y2O3-ZrO_2$ is the key system | |
| | Prototypical salt: $Na_2SO_4$ ($T_{melt} = 881$ °C) | Fluxing process (Type I) | ~ 900 °C |
| | | Salt-component processes (Type II) | ~ 700 °C |

**Table 17**: Characterization of the reaction space for CMAS and calcium sulfate attacks, with and without sea salt [61].

8. List of Features Jointly Characterizing CMAS and Calcium Sulfate Corrosion Attacks

Table 18 summarizes a unified list of features characterizing CMAS and calcium sulfate corrosion attacks, both in air and sea water. In sea water, sodium chloride or sodium sulfate tend to be prevalent, and the corrosion reactions may occur at lower temperature. But the same feature list still applies (with adjusted input values) [61].

| I/O | Parameter of Feature List | Susceptible to Attack |
|-----|---------------------------|-----------------------|
| Inputs | Composition of CMAS | CMAS |
| | Composition of TBC | Both CMAS and $CaSO_4$ |
| | Melting / solidification temperature | CMAS |
| | Modulus of TBC | CMAS |
| | Modulus of CMAS | CMAS |
| | Particle size | CMAS |
| | Surface temperature | CMAS |
| | Wall shear velocity | CMAS |
| | Air-fuel ratio | $CaSO_4$ |
| | Base alloy composition | $CaSO_4$ |
| | $CaSO_4$ concentration | $CaSO_4$ |
| | Dew point | $CaSO_4$ |
| | Surface pressure | $CaSO_4$ |
| | Time | Both CMAS and $CaSO_4$ |
| Outputs | Chemical reactions | CMAS |
| | Coating modulus after infiltration | CMAS |
| | Particle temperature | CMAS |
| | Particle velocity | CMAS |
| | Deposit build-up rate | CMAS |
| | Depth of attack | $CaSO_4$ |
| | Metal loss | $CaSO_4$ |
| | Weight change | $CaSO_4$ |

**Table 18**: Unified feature list for analysis of CMAS and calcium sulfate hot corrosion, *both* covering attacks in air and sea water [61].

9. Canonical Component Analysis for Deriving Distinguishing Characteristics between CMAS and Calcium Sulfate Hot Corrosion Attacks

In order to evolve distinguishing characteristics between CMAS and calcium sulfate hot corrosion, with or without the influence of sea salt, for the purpose of developing coatings resistant to CMAS and calcium sulfate hot corrosion, we apply canonical component analysis (CCA), as qualitatively shown in Figure 37 [61]. As an alternative to canonical component analysis, one also can conduct correlation between the input and output features, using regression analysis, as shown in [101].

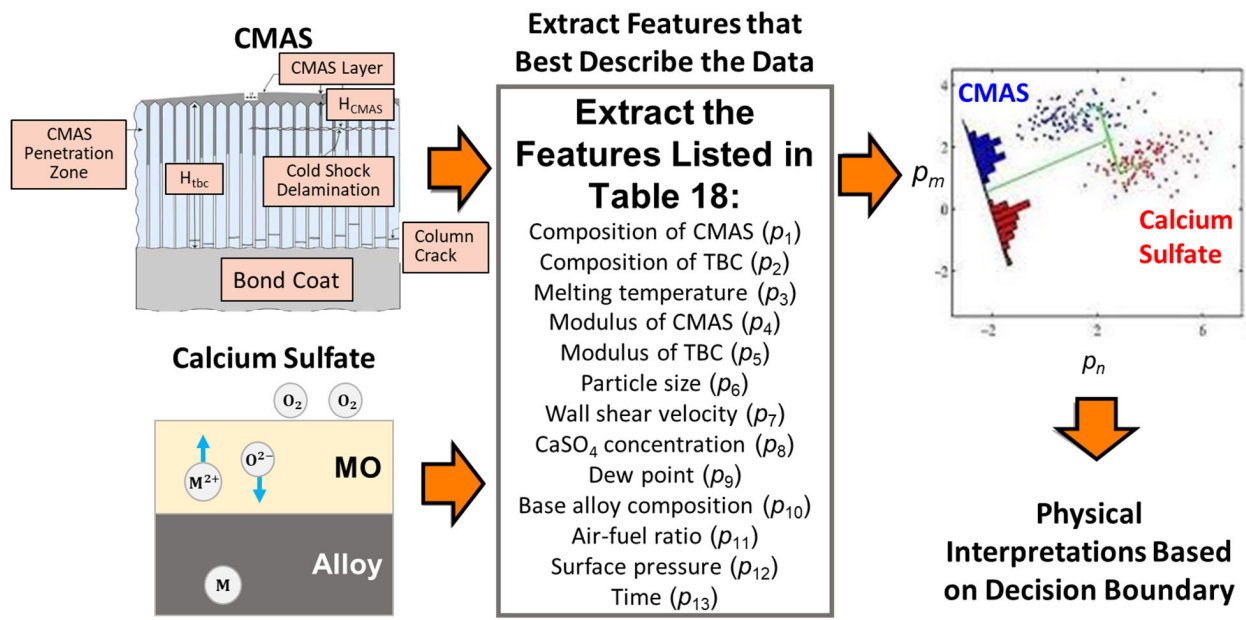

**Figure 37**: The Figure shows how canonical correlation analysis can be applied to analysis of feature sets corresponding to CMAS and calcium sulfate hot corrosion attacks [61].

## 10. Joint Optimization

To accurately identify a (TBC, base alloy) combination, that is likely to both protect against a CMAS and calcium sulfate attack, assuming the CMAS protection is measured through deposit build-up rate, but that the calcium sulfate protection is measured through weight change, metal loss and depth of attack, we apply joint optimization. One can optimize a weighted objective function of the form [61]

$$\text{Objective} = w_1 * \text{deposit\_build\_up\_rate} + w_2 * \text{weight change} + w_3 * \text{metal loss} + w_4 * \text{depth\_of\_attack}. \quad (81)$$

## 4.10. Example 4: Optimization of Tensile Strength for CMCs or PMCs

### 1. Motivation

There is an ever-changing, constant need for designing novel composite materials for aerospace applications that offer a broad gamut of multi-functionality (sensing, electrical and thermal properties, characteristics, adhesion, energy storage or harvesting, low density, etc.) and structural stability (mechanical behavior, stiffness or compliance, fracture toughness, high temperature stability, minimal physical aging, etc.). The eventual goal is to minimize costs and maximize operational performance and efficiency. Machine learning and data analytics can be employed to investigate and develop understanding of structure-property-performance relationships, together with well-formalized modeling methods involving quantum chemistry, molecular dynamics or coarse-grain simulations. The machine learning and data analytics can:

1. Complement experimental efforts towards better appreciation of molecular origins of structure-property-performance relationships, and
2. Facilitate accelerated materials design for next-generation multi-functional composites, geared towards aerospace applications.

### 2. Approach

For the purpose of optimizing the tensile strength, one can employ ML models similar to the ones in Example 1 (Section 4.7). More generally, for investigating structure-property-performance relationships, one can define dependencies between inputs (in part, structure related) and outputs (observed properties), similar to Figure 1. In addition to the ML models presented in Example 1, one can in general also employ forward-backward prediction, such as in Figure 15.

### 3. Physics-Based Models Available

Curtin proposed a model to characterize the UTS of a brittle fiber-reinforced ceramic matrix composites, $\sigma_u$, as follows [165]:

$$\sigma_u = V_f \left( \frac{2}{m+2} \right)^{\frac{1}{m+1}} \left( \frac{m+1}{m+2} \right) \left( \frac{\sigma_f^m \tau L_g}{R \ln(2)} \right)^{\frac{1}{m+1}}, \quad (82)$$

where $m$ and $L_g$ denote Weibull modulus and gauge length of fibers, respectively. The volume fraction of broken fibers, $V_{fb}$, and the load carrying capacity of intact fiber, $\sigma_{fi}$, can be expressed as [165]:

$$V_{fb} = \frac{2}{m+2} \tag{83}$$

$$\sigma_{fi} = \left(2L_g \tau \sigma_f^m / ((m+2)R\ln(2))\right)^{\frac{1}{m+1}} (1 - V_{fb})V_{fb}. \tag{84}$$

Deng et al. established a model to forecast the fracture strength, $\sigma(T)$, at different temperatures for ceramics [166]:

$$\sigma(T) = \sigma(T_0)\left[\frac{E(T)}{E(T_0)} \cdot \frac{T^M - T}{T^M - T_0}\right]^{0.5} \tag{85}$$

Here, $E(T)$ denotes the Young's modulus of ceramics at temperature $T$, $T_M$ denotes the melting point of the ceramic material and $T_0$ denotes an arbitrary reference temperature.

According to an analytical model by Zhang et al. [167], the volume fraction of broken fibers, $V_{fb}(T)$, and the load carrying capacity, $\sigma_{fb}(T)$, at different temperatures, can be modeled as

$$V_{fb}(T) = \frac{2}{m(T)+2} \tag{86}$$

$$\begin{aligned}\sigma_{fb}(T) = \ & \sigma_{UCP}(T) - \eta\left(2L_g\tau(T)\sigma_f^{m(T)}(T_0)\left(\frac{E_f(T)}{E_f(T_0)} \cdot \frac{T_f^M - T}{T_f^M - T_0}\right)^{0.5m(T)}/((m(T)+2)R\ln(2))\right)^{\frac{1}{m(T)+1}} \\ & (1 - V_{fb}(T))V_f \end{aligned} \tag{87}$$

where

$$\begin{aligned}\sigma_{UCP}(T) = \ & \eta\, V_f\left(\frac{2}{m(T)+2}\right)^{\frac{1}{m(T)+1}}\frac{m(T)+1}{m(T)+2} \\ & \left(\sigma_f^{m(T)}(T_0)\left(\frac{E_f(T)}{E_f(T_0)} \cdot \frac{T_f^M - T}{T_f^M - T_0}\right)^{0.5m(T)}\frac{(\sigma_{mcs}(T)+AP(T))^3(1-V_f)E_m^2(T)L_g}{12\ln(2)\gamma_m(T_0)E_f(T)V_f^2E_c^2(T)} \times \frac{T_m^M - T_0}{T_m^M - T}\right)^{\frac{1}{m(T)+1}}\end{aligned} \tag{88}$$

For detailed explanations of the terms, refer to [167].

Evan presented a model for predicting the composite ultimate strength based on weakest link statistics, incorporating the fiber Weibull modulus, $m$ [168]. The model is a modified bundle failure analysis which assumes failed fibers have no load bearing ability. The model of the modified bundle failure theory is presented in the equation below [168]:

$$\sigma_{cubm} = V_f\, \sigma_{fub}\, \exp\left(\frac{1 - \left(1 - \frac{\tau x}{R\,\sigma_{fub}}\right)^{m+1}}{(m+1)\left[1 - \left(1 - \frac{\tau x}{R\,\sigma_{fub}}\right)^m\right]}\right). \tag{89}$$

The fiber bundle strength, $\sigma_{fub}$, is determined by iteratively solving the following equation [168]:

$$\left(\frac{R\,\sigma_{fub}}{\tau x}\right)^{m+1} = \frac{A_o}{2\pi R\,L_c}\left(\frac{R\,\sigma_o}{\tau x}\right)^m\left[1 - \left(1 - \frac{\tau x}{R\,\sigma_{fub}}\right)^m\right]^{-1}. \tag{90}$$

Here, $\tau$ is the interfacial shear stress, $x$ is the saturated matrix crack spacing, $R$ is the fiber radius, $L_c$ is the composite gauge length, and $A_o$ is an area normalizing factor.

Cao and Thouless made an attempt to predict the ultimate strength of a ceramic composite with the application of two parameter Weibull statistics [168]. Their theory assumed that the matrix was saturated with cracks. As a result, the initial linear elastic behavior and the nonlinear deformation associated with matrix cracking were not incorporated. Another simplifying assumption was that upon fracture of a fiber anywhere within the gage length of the composite, the fiber was not able to carry any load. Given these assumptions, the following equation was used to predict the ultimate strength of a ceramic composite [168]:

$$\sigma_{cu} = V_f \Sigma\left(\frac{\Sigma\,R}{m(m+1)\tau L_c}\right)^{1/m} e^{-1/m} \tag{91}$$

where

$$\Sigma = \left[\frac{A_o \sigma_o^m \tau(m+1)}{2\,\pi\,R^2}\right]^{\frac{1}{m+1}}. \tag{92}$$

These analytical models account for the underlying physical dependencies, and help greatly in terms of developing understanding of the structure-property-performance relationships in CMCs, at least as far as the UTS is concerned.

## 4. Data Available

Table 19 summarizes key material properties of CMCs and PMCs [165]. For additional data, refer to [167-170].

| Fibre | Matrix | $f$ | Fibre strength (MPa) | $m$ | $r$ (μm) | $\tau$ (MPa) |
|---|---|---|---|---|---|---|
| Nicalon | LAS-II | 0.46 | 1580 | 3.8 | 8 | 2.5 |
| | | 0.46 | 1520 | 2.7 | 8 | 2.5 |
| | | 0.44 | 1470 | 3.9 | 8 | 2.5 |
| | | 0.44 | 1450 | 3.1 | 8 | 2.5 |
| SCS-6 | LAS-V | 0.20 | 3500 | 8 | 70 | 10 |
| Nicalon | LAS-III, aluminosilicate | 0.45 | 2800 | 2.0 | 8 | - |
| | Soda lime | 0.45 | 1440 | 3.0 | 8 | - |
| | Carbon | | | | | |
| | 1 | 0.22 | 2200 | 4.5 | 8 | 10 |
| | 2 | 0.20 | 2300 | 4.5 | 8 | 14 |
| | 3 | 0.21 | 2500 | 4.5 | 8 | 90 |
| | CAS | 0.37 | 2000 | 3.6 | 8 | 10 |
| SCS – 6 | Ti – 6 – 4 | Variable | 3650 | 9 | 70 | 40 |
| | Ti – 24 – 11 | | | | | |
| | 1 | Variable | 3850 | 17 | 70 | 56 |
| | 2 | 0.26-0.27 | 4633 | 24 | 70 | 56 |

**Table 19:** Material parameters for CMC and MMC systems at room temperature [165].

## 4.11. Example 5: Joint Optimization of Material Strength, Ductility and Oxidation Resistance

1. Motivation

For operating turbines or other energy conversion devices at higher temperature, and with improved efficiency, there is need for materials yielding good strength at higher temperature, preferably without sacrificing ductility at room temperature. The design goals involve joint optimization:
1. Achieving the required material strength at higher temperature.
2. Improving room temperature ductility (decreasing the brittle-to-ductile transition below room temperature).
3. Offering acceptable oxidation resistance.

Achieving excellent combination of both high strength and good ductility is the ultimate goal of material scientists who work on mechanical properties. This has been a significant and a long-standing challenge.

2. High-Level Approach

In principle, there are two, primary routes for formulating, such joint optimization problems:
1. Through maximization of a joint objective function, one accounting both for strength and ductility.
2. Through maximization of an objective function only capturing the strength, but where the ductility is accounted for in the constraints.

Traditionally, electrons structure calculations, based on DFT, have been used to study the reactivity and oxidation resistance for down-selected alloys by multiscale modeling and an experimental approach.

Proper approaches for addressing the strength-ductility trade-off need to account for the underlying physics. According to [171], recent decades have seen the emergence of several strategies for producing materials with ever-greater strength, such as grain size refinement, solid solution alloying and plastic straining. So far, most of the existing material-strengthening techniques result in decreased ductility [171]. The strength-ductility trade-off can be circumvented by several complementary approaches that involve sophisticated microstructure design, including metals with bimodal grain sizes, composites with an amorphous phase and nanograins, as well as laminated materials [171]. More recently, engineering coherent internal boundaries at the nanoscale has been viewed as an efficient way to achieve high strength while maintaining substantial ductility [171].

3. Design Objectives Presented to Alloy Designers

The design objectives, that provide guidance to the development of novel chemistries, can have 3 categories:
1. Alloy properties;
2. System compatibility with coatings;
3. Processability.

While material properties (strength and oxidation resistance) are of prime importance, the coating capabilities need to be evaluated without degrading properties to meet the turbine efficiency and cooling air reduction targets. Processability plays a critical role to obtain defect free near net shape parts with casting or AM for advanced designs. This is also critical for joining technologies that determine the repairability aspect of the components.

The starting point is to define the window for solvus temperature that will define the phases in these multicomponent alloy systems and their stability in the desired temperature regime. This is mainly due to significant microstructure evolution and mechanical property changes observed in service. These calculations of phase-equilibria, diffusion fields, and precipitation events form an important foundation for designing high-temperature-capability HEAs. Following this, the necessary structure-property models need to focus on strength, creep, fatigue, and oxidation. An emphasis will be on high-temperature creep/fatigue resistance, minimizing vacancy diffusivity, maximizing elevated-temperature solution hardening, and maximizing high-temperature oxidation resistance.

The system compatibility studies needed are shown in Figure 38. There is a need for simulations and experimental testing to optimize the HEA, bond coat and ceramic TBC compositions for thermo-mechanical stability in gradient turbine environments. These include diffusion kinetics/interface reactions for better oxidation performance, strain to crack, thermo-mechanical fatigue, TBC spallation and high heat flux testing for performance on HEA alloys. Follow there, the development of materials-processing science (casting/additive manufacturing technologies) is necessary for component-level scale-up.

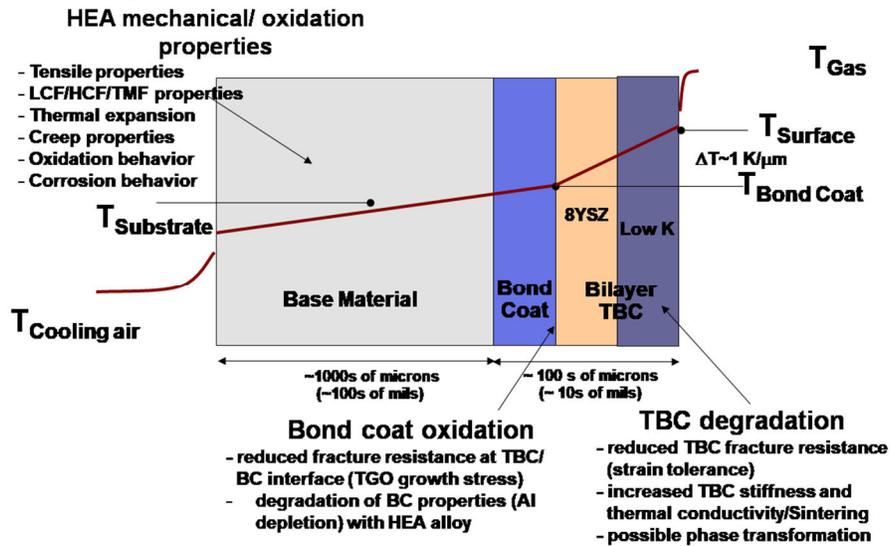

**Figure 38:** System compatibility studies for HEAs for turbine components.

## 4. Pareto Optimality: Concept of Key Importance for Multi-Objective Optimization

*Pareto efficiency* or *Pareto optimality* is a situation where no individual or preference criterion can be better off without making at least one individual or preference criterion worse off [172]. The *Pareto frontier* is the set of all Pareto efficient allocations [172]. As there often exist multiple Pareto optimal solutions for multi-objective optimization problems, what it means to "solve" such a problem is not as straightforward as for a conventional, single-objective optimization problem [173].

## 5. Primary Approaches to Multi-Objective Optimization

Primary approaches to multi-objective optimization consist of [174]:
1. Scalarization;
2. No-preference methods;
3. A priori methods;
4. A posteriori methods;
5. Interactive methods;
6. Hybrid methods.
For an example of a scalarization method, refer to Eq. (81).
The method of global criterion is an example of a no preference method. Here a problem of the form

$$\min \| f(\mathbf{x}) - z^{ideal} \|$$
$$\text{such that } \mathbf{x} \in \mathbf{X} \tag{93}$$

is solved, where the norm $\|\cdot\|$ can represent any $L_p$ norm, such as the Frobenius norm, and $z^{ideal}$ denotes the ideal objective vector, defined as $z_i^{ideal} = \inf_{\mathbf{x} \in \mathbf{X}} f_i(x) \ \forall \ i$, with $f_i(\mathbf{x})$ representing multi-objective function $i$ [173].

A posteriori methods may consist of ε-constraint methods, multi-objective branch-and-bound methods, normal boundary intersection methods, modified normal boundary intersection methods, modified normal constraint methods, successive Pareto optimization methods, directed search domain methods, Pareto surface generation for convex multi-objective instances, indirect optimization on the basis of self-organization, S-metric selection evolutionary multi-objective algorithms, approximation-guided evolution, reactive search optimization, Benson's algorithm for multiple objective linear programs and for multiple objective convex programs, multi-objective particle swarm optimization and subpopulation algorithm based on novelty [173].

The a priori methods require that sufficient preference information is expressed before the solution process [173]. One for example may be looking maximize the strength at high temperature, but the material designer (and the customer) are satisfied, as long as ductility at room temperature exceeds a given threshold.

An example of an interactive method involves maximizing

$$\text{Max [tensile strength } (T), \text{ yield strength } (T)] \tag{94}$$

This may give rise to multiple strength values that are optimal. Then based on the ductility, the designer could pick the most optimal one.

A hybrid method, in context with multi-objective optimization, consists of a combination of hybridized multi-criteria decision making and evolutionary multi-objective optimization [173].

6. <u>Sample Frameworks for Joint Optimization of Strength and Ductility for Low-Carbon Steels</u>

Reference [175] presents a comparative study between conventional goal attainment strategy and an evolutionary approach using a genetic algorithm conducted for multi-objective optimization of strength and ductility of low-carbon ferrite-pearlite steels. The optimization is based upon composition and microstructural relations of the mechanical properties established earlier through regression analyses. Upon finding that a genetic algorithm is more suitable for such a problem, some nice Pareto fronts are presented which provide a range of strength and ductility combinations useful for alloy design.

7. <u>Sample Joint Optimization Framework for Single-Crystal Nickel-Based Superalloys</u>

Reference [176] introduces a method for finding the optimum alloy compositions considering a large number of property requirements and constraints by systematic exploration of large composition spaces. The method is based on a numerical multi-criteria global optimization algorithm (a multi-start solver employing sequential quadratic programming), which is reported to deliver the exact optimum considering all constraints.

8. <u>Extensions to HEMs: Improvements through Incorporation of Physics-Based Models</u>

Metallurgists have studied the relationship between strength and ductility for HEAs as well as other alloys fairly extensively. There are models available in the literature not only for predicting the yield strength and UTS, but also for predicting the intrinsic ductility of alloys. And processing can be employed to control the extrinsic ductility.

As suggested by Table 8, it appears that an accurate model for joint optimization of strength and ductility may incorporate physics-based models (strengthening mechanisms), such as the grain boundary strengthening mentioned in Section 4.5, and possibly account for heat treatment effects.

According to Reference [177] , most methods for producing stronger materials are based on controlling defects to impede the motion of dislocations. Dislocation processes govern the deformation regime in which the quintessential attributes of strength and ductility of alloys can be tuned. The authors present dual-phase nano-structuring as a route to designing magnesium alloys of high strength.

Similarly, according to Reference [178], strengthening mechanisms for alloys traditionally involve the controlled creation of internal defects and boundaries so as to obstruct dislocation motion. Such approaches invariably compromise ductility. The authors outline an approach to optimize strength and ductility by identifying three essential structural characteristics for boundaries: coherency with surrounding matrix, thermal and mechanical stability, and smallest feature finer than 100 nanometers.

Reference [179] shows that an FeAl-type brittle but hard intermetallic compound (B2) can be effectively used as a strengthening second phase in high-aluminum low-density steel, while alleviating its harmful effect on ductility, by controlling its morphology and dispersion. The authors found that alloying of nickel catalyzes the precipitation of nano-meter-sized B2 particles in the face-center cubic matrix of high-aluminum low-density steel during heat treatment of cold-rolled sheet steel.

Reference [180] reveals that the precipitation of hard $\sigma$ and $\mu$ intermetallic compounds tremendously strengthens $CoCrFeNiMo_{0.3}$ HEA, without causing serious embrittlement. The hard intermetallic phases strengthen the ductile $CoCrFeNiMo_x$ HEAs.

Nanocrystalline metals, i.e., metals with grain sizes less than 100 nm, have strengths exceeding those of coarse-grained or even alloyed metals. But nano crystalline materials often exhibit low tensile ductility at room temperature, which limits their practical utility. Reference [181] describes a thermomechanical treatment of Cu that results in a bimodal grain size distribution, with micro-meter-sized grains embedded inside a matrix of nanocrystalline and ultrafine (< 300 nm) grains. The inhomogeneous microstructure induces strain-hardening mechanisms, which stabilize tensile deformation, leading to high tensile ductility.

## 5. Statistical Models

### 5.1. Key Take-Aways

1. Due to the complexity of fatigue behavior of an AM HEA part, a comprehensive toolset, based on an ICME framework, may be needed to predict fatigue strength and fatigue life in AM metallic components.
2. In an effort to account for the complexity in the underlying process, we present a ML framework for predicting fatigue properties. The framework is a generalization (augmentation) of a Statistical Fatigue Life model of [63].
3. We present a prediction framework capable of predicting the fatigue life of HEA components fabricated by additive manufacturing with accuracy superior to that of existing tools.
4. The prediction framework systematically accounts for the wide variety of sources (over 100) that can impact the fatigue life of AM HEA components.
5. Through sensitivity analysis, provided by the augmented Statistical Fatigue Life model, the framework has the ability to provide feedback for manufacturing. The framework is capable of providing feedback on the impact that variations in given input parameters have on the fatigue life.
6. We also present a methodology for predicting fatigue properties of AM HEA components, one relating material and processing induced imperfections. The predictive methodology for fatigue properties of AM HEA components relates material and processing induced imperfections.

### 5.2. Key Advantages of ML for Prediction of Fatigue Life of AM HEA Components

1. *ML can help address (avoid) inaccuracies in fitting traditional models to real-world fatigue data. Traditional models are prone to under- or over-fitting, and can lead to error magnification during integration* [61].
2. *ML can help in terms of accounting for all the sources that can impact fatigue life of AM HEA components.* As noted above, it has been reported that *over 100 different process parameters* can affect the fatigue life of AM HEA components [63]. Traditionally, parametric models have been designed to account for key sources, but not for all the sources [61].
3. To our knowledge, existing software tools cannot predict the material properties or account for AM [61].
4. Given the Statistical Fatigue Life model, or its augmented version, one can play with the input parameters such as to maximize fatigue life (optimize for fatigue life), and provide feedback on manufacturing process [61].

### 5.3. Statistical Modeling Compared and Contrasted to Physics-Based Modeling

For certain applications, the selection of the prediction model may be based on a probabilistic, not physics-based, derivation. The assumption of a constant failure rate, i.e., of $1/\lambda$ representing the time to failure, results in an exponential function (a standard Weibull distribution) [61].

At times, one may assume independence between events, and invoke the Central Limit Theorem of statistics. In case of alloy design, the parameters may tend to be inter-related, and hence, we may not be able to invoke the Central Limit Theorem. One may need to derive dependent distributions, based on the dependency relations established, per Figure 23 [61].

### 5.4. Statistical Fatigue Life Model for Analytical Representation of Stress/Life (S/N) Curves

A commonly used analytical representation of S/N curves, like the ones presented in Figure 39, is given by [24, 61, 63]:

$$N(\sigma) = c\sigma^{-d}. \qquad (95)$$

Here, $\sigma$ represents the applied stress, $N(\sigma)$ is the expected cycles to failure at the stress, $\sigma$, and $c$ and $d$ are positive material parameters. Taking the logarithm of the S/N relation given by Eq. (95) yields [24, 61, 63]:

$$\log[N(\sigma)] = a + b * \log(\sigma) \qquad (96)$$

where $a = \log(c)$ and $b = -d$. Eq. (96) describes the relationship between the mean-log of the fatigue life and the applied stress. In order to account for scattering observed in the fatigue-life experiments, a regression model is formulated by adding a random error term [24, 61, 63]:

$$\log[N_{ij}] = \mu_i(\sigma_{ij}) + \varepsilon_{ij} = a_i + b_i * \log(\sigma_{ij}) + \varepsilon_{ij} \qquad (97)$$

Here, $i$ indexes the process categories available, but $j$ the data points available for each process category. $N_{ij}$ is the $j$-th data point under condition $i$ collected at the stress, $\sigma_{ij}$. $\varepsilon_{ij}$ is a random error term, which is assumed to follow normal distribution with mean zero and standard deviation, $s_i$. $\mu_i(\sigma)$ is the mean (also the median) logarithm transformed fatigue life at stress, $\sigma$, under condition $i$. The collected fatigue life data is denoted by $\{N_{ij}, \sigma_{ij}, \delta_{ij}\}$, where $\delta$ is a runout indicator, defined as $\delta = 1$ for a failure observation and $\delta = 0$ for a runout. The likelihood function for the observed fatigue life data under condition $i$ is then given by [24, 61, 63]:

$$L_i(a_i, b_i, s_i) = \prod_{j=1}^{m_i} \left[ \phi \left( \frac{\log(N_{ij}) \cdot a_i \cdot b_i \cdot \log(\sigma_{ij})}{s_i} \right) \right]^{\delta_{ij}} \times \left[ 1 - \phi \left( \frac{\log(N_{ij}) \cdot a_i \cdot b_i \cdot \log(\sigma_{ij})}{s_i} \right) \right]^{1-\delta_{ij}} \tag{98}$$

where $\phi(\cdot)$ and $\Phi(\cdot)$ represent the probability density function and the cumulative distribution function of the standardized normal distribution, respectively [61].

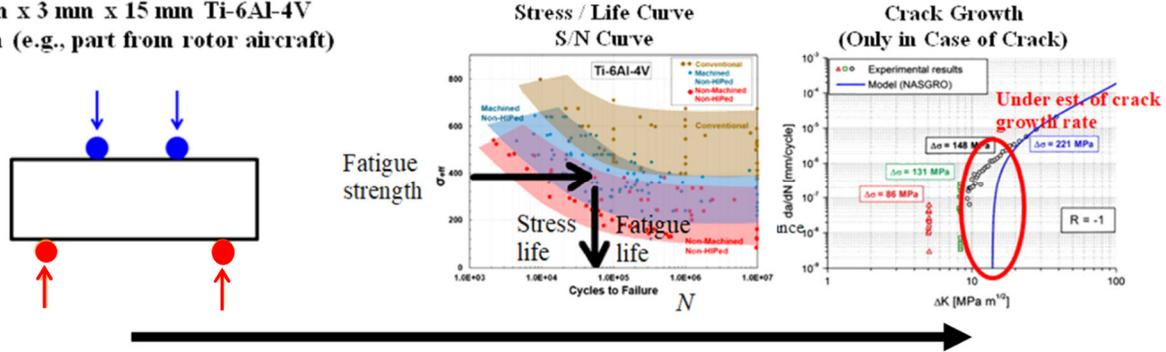

**Figure 39**: The Figure outlines a classical approach for analyzing fatigue life. The figure to the left shows the stress applied (from 4-point testing). The figure in the center shows the fatigue life obtained from the stress applied. The figure to the right shows the fatigue life remaining (curve fit to data). The classical NASGRO model can yield over- or under-estimation of the fatigue life, due to over- or under-fitting [182]. This can result in error magnification during integration. Machine learning can account for all the parameters affecting fatigue life of AM components, not only those modeled by the NASGRO equation, and hence yield more accurate results [61].

### 5.5. Augmentation of the Statistical Fatigue Life Model

The Statistical Fatigue Life Model can be augmented such as to include additional input parameters. In case of $N$ independent events, the probability of failure, $P(\text{fail})$, can be formulated using Poisson distribution [61]:

$$P(fail) = \prod_{i=1}^{N} P_i(fail) = \prod_{i=1}^{N} f_i exp(-c_i) = f_1 * f_2 * \cdots f_N * exp(-\prod_{i=1}^{N} c_i). \tag{99}$$

With this in mind, it makes sense to model the augmented version of the Statistical Fatigue Life model (Eq. (100)) as [61]:

$$N(\sigma, p_1, p_2, p_3, \cdots, p_N) = f_1(p_1, p_2, p_3, \cdots, p_N) \sigma^{-f_2(p_1, p_2, p_3, \cdots, p_N)}. \tag{100}$$

Here, $p_1$, $p_2$, $p_3$, ..., $p_N$ model the input parameters impacting the fatigue life of AM metallic components. For a specific parameter selection, refer to Table 1. We are assuming that multiple effects cause failure and that these effects are close to independent. The function, $f_1(\cdot)$, models a prior knowledge and the function $f_2(\cdot)$ conditional probabilities. For dependent events, one can apply a Bayesian model, with the same definition of $f_1(\cdot)$ and $f_2(\cdot)$ [61].

In case of independent events, it makes sense to apply the direct linear regression to assess $f_1(\cdot)$ and $f_2(\cdot)$. But in case of coupled failure modes, $f_1(\cdot)$ and $f_2(\cdot)$ may consist of complex Bayesian functions. We may not know these functions, and one may need to apply regression analysis. These functions can be hard to derive. *Therefore, this is where ML comes in.* One can apply neural networks or support vector machines to effectively deduce these functions from the data. Even if $100 - 200$ parameters impact the fatigue life of AM metallic components, this is still a relatively small set by the standards of ML [61].

The model of Eq. (100) should be able to predict the time to failure with better accuracy, for reasons similar to the quadratic regression model in Figure 28 providing accuracy superior to that of the linear regression. The main objective is to quantify the accuracy and reduce the error bars (improve the accuracy of the fatigue prediction) [61].

One of the primary advantages of the model in Eq. (100) involves the ability to:

1. Determine the top factors that contribute to fatigue life, and
2. Provide feedback, through sensitivity analysis.

By estimating [61]

$$\frac{\partial N(\sigma, p_1, p_2, p_3, \cdots, p_N)}{\partial p_i} \qquad i = 1, 2, 3, ..., N \tag{101}$$

one can assess the contribution of the input parameter, $i$, to the fatigue life. Such feedback may yield significant, tangible benefits. The dominant factors contributing to the reliability may involve something in manufacturing. Maybe it is something involving the material properties. Maybe a smaller grain size will improve the fatigue life. Our model may be able to quantify for how long to expose the laser, at which temperature, with which grain size, and translate into fatigue life. Such information can be of great value, and may lead to iterative refinements. The fatigue-prediction toolset can advise on, say, how to change a given material property. Once the property has been changed, and new data generated, the data can be fed back into the model and the impact assessed. It is our understanding that small variations in the atomic % of Sulphur or Carbon can have significant impacts on fatigue life of the stainless steel [61].

## 5.6. Example 6: Prediction of Fatigue Life (Stress Life or Strain Life) and Crack Growth

1. <u>Prediction of Fatigue Life Remaining (S/N Curve), in Absence of Cracks</u>

*1.1 Objective*

For the triplet [61]

$$\text{(process, stress applied, cycles to failure),} \tag{102}$$

the goal is to accurately infer the "process" parameter from the combination [61]

$$\text{(stress applied, cycles to failure),} \tag{103}$$

in order to properly differentiate between "process" categories [61].

Similarly, we can look to estimate "cycles to failure" (fatigue life) from the combination (process, stress applied), or "stress applied" (fatigue strength) from the combination (process, cycles to failure) [61].

*1.2 Preferred Method*

Our preferred approach to predicting the fatigue life is based on the augmented Statistical Fatigue Life model [Eq. (100)]. It consists of the following steps:

1. Inferring the parameters, that impact the fatigue life ($p_1, p_2, p_3, \ldots, p_N$), from the input data, using a model similar to the one used for estimating the endurance limit or UTS [see Eq. (80) or Eq. (81)], for example multi-variate linear regression;

2. Predicting variations for the individual parameters, using a model such as

$$p_1 = g_1[\text{UTS, process, defect (process), grain (process), microstructure (process)}, T. \ldots],$$
$$p_2 = g_2[\text{UTS, process, defect (process), grain (process), microstructure (process)}, T. \ldots],$$
$$p_3 = g_3[\text{UTS, process, defect (process), grain (process), microstructure (process)}, T. \ldots], \tag{104}$$
$$\ldots$$
$$p_N = g_N[\text{UTS, process, defect (process), grain (process), microstructure (process)}, T. \ldots];$$

3. Comparing and contrasting the new parameter set with physics-based intuitions (see Figure 40); and

4. Reconstructing the S/N curve based on the new set of parameters, again using [Eq. (100)].

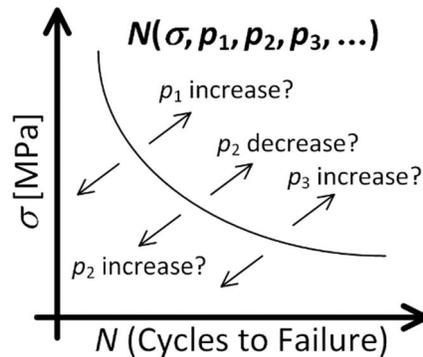

**Figure 40**: Prediction of a stress / life curve.

*1.3 Alternative Method*

The alternative method, motivated by Figure 39, is analogous to the method used for prediction of the endurance limit (see Eq. (76) – Eq. (80)). The main difference is that here we are taking the stress applied as an additional input parameter. The method will essentially be the same as for predicting the endurance limit, except here we are doing separate prediction at each stress level [61].

*1.4 Expected Results*

Defects are easier to introduce with AM than with casting. Hence, we expect more scatter in S/N data for AM than casting [61].

2. <u>Prediction of Fatigue Life Remaining, Given a Crack</u>

Fatigue life, defined in terms of the number of cycles, *N*, in the presence of cracks, is usually estimated as [61]:

$$N = \int_{a_0}^{a_f} \left(\frac{dN}{da}\right) da. \tag{105}$$

According to the NASGRO equation by Forman and Mettu [183],

$$\frac{da}{dN} = (C*F)^{-1}*\Delta K^{-m}*\frac{\left(1-\frac{K_{max}}{K_c}\right)^q}{\left(1-\frac{\Delta K_{th}}{\Delta K}\right)^p} = (C*F)^{-1}*\Delta K^{p-m}K_c^{-q}\frac{(K_c-K_{max})^q}{(\Delta K-\Delta K_{th})^p}. \tag{106}$$

Here, $C$ represents a crack growth constant, $F$ a crack velocity factor, $\Delta K$ a stress intensity factor range, and $K_{max}$ a maximum stress intensity factor. Furthermore, $m$ denotes Paris exponent and $\Delta K_{th}$ the threshold stress intensity factor range for crack propagation [61].

*By applying ML to estimating the number of cycles, N, directly, one may avoid error magnification that otherwise could occur during the integration process, due to over- or under-fitting* [61].

## 6. Other Applications of ML, AI and Data Analytics to Material Science or Manufacturing

The coverage presented here is the continuation of the exposition from Section 4.

### 6.1. ML or AI for Real-Time Quality Control in Powder Bed AM – Multi-Beam Approach

1. <u>Motivation Revisited</u>

As noted in Section 1.2, ML can be used to better distribute, monitor and control the processing energy in a laser metal powder bed fusion AM systems, for the purpose of real-time process monitoring and control towards producing high-quality, defect-free HEA parts with build periods comparable or shorter than present ones.

Despite continued progress in AM technologies, AM HEA parts still require several trial and error runs with post-processing treatments and machining to optimize builds, reduce defects and residual stresses, and meet tolerances. AM still lacks a stable process that can produce consistent, defect-free parts on a first time basis due to inability to reliably predict the optimal trajectory in the multidimensional process parameter space due to the inherent spatiotemporal variability in process parameters and the chaotic nature of the AM process.

2. <u>General Approach to AM Parameter Optimization</u>

Key steps can include systematically changing the key parameters (the laser power and travel speed during deposition, powder feed rate and increment, number of laser tracks for each patch and overlap value, repeat times and laser power during re-melting, the powder size, powder shape, powder distribution and purity), as shown in Figure 41.

3. <u>Challenges with Powder Bed AM</u>

Factors that make AM, and in particular laser powder bed fusion AM, such a challenging manufacturing process are:
1. The smallness of the laser processing volume and rapid melt time when compared to the final part size and build time, respectively, and the associated process variabilities that result from them.
2. The intrinsic variability of all the powder bed physical (mass, heat capacity, thermal conductivity, emissivity, reflectivity) and chemical (composition, oxidation state, wetting angle) properties that compound to the above-mentioned process variabilities.
3. The large power densities required to process the powder bed and the associated large heating rates and thermal gradients, which when combined with the above-mentioned variability makes it difficult to control the microstructure of the processed volume.

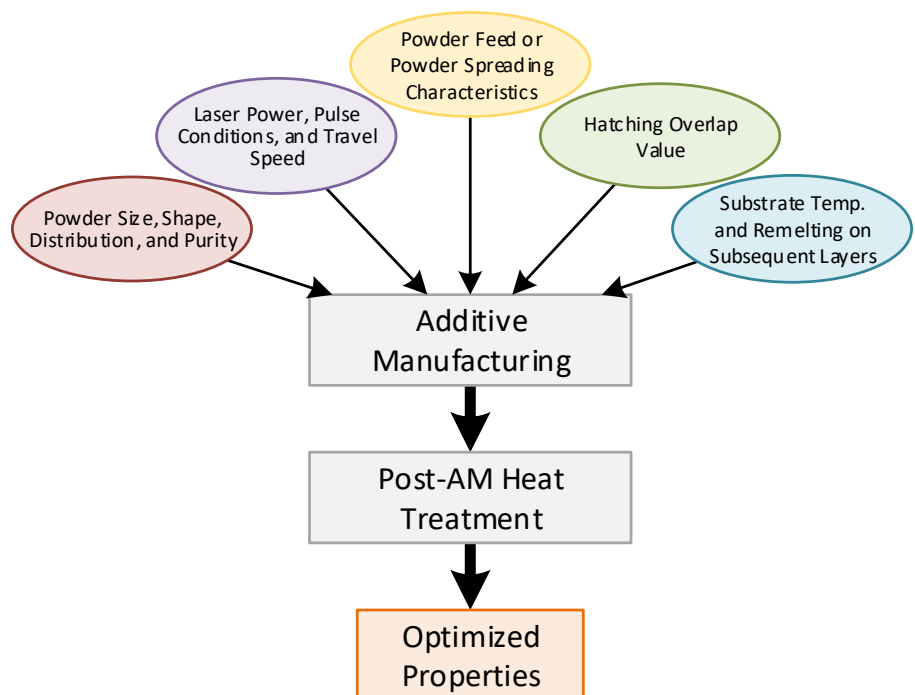

**Figure 41**: High-level overview over optimization of an AM process parameters for HEAs [61].

4. The chaotic nature of the AM process that results from combining the small spatial and temporal scales described above with the high energy densities required for melting the powder, which makes it difficult to reliably predict the process trajectory in the multi-parameter process space before the build process starts and virtually impossible to control it in real time.

5. The large number of process parameters (in some cases over 100) that can affect the outcome of the AM process and make it almost impossible to model with physics-based models.

6. The non-symmetric deposition of the processing energy that results from raster scanning a single laser beam over the powder bed which leads to non-uniform heating/cooling rates, thermal gradients, residual stresses and part defects and distortion.

4. <u>Addressing the Challenges with a Real-Time AI Controller and a Multi-Beam Approach – Intelligent AM</u>

Most of these challenges can be alleviated by better controlling and distributing the laser energy at and around the melt pool area and/or the processing part surface area combined with real-time monitoring of the same area or beyond and by intelligently linking the laser energy control parameters with the process monitoring sensors to learn and adapt to the continuously evolving environment. Distributing the process energy intelligently at and around the melt pool will help reduce the process variability, the powder bed physical property variability, the heating/cooling rates and the thermal gradients. For example, it may make sense to pre-heat the powder ahead of the melt-pool without melting it, to reduce the heating rates and thermal gradients later during melting. Doing so may allow processing the powder faster, and reducing the build time, while at the same time reducing evaporative recoils, ejecta and denudation effects (which induce defects in the final part). Monitoring the temperature profile around the melt-pool area can be used to adjust the distributed laser energy control parameters (power levels and distribution) in real time in a system where the temperature profile is directly linked to the heating source control parameters via an AI processor. Similar improvements can be achieved by intelligently distributing the laser processing energy over the entire part surface, while monitoring the temperature evolution over the same area [61].

The intelligent AM system links the actuators controlling the laser energy distribution over the powder bed with the sensors that monitor the temperature distribution and/or other relevant process parameters over the powder bed using a real-time AI controller. Specifically, the intelligent AM system employs a multi-beam strategy to customize the melt pool temperature and the energy distribution over the powder bed, as shown in Figure 42 [61].

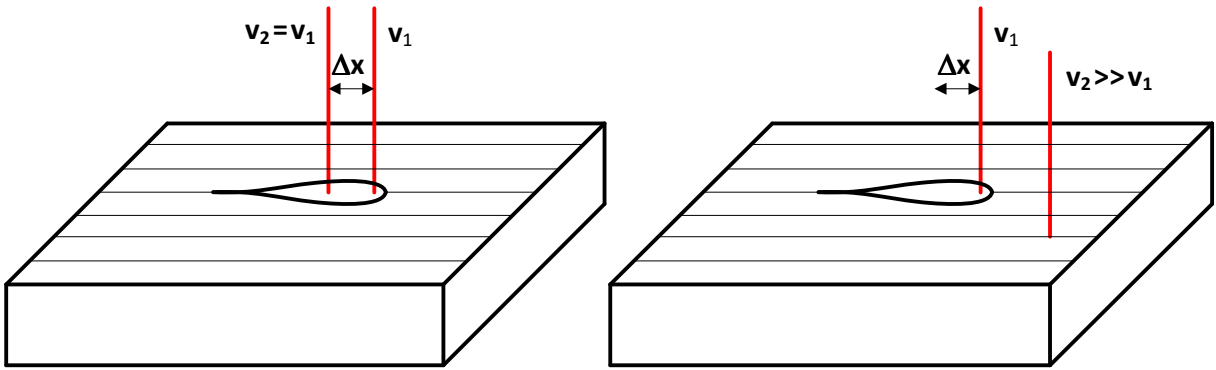

**Figure 42**: The Figure to the left shows an offset strategy, using two beams in a defined offset at the same speed. The Figure to the right shows a heating strategy, one that uses a melting beam of scan speed v1 and a heating beam of v2 which is much faster and irradiates the layer multiple times while beam 1 melts the tracks [61].

## 6.2. ML or AI for Real-Time Quality Control in a Generic AM System – Reinforcement Learning

1. <u>Reinforcement Learning in Context with Robotics and Aerostructure Manufacturing</u>

In the aerospace industry, robotics has been one of the critical workforces for aerostructure manufacturing, because of precision and accuracy. Digital factories have been implemented to design and test the manufacturing process virtually. As noted in Section 2.2, reinforcement learning is a type of machine learning that enables software agents to learn and take actions in an environment by maximizing rewards. RL has been employed to learn ATARI games by Google Deepmind [184], to AM robotic hands [185], or to in-situ AM quality control [186], which is the primary application of interest in this sub-section.

2. <u>More on the Challenge Addressed</u>

The complexity of manufacturing processes varies a lot, depending on the aerostructure manufacturing task involved. Many manufacturing processes still require subject matter experts to design the process and program (optimize). Additive manufacturing typically requires designers and operators to manually go through the details of figuring out which AM process configuration is needed, which configuration parameters are needed, and identify if there are conflicts. Through incorporation of machine (reinforcement) learning, the system can identify the best parameter configuration for additive manufacturing of HEA parts by itself.

3. Application of Reinforcement Learning to In-Situ Quality Control of a Generic AM System

Figure 43 presents a system for designing and testing RL in an industrial digital factory environment, where the RL model controls the robotics to learn the best way to manufacture the aerostructure. Upon proper implementation, the model can be deployed to the aerostructure manufacturing process to support automation and reduce human engagement demand.

In Figure 43, the RL system is trained in a supervised manner [187] with data collected from previous tasks, from industry or from the literature. The intent is to train and test the reinforcement learning for classification to determine the quality by exploring and exploiting the space to maximize its reward. RL allows for interpretation of the received data in terms of quality. The RL agent starts at an initial state and searches through the space with policies that map the states to actions. This enables RL to self-learn with minimum supervision and to have efficient adaption to new AM conditions for real-time in-situ quality monitoring. Upon proper implementation, the model can be deployed to the aerostructure manufacturing process to support automation and reduce human engagement demand.

Assuming a WAAM system in Figure 43, different wire feed rates, robot arm speeds, arc characteristics, and other parameters can potentially affect the quality of the part produced. Thus, implementing real-time in-situ quality control can help reduce the number of defects as well as the cost of the production.

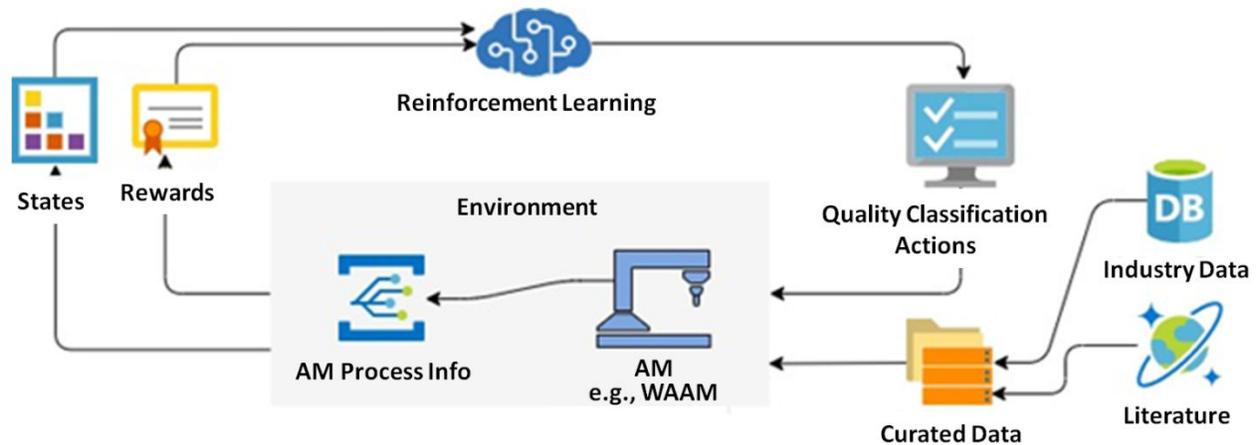

**Figure 43**: Reinforcement learning for the in-situ real-time AM quality control.

## 6.3. ML in Low-Data Environments

1. Estimates of Sample-Complexity

In [104], the authors show that the sample-complexity to learn convolutional neural networks (CNNs) and recurrent neural networks (RNNs) scales linearly with their intrinsic dimension and this sample-complexity is much smaller than for their Fully-Connected Neural Network (FNN) counterparts. For both CNNs and RNNs, the authors also present lower bounds showing their sample complexities are tight up to logarithmic factors.

2. Surrogate Modeling

Surrogate modeling is a method for iterative model refinement, as additional data becomes available. It is useful, when the function in Eq. (6) is costly to evaluate, when the $y$ values are not easily measured, and hence when one does not have many of these values. Figure 44 demonstrates a simple example of iterative refinement through surrogate modeling, where $y$ for example can be taken to represent tensile strength but $x$ a given composition. Initially, when one only has two (2) data points for $y$, as shown in the figure to the left, the best estimate for the function $f(\cdot)$ is a straight line. This is the surrogate model, in this case. It advises on the next composition to try ($x = (a + b)/2$). Next, when one has three (3) data points for $y$, the best estimate (surrogate model) for the function, $f(\cdot)$, is the polynomial fit shown in the middle figure. Again, it advises on the next composition to try ($x = d$). Then, when one has four (4) data points for $y$, the best estimate (surrogate model) of the function $f(\cdot)$ is the polynomial fit shown in the figure to the right. Yet again, it advises on the next composition to try ($x = e$). The iterative refinement of surrogate modeling lends itself well to the sequential learning shown in Figure 8.

To address data limitations, in application of ML to material science, authors such as Joly et.al. [188] have suggested using multi-source learning or surrogate assisted optimization. These authors have suggested arranging the data in a hierarchy of fidelity. They have claimed that one can sample low-fidelity data and use to improve a prediction model (reduce variance, but not improve a mean).

In [189], the authors combine a machine learning surrogate modeling with experimental design algorithms to search for compositions with large hardness in a model AlCoCrCuFeNi system. The authors fabricated several alloys with hardness 10% higher than the best value in the original training data set. The authors found, that a strategy using both the composition and descriptors, based on knowledge of the properties of HEAs, outperformed a strategy based on the compositions alone.

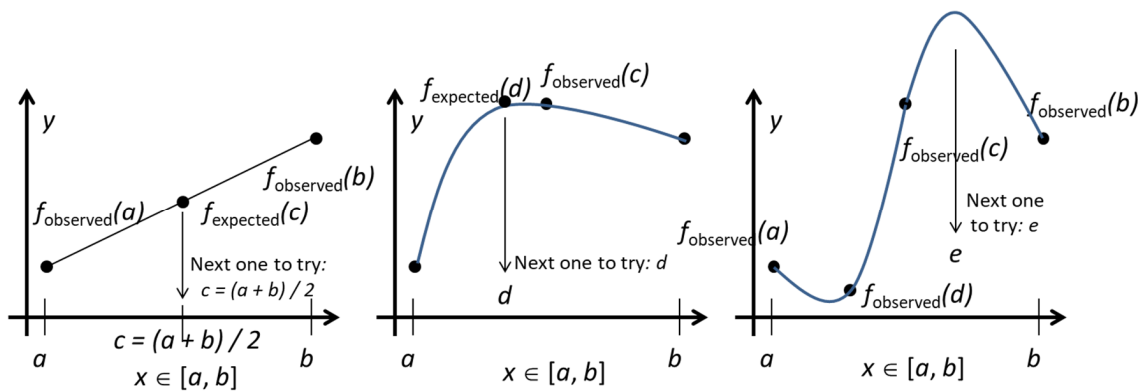

**Figure 44**: Simple example of iterative refinement through surrogate modeling.

## 6.4. ML for Prediction of Solid Solution Phases

In Reference [190], Qi, Cheung and Poon present a phenomenological method for analyzing binary phase diagrams, for the purpose of predicting HEA phases. Their underlying hypothesis is that the HEA structural stability is encoded within the phase diagrams. Accordingly, the authors introduce several phase-diagram inspired parameters and employ machine learning to classify 600+ reported HEAs based on these parameters. Compared to other large database statistical prediction models, this model gives more detailed and accurate phase predictions. Both the overall HEA prediction and specifically single-phase HEA prediction rate are above 80%. To validate their method, the authors demonstrated its capability in predicting HEA solid solution phases with or without intermetallics in 42 randomly selected complex compositions, with a success rate of 81%. Their search approach can be employed with high predictive capability to interact with and complement other computation-intense methods such as CALPHAD in providing an accelerated and precise HEA design.

In [191], Pei, Yin, Hawk, Alman and Gao show that ML using Gaussian Process Classification can predict new, important features, such as molar volume, bulk modulus and melting temperature, that impact materials properties. The example in Figure 45 shows how ML can predict solid solution formation, based on 1252 alloy datasets. According to Figure 45 (left), simple correlation between elemental properties and alloy phases does not seem to capture the mapping from the elemental properties to the alloy phases that well. According to Figure 45 (right), ML is capable of reliably predicting solid solution formation among millions of combination of MPEAs only using properties of pure elements.

Other work on application of ML for phase prediction of HEAs, or identification of Heusler compounds, includes [83, 191-193].

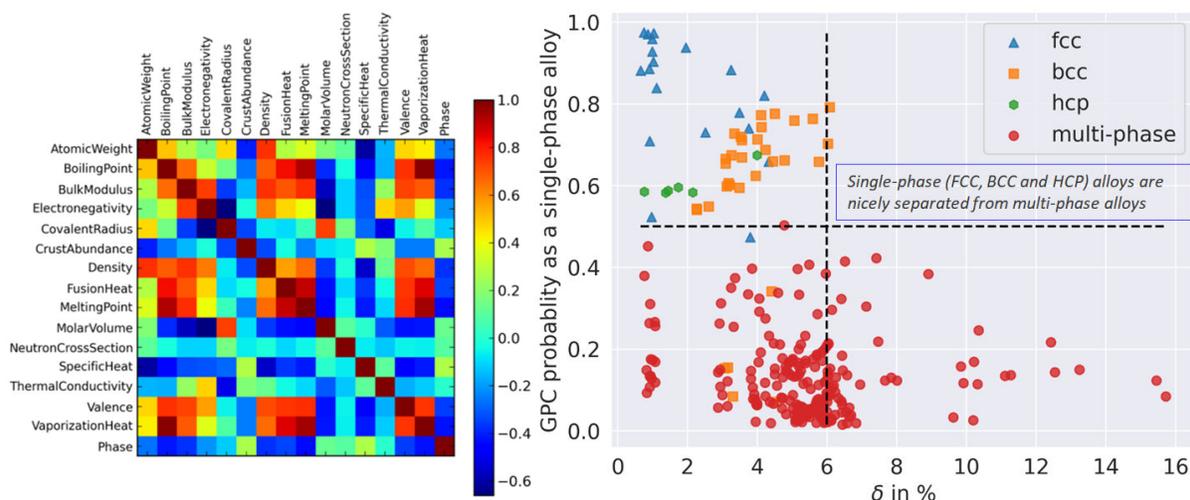

**Figure 45**: Machine learning using Gaussian process classification for the prediction of solid-solution formation [191]. Reproduced with permission.

## 6.5. Other Applications of Machine Learning to Optimization of Mechanical Properties

1. <u>Sequential Learning through Random Forrest Modeling with Well-Calibrated Uncertainty Estimates</u>
In [76], Ling et al. present a methodology for identifying materials exhibiting desirable characteristics, one that can guide the practitioner to test the most promising candidates early. The data-driven models presented incorporate uncertainty analysis, so that new experiments can be proposed based on a combination of exploring high-uncertainty candidates and exploiting high-performing regions of the parameter space.
In [87], Ling et al. demonstrate the effectiveness of machine learning methods for three different alloy classes: aluminum alloys, nickel-based super-alloys, and shape memory alloys. For the first two alloy classes, machine learning models are built to predict the mechanical properties, based on the composition and processing information. In the case of shape memory alloys, a model is trained to predict the austenite to martensite transformation temperatures. The authors also provide well-calibrated, heteroscedastic uncertainty estimates with each prediction.

2. <u>Canonical Component Analysis and Genetic Algorithms for Prediction of Hardness</u>
In [194], Rickman et al. outline a supervised learning strategy for efficient screening of HEAs, one that combines two complementary tools: (1) a multiple regression analysis and CCA and (2) a genetic algorithm with a CCA-inspired fitness function. The authors implement this procedure using a database for which mechanical property information exists and highlight new alloys having high hardness. Their methodology is validated by comparing predicted values of hardness with alloys fabricated by arc melting, identifying alloys exhibiting very high values of measured hardness.

3. <u>Artificial Neural Networks for Prediction of Material Hardness</u>
In [195], the authors utilize an ANN to predict the composition of HEAs based on non-equimolar AlCoCrFeMnNi in order to achieve the highest hardness in the system. A simulated annealing algorithm is combined with the ANN to optimize the composition. A bootstrap approach is adopted to quantify the uncertainty of the prediction. The authors demonstrate that, by applying ML, new compositions of AlCoCrFeMnNi-based HEAs can be obtained, exhibiting hardness values higher than the best literature value for the same alloy system.

4. <u>Gradient Boosting Tree Algorithm for Estimating Elastic Moduli</u>
In [196], machine learning modeling was implemented to explore the use of fast and computationally efficient models for predicting elastic moduli of HEAs. ML models, that were trained on a large data set of inorganic structures, were shown to make accurate predictions of elastic properties for the HEAs. The ML models also demonstrated that the bulk and shear moduli exhibited dependence on several material features, which can act as guides for tuning the elastic properties of HEAs.

## 6.6. Other Applications of Machine Learning to Optimization of AM Processes

Reference [197] addresses non-dimensionalization of variables to enhance machine learning in additive manufacturing processes. The method includes applying a transform to values of at least two variables of the process data to generate a dimensionless parameter having a parameter value corresponding to each measurement of the physical system for at least two variables.

Other related work specifically addressing the prediction of properties of additively manufactured components includes [198-200]. For relevant papers on HEA design for additive manufacturing, refer to [201-203]. Related work specifically addressing applications of ML to real-time quality control for AM systems includes [199, 204, 205].

## 7. Future Work

The framework presented in this chapter assumes the selection of an optimization technique suitable for the application at hand and the data available. The central themes, captured in Figure 46, pertain to:

1. Joint optimization of material design and manufacturing;

2. Applying machine learning to efficiently explore the huge composition (feature) space involved;

3. Using physics-based insights for guiding the optimization process, as opposed to running brute-force optimization. To this end, there are additional specifics to be formulated. We have presented initial, joint feature lists in Table 6 and Table 7. But the joint optimization is yet to be carried out, and the feature lists are yet to be refined. Once the joint optimization problem has been formulated in a closed form, and the pertinent input data has been collected, organized, and curated, standard packages explore the multi-objective optimization, such as matlab [206], jMetal [207], openGA [208], PAGMO (or pygmo) [209], LINGO [210], GAMS [211], or MultiJuMP [212], may be employed for the initial solution. But a key to success may involve the specifics of how physics-based insights, such as from [190], are incorporated into the general problem statement, for purpose of providing initial compositions (starting points), as the closed-form (application specific) problem formulations are crafted.

Furthermore, specifically in regards to the physics-based models (Examples 1 and 2), one can develop a more accurate prediction model for the tensile strength, by explicitly accounting for process conditions and defect levels, in the prediction model, in addition to the compositions. Table 13 lists some specifics related to the heat-treatment process applied. Hence, one already has access to much of the input data needed for extending the prediction model and improving its accuracy. The physics-based models, together with the information related to the nearest neighbors in the parameter space, can also be utilized for controlling the step size used to for the regression analysis (e.g., for the extrapolation in Figure 29 - Figure 31). Along these lines, the prediction model can be extended such as to accommodate explicitly the parametrized representation of the heat-treatment process, defect levels, grain characteristics or microstructure.

The variance analysis for the extended prediction model can be correspondingly enhanced in accordance with Eq. (29). One can account for the measurement errors for the UTS and accordingly devise an approach for the data curation. As long as enough strength data is available, for a range of compositions and temperature, the prediction model can be extended such as to predict the tensile strength, as a function of temperature, for a given composition. Furthermore, assuming that one has enough input data, and data of the right type, there are many questions that one can address, such as how to apply suitable optimization techniques to find the heat-treatment process, along with the composition, which yields the best fatigue performance.

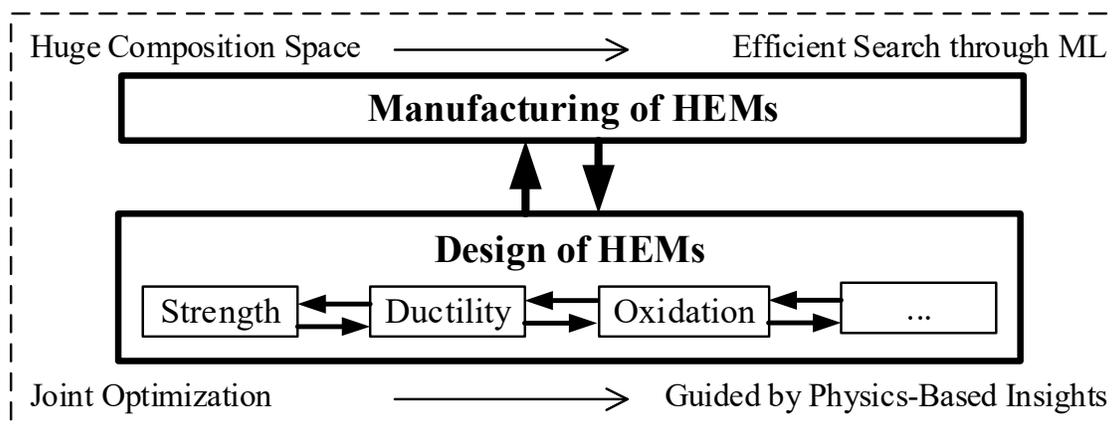

**Figure 46**: Machine learning applied to the design and manufacturing of HEMs through the joint optimization.

## 8. Conclusions

A fundamental challenge in the design or manufacturing of high-entropy materials involves the huge size of the feature (composition) space. In this chapter, we have reviewed techniques for effectively exploring the large feature space in search for HEMs exhibiting specific properties of interest. We also looked at associated methods for manufacturing and for quality control of AM HEA components.

In Section 2, we presented machine learning as a non-linear correlation technique. As opposed to applying ML, narrowly defined in terms of neural networks, Bayesian graphical models, support vector machines or decision trees, to the identification of high-entropy alloys or composites of interest, we reformulated the task of deriving the system model in the broader context of engineering optimization. For determining the association between the source factors and observed properties, we recommended selecting an optimization technique suitable for the application at hand and the data available.

For accurate prediction, it is of great importance to properly understand and account for, the sources contributing to variations in the quantity predicted. In Section 3, we presented a framework for "forward" predicting the observed properties. "Backward" prediction (identification of candidate compositions yielding the specific properties of interest) was accomplished through an "inverse" design framework, one that identifies the candidate compositions, for the next round of testing, based on the property specifications and design goals.

Section 4 outlined methods for employing physics-based models, i.e., models that account for physical dependencies, and factor in the underlying physics as *a priori* information, during the prediction process. There can be significant benefits derived from combining ML with physics-based modeling approaches for alloys and composites, for improved prediction accuracy. Such modeling approaches, which may rely on thermo-dynamics, first principle effects, empirical rules or mesoscale models, can offer physical insight as unexplored realms of the composition space are investigated.

Due to the complexity of fatigue behavior of AM HEA components, a comprehensive toolset, based on an ICME framework, may be needed to predict fatigue strength and fatigue life of such components. In an effort to account for the complexity in the underlying process, we presented in Section 5 a framework, based on machine learning, for predicting the fatigue properties. The framework was a generalization of the Statistical Fatigue Life model of [63]. The ML prediction framework can systematically account for the great variety of sources that can impact the fatigue life of AM HEA components.

## 9. Acknowledgement

XF and PKL very much appreciate the supports from (1) the U.S. Army Office Project (W911NF-13-1-0438 and W911NF-19-2-0049) with the program managers, Drs. Michael P. Bakas, David M. Stepp, and S. Mathaudhu, and (2) the National Science Foundation (DMR-1611180 and 1809640) with the program directors, Drs. Judith Yang, Gary Shiflet, and Diana Farkas. BS very much appreciates the support from the National Science Foundation (IIP-1447395 and IIP-1632408) with the program directors, Drs. G. Larsen and R. Mehta.

Prof. Duckbong Kim of Tennessee Technological University is thanked for sharing valuable information related to the interrelationship between sources impacting properties of AM components and the underlying physics. Prof. Wen Chen of the University of Massachusetts furthermore contributed a diagram which became the precursor to Figure 5.